\documentclass[prd,twocolumn,aps,letterpaper,amsmath,amssymb,preprintnumbers,showpacs,superscriptaddress,floatfix,longbibliography,nofootinbib]{revtex4-2}
\usepackage{array}
\usepackage{graphicx}
\usepackage[export]{adjustbox}
\usepackage[caption=false]{subfig}
\usepackage{tikz}
\usepackage[normalem]{ulem}
\usepackage{multirow}
\usepackage{dsfont} 
\usepackage{ytableau}
\usepackage{diagbox}
\ytableausetup{smalltableaux,aligntableaux=center}
\usetikzlibrary{tikzmark}
\usetikzlibrary{calc}
\usepackage{lipsum}
\allowdisplaybreaks 
\usepackage[pdfencoding=auto,hypertexnames=true]{hyperref}
\usepackage[capitalize]{cleveref}
\hypersetup{
    colorlinks=true,       
    linkcolor=blue,          
    citecolor=blue,        
    filecolor=blue,      
    urlcolor=blue           
}

\usepackage{slashed}
\usepackage{mathtools}
\usepackage{listings}
\usepackage{pgf}
\usepackage{fancyvrb}
\usepackage{longtable}
\usepackage{bm}
\usepackage{tikz-cd}

\definecolor{darkgreen}{rgb}{0.0, 0.545098, 0.0}

\DeclareMathOperator{\tr}{tr}

\DeclareMathOperator{\vecspan}{span}
\DeclareMathOperator{\Dic}{Dic} 
\newcommand {\R}{\mathbb{R}} 
\newcommand {\Z}{\mathbb{Z}} 

\newcommand{\subgroupeq}{\leq}

\newcommand{\basis}[0]{B} 
\newcommand{\degree}[0]{d}
\newcommand{\wvnum}[0]{\bm{n}} 
\newcommand{\fiducialmom}[0]{\bm{i}}
\newcommand{\internallabel}[0]{\varepsilon}
\newcommand{\bra}[1]{\ensuremath{\langle #1 |}}   
\newcommand{\ket}[1]{\ensuremath{| #1 \rangle}}   
\newcommand{\Ket}[1]{\ensuremath{\left| #1 \right\rangle}}   
\newcommand{\braket}[2]{\ensuremath{\langle #1 | #2 \rangle }}
\newcommand{\amp}[3]{\ensuremath{\left\langle #1 \left| #2 \right| #3 \right\rangle}}  

\newcommand{\xddots}{%
  \raise 4pt \hbox {.}
  \mkern 6mu
  \raise 1pt \hbox {.}
  \mkern 6mu
  \raise -2pt \hbox {.}
}

\newcommand{\orbit}[0]{K}

\newcommand{\permutation}[0]{\sigma}
\newcommand{\id}[0]{\mathds{1}}
\newcommand{\littlegroup}[0]{G_{\bm{P}}}
\newcommand{\doublelittlegroup}[0]{G_{\bm{P}}^D}
\newcommand{\Littlegroup}[1]{G_{\bm{P}_{#1}}}
\newcommand{\stabilizerS}[1]{H_{\bm{P}}^{(#1)}}
\newcommand{\doublestabilizer}[0]{H_{\bm{P}}^{D(s)}}
\newcommand{\stabilizer}[0]{\stabilizerS{s}}

\newcommand{\exchangegroup}[0]{S}
\newcommand{\ytproj}[0]{\scalebox{1.5}{$\Pi$}}
\newcommand{\ytsub}[1]{\scriptsize \begin{ytableau} #1 \end{ytableau}}

\makeatletter
\newsavebox{\@brx}
\newcommand{\llangle}[1][]{\savebox{\@brx}{\(\m@th{#1\langle}\)}%
  \mathopen{\copy\@brx\kern-0.5\wd\@brx\usebox{\@brx}}}
\newcommand{\rrangle}[1][]{\savebox{\@brx}{\(\m@th{#1\rangle}\)}%
  \mathclose{\copy\@brx\kern-0.5\wd\@brx\usebox{\@brx}}}
\makeatother
\newcommand{\tensornorm}[2]{\llangle #1, #2 \rrangle}

\begin{document}

\title{Multi-particle interpolating operators in quantum field theories with cubic symmetry}

\author{William~Detmold}
\affiliation{Center for Theoretical Physics, Massachusetts Institute of Technology, Cambridge, MA 02139, USA}
\affiliation{The NSF AI Institute for Artificial Intelligence and Fundamental Interactions}

\author{William~I.~Jay}
\email{willjay@mit.edu}
\affiliation{Center for Theoretical Physics, Massachusetts Institute of Technology, Cambridge, MA 02139, USA}

\author{Gurtej~Kanwar}
\email{kanwar@itp.unibe.ch}
\affiliation{Albert Einstein Center, Institute for Theoretical Physics, University of Bern, 3012 Bern, Switzerland}
\affiliation{The NSF AI Institute for Artificial Intelligence and Fundamental Interactions}

\author{Phiala~E.~Shanahan}
\affiliation{Center for Theoretical Physics, Massachusetts Institute of Technology, Cambridge, MA 02139, USA}
\affiliation{The NSF AI Institute for Artificial Intelligence and Fundamental Interactions}

\author{Michael~L.~Wagman}
\email{mwagman@fnal.gov}
\affiliation{Fermi National Accelerator Laboratory, Batavia, IL 60510, USA}

\preprint{FERMILAB-PUB-24-0101-T}
\preprint{MIT-CTP/5685}

\begin{abstract}
Numerical studies of lattice quantum field theories are conducted in finite spatial volumes, typically with cubic symmetry in the spatial coordinates.
Motivated by these studies, this work presents a general algorithm to construct multi-particle interpolating operators for quantum field theories with cubic symmetry.
The algorithm automates the block diagonalization required to combine multiple operators of definite linear momentum into irreducible representations of the appropriate little group.
Examples are given for distinguishable and indistinguishable particles including cases with both zero and non-zero spin.
An implementation of the algorithm is publicly available at \href{https://github.com/latticeqcdtools/mhi}{\texttt{https://github.com/latticeqcdtools/mhi}}.

\end{abstract}

\maketitle


\section{Introduction}

The determination of energy spectra is a central task in numerical studies of lattice quantum field theories (QFTs) and is the precursor to more complex studies of the properties and interactions of the states in the theory.
In strongly-coupled field theories, very little is known about the spectra a priori. 
However, analysis of the Euclidean-time dependence of two-point correlation functions between operators with the quantum numbers of the states of interest provides an avenue for first-principles determinations of spectra.
This approach has been used to explore many different field theories, most notably to determine the low-energy excitations in the hadronic and nuclear spectra in Quantum Chromodynamics (QCD), the theory of the strong interactions.
The wide-reaching goals and achievements of lattice QCD are summarized, e.g., in Refs.~\cite{Detmold:2019ghl,Bazavov:2019lgz,Kronfeld:2019nfb,USQCD:2019hyg,Cirigliano:2019jig,USQCD:2019hee}.

To construct correlation functions sensitive to the eigenstates of a strongly-interacting theory such as QCD requires the use of \emph{interpolating operators}, composite objects built from products of the elementary fields of the theory.
In numerical studies, the behavior of correlation functions depends sensitively on the choice of interpolating operators used in the calculation.
To determine the energies of the eigenstates most effectively, it is advantageous to project these interpolating operators to particular symmetry sectors, thereby reducing the number of states that contribute to the corresponding correlation functions. 

Numerical studies in lattice QFTs are usually performed in a finite cubic spatial volume.
In this setting, continuous rotational symmetry is broken down to a discrete subgroup, which (combined with spatial inversions) is the cubic group, $O_h$.
Moreover, provided translational symmetry is preserved (e.g., via the use of periodic boundary conditions), the total momentum of an energy eigenstate is a conserved quantity which further reduces the spatial symmetry.
Specifically, the \textit{little group} of rotations that leave a given total momentum $\bm{P}$ invariant is the subgroup $\littlegroup \subgroupeq O_h$.
Understanding the transformation properties of states---and interpolating operators---under the cubic group and its subgroups is thus critical in analysis of lattice QFT calculations~\cite{Mandula:1982us,Mandula:1983ut}.

Multi-particle interpolating operators constructed in coordinate space or through momentum projection typically transform \textit{reducibly} under these finite groups, obscuring the transformation properties of energy eigenstates.
However, it is possible to decompose the operators into linear subspaces that do not mix with each other and only transform internally.
A complete decomposition into these \textit{irreducible representations} (irreps) is possible for any set of operators that is closed under the little group~\cite{Dresselhaus:2008}. 
For example, suppose an interpolating operator transforms under a reducible representation labelled by $s$, with representation matrices $D^{(s)}(R)$ for $R \in \littlegroup$. Decomposing the reducible representation into a direct sum of irreps $\Gamma^{(s)} = \Gamma_a \oplus \Gamma_b \oplus \cdots$, 
there exist block-diagonalization (or change-of-basis) matrices $U^{(s)}$ specific to $\Gamma^{(s)}$ that enact the decomposition into irrep matrices $D^{(\Gamma)}(R)$ according to 
\begin{align}
    [U^{(s)}]^\dagger
    D^{(s)}(R) U^{(s)} \label{eq:block_diag}
    = \bigoplus_{i} D^{(\Gamma_i)}(R).
\end{align}
After changing to the irrep basis, the transformation properties of multi-particle operators---and the states that they create---are then simpler to understand.
Previous work has addressed the construction of lattice interpolating operators for single baryons~\cite{Basak:2005aq,Basak:2005ir}, single mesons~\cite{Thomas:2011rh}, two-hadron systems~\cite{Luu:2011ep,Amarasinghe:2021lqa} with arbitrary spin and momenta~\cite{Morningstar:2013bda,Prelovsek:2016iyo}, and three-boson systems~\cite{Hansen:2020zhy}. These results cover the construction of local and extended operators with definite cubic transformation properties, as well as their combination into irreps of the relevant little group in cases of up to three local operators.

The present work provides a concrete algorithm and a numerical implementation~\cite{code} that carries out the block diagonalization for any product of $N$ operators with definite momentum, spin, and permutation properties. Each operator can be a point-like operator, a smeared or extended operator, or an even more general construction.
It is further shown that the block-diagonalization matrices can be determined for all $N$ by enumerating a small set of examples.
For spin-zero operators, only examples from $N=1$ and $N=2$ operators are required to specify the decomposition for general $N$.
Additional internal symmetries such as flavor, as well as any combination of fermionic and bosonic operator-exchange symmetries, can be incorporated with a simple extension of the formalism that is also described herein.

The remainder of this article is organized as follows. 
Section~\ref{sec:spinzero} lays out the formalism for building the block-diagonalization matrices for the simple case of $N$ distinguishable, spin-zero operators.
Section~\ref{sec:spin} addresses distinguishable operators with spin, including a generic method for calculating block-diagonalization matrices; concrete results are given for several examples.
Section~\ref{sec:identical} presents the generalization of this construction to operators involving internal symmetries or identical particles.
\Cref{sec:algorithm} collects the formalism of the preceding sections to give the complete block-diagonalization algorithm.
Finally, \Cref{sec:outlook} provides an outlook.
The appendices specify group-theoretical conventions, discuss the method of polarization tensors for evaluating irrep matrices, and give  explicit examples of the block-diagonalization matrices appearing in \cref{eq:block_diag}.

\section{Distinguishable spin-zero operators}\label{sec:spinzero}

Consider a lattice QFT defined on a
geometry whose spatial structure is a periodic cubic lattice
$\Lambda$ with volume $V=L^3$, lattice spacing $a$, and with associated Hilbert space $\mathcal{H}$ and vacuum state $\ket{\Omega}_{\mathcal{H}} \in \mathcal{H}$.\footnote{The temporal geometry (discrete or continuous) and the spacetime metric (Euclidean or Minkowski) are left unspecified because the classification of operator representations under spatial rotations/translations is insensitive to these choices.}
The spatial symmetry group is therefore the cubic group, $O_h$.
In the following, all quantities will be given in lattice units with $a=1$, where $\Lambda \sim (\Z_L)^{3}$.

Multi-particle states can be created by acting on the vacuum with interpolating operators with the quantum numbers of the desired states.
Often, products of $N$ local operators serve as useful interpolating operators for multi-particle states.
Such a product may generically be written as $\mathcal{O}_1(\bm{x}_1) \dotsb \mathcal{O}_N(\bm{x}_N)$, 
where the labels $\bm{x}_1, \dots, \bm{x}_N \in \Lambda$ indicate three-vector coordinates of spatial lattice sites. 
The operators $\mathcal{O}_i$ are built from the fundamental fields in the theory and need not be distinguishable.
For simplicity, the initial discussion will focus on distinguishable operators transforming as scalars under spatial rotations, in which case only the coordinates $\bm{x}_1,\ldots,\bm{x}_N$ of the operators transform under the spatial symmetry group.
The extension to operators with non-zero spin or composite operators with non-trivial $O_h$-transformation properties is addressed in Sec.~\ref{sec:spin}. 
The extension to indistinguishable operators is addressed in Sec.~\ref{sec:identical}.

For a fixed list of $N$ local operators, $\mathcal{O}_1, \dots, \mathcal{O}_N$, this work investigates the transformation properties of linear combinations of products constructed via
\begin{equation}
    \sum_{\bm{x}_1 \in \Lambda}
    \dotsb
    \sum_{\bm{x}_N \in \Lambda} c(\bm{x}_1, \dots, \bm{x}_N) \, \mathcal{O}_1(\bm{x}_1) \dotsb \mathcal{O}_N(\bm{x}_N), \label{eq:generic_interpolating_operators}
\end{equation}
where the $c(\bm{x}_1, \dots, \bm{x}_N) \in \mathbb{C}$ are $V^N$ arbitrary coefficients.\footnote{In a continuum theory, $\sum_{\bm{x}_i \in \Lambda}$ is replaced with $\int \prod_i d^3 \bm{x}_i$ and $c(\bm{x}_1, \dots, \bm{x}_N)$ should be replaced by a continuous field over $N$ coordinates, but otherwise the formalism developed in this work applies.}
Acting on the vacuum with any such linear combination yields a state
\begin{equation} \label{eq:state-creation}
     \begin{split}
     &\ket{\psi}_{\mathcal{H}} \equiv \\
     &\sum_{\bm{x}_1 \in \Lambda}\dotsb
     \sum_{\bm{x}_N \in \Lambda} c(\bm{x}_1, \dots, \bm{x}_N) \, \mathcal{O}_1(\bm{x}_1) \dotsb \mathcal{O}_N(\bm{x}_N) \ket{\Omega}_{\mathcal{H}}.
     \end{split}
\end{equation}
Invariance of the vacuum state under spatial rotations means that any such $\ket{\psi}_{\mathcal{H}}$ inherits the transformation properties of the multi-particle operator itself.

The space defined by the coefficients $c(\bm{x}_1, \dots, \bm{x}_N)$ forms a $V^N$-dimensional vector space which will be denoted~$\mathcal{V}$.
A symbolic (position-space) basis for $\mathcal{V}$ consists of the vectors
$\ket{\bm{x}_1, \dots, \bm{x}_N}$ for each choice of $\{ \bm{x}_i \in \Lambda \}$. These basis vectors are defined to be
orthonormal with respect to the inner product on $\mathcal{V}$,
\begin{equation}\label{eq:innerprod}
    \left< \bm{x}'_1, \dots, \bm{x}'_N | \bm{x}_1, \dots, \bm{x}_N \right>
    = \delta_{\bm{x}'_1,\bm{x}_1} \dotsb \delta_{\bm{x}'_N,\bm{x}_N}.
\end{equation}
To avoid ambiguity, elements of the abstract space $\mathcal{V}$ are written with unadorned kets, while quantum states such as $\ket{\Omega}_{\mathcal{H}}$ or $\ket{\psi}_{\mathcal{H}}$ carry a subscript.
Since the list of operators is fixed, each vector in $\mathcal{V}$ is associated uniquely with an interpolating operator by the linear map $\mathcal{L}:\ket{\bm{x}_1, \dots, \bm{x}_N} \mapsto \mathcal{O}_1(\bm{x}_1) \cdots \mathcal{O}_N(\bm{x}_N)$.

A useful starting point for the decomposition of a state $\ket{\psi}_{\mathcal{H}}$ into subspaces that transform irreducibly under spatial symmetries is the plane-wave basis of $\mathcal{V}$, defined by vectors
\begin{equation} \label{eq:plane-wave-basis}
\begin{aligned}
    &\ket{\bm{n}_1, \dots, \bm{n}_N} \equiv \\
    &
    \sum_{\bm{x}_1 \in \Lambda}
    \cdots \sum_{\bm{x}_N\in \Lambda}
    e^{i \frac{2\pi}{L} \bm{n}_1 \cdot \bm{x}_1} \cdots e^{i \frac{2\pi}{L} \bm{n}_N \cdot \bm{x}_N} \ket{\bm{x}_1, \dots, \bm{x}_N},
\end{aligned}
\end{equation}
for each possible (ordered) set of wavevectors $\{ \bm{n}_i \in (\Z_L)^{ 3}\}$.\footnote{In a continuum theory, $\Z_L$ is replaced with $\mathbb{Z}$.} 
Here, the distinction between the position-space and plane-wave bases is made by the use of the letters $\bm{x}$ and $\bm{n}$, respectively.
Since there is a one-to-one mapping between wavevectors $\bm{n}_i$ and momenta $(2\pi/L) \bm{n}_i$, these terms will be used interchangeably for the wavevectors $\bm{n}_i$.
Each vector $\ket{\bm{n}_1, \dots, \bm{n}_N}$ is associated with a plane-wave interpolating operator,
\begin{equation}
\begin{aligned}
    &\widetilde{\mathcal{O}}_1(\bm{n}_1) \cdots \widetilde{\mathcal{O}}_N(\bm{n}_N) \equiv \\
    &
    \sum_{\bm{x}_1 \in \Lambda}
    \cdots \sum_{\bm{x}_N\in \Lambda}
    e^{i \frac{2\pi}{L} \bm{n}_1 \cdot \bm{x}_1} \mathcal{O}_1(\bm{x}_1) \cdots e^{i \frac{2\pi}{L} \bm{n}_N \cdot \bm{x}_N} \mathcal{O}_N(\bm{x}_N).
    \label{eq:plane-wave-interpolator}
\end{aligned}
\end{equation}

The position-space basis vectors $\ket{\bm{x}_1, \dots, \bm{x}_N}$ and plane-wave basis vectors $\ket{\bm{n}_1, \dots, \bm{n}_N}$ have simple transformation properties under $O_h$ and its subgroups.
Each group element $R\in O_h$ can be specified by its action on arbitrary three-vectors $\bm{x} \in \Lambda$, i.e., with each group element defined as a unique orthogonal $3\times3$ matrix, $R\in O_h \subgroupeq O(3)$. 
A concrete specification of these matrices is presented in Appendix~\ref{app:group_conventions}. 
When acting on vectors in $\mathcal{V}$, the abstract operator associated with rotation $R$ will be denoted $\mathcal{D}(R)$.
Its action on the position-space basis vectors is given by
\begin{align}
    \mathcal{D}(R) \ket{\bm{x}_1, \dots, \bm{x}_N}
    &= \ket{R \bm{x}_1, \dots, R\bm{x}_N} \label{eq:n_states}
\end{align}
and extends linearly to the rest of $\mathcal{V}$, including to the plane-wave basis vectors, which transform as
\begin{equation}
\mathcal{D}(R) \ket{\wvnum_1, \dots, \wvnum_N}
    = \ket{R \wvnum_1, \dots, R \wvnum_N}
\end{equation}
based on their definition in Eq.~\eqref{eq:plane-wave-basis}.
Through the map ${\cal L}$ (defined following Eq.~\eqref{eq:innerprod}), this defines the action of rotations on the corresponding interpolating operators.

The remainder of this section proceeds as follows.
First, \cref{ssec:momentum_orbit} constructs the reducible representations associated with interpolating operators in the form of \cref{eq:plane-wave-interpolator} for distinguishable spin-zero operators. 
Second, \cref{subsec:irrep-bases} reviews the irreps of the cubic group and its subgroups.
\Cref{subsec:cob-construction} presents the algorithm for constructing the block-diagonalization matrices that change basis from the interpolating operators in \cref{eq:plane-wave-interpolator} into operators that transform irreducibly.
\Cref{ssec:stabilizer_classification} discusses a classification of cases with identical block-diagonalization matrices using the stabilizer groups of specific momenta, and finally \cref{subsec:examples} presents examples of the block-diagonalization that are sufficient to implement this process for any number of spin-zero operators.

\subsection{Momentum orbits and their representations \label{ssec:momentum_orbit}}

To construct the reducible representations associated with plane-wave interpolating operators, the space $\mathcal{V}$ is systematically decomposed into subspaces associated with \emph{momentum orbits}, i.e., sets of plane-wave momenta closed under little-group transformations. 
These subspaces, labelled by $s$, each correspond to a definite total momentum $\bm{P}$ and carry a reducible representation $\Gamma^{(s)}$ of the little group $G_{\bm{P}}$. 
The representation matrices $D^{(s)}(R)$ take a simple form and are explicitly constructed below.

Given a plane wave $\ket{\bm{n}_1, \dots, \bm{n}_N} \in \mathcal{V}$, the total momentum is $\bm{P} = \frac{2\pi}{L} \sum_i \wvnum_i$. 
The ordered list of wavevectors defining the basis state $\ket{\bm{n}_1, \dots, \bm{n}_N}$ is denoted by $[\bm{n}_1, \dots, \bm{n}_N]$ and will be used to define momentum orbits below.
The set of such lists of wavevectors can be partitioned into disjoint subsets $\orbit_{\bm{P}}$ of fixed total momentum $\bm{P}$, 
\begin{equation}
    \orbit_{\bm{P}} \equiv \Big\{ [\bm{n}_1, \dots, \bm{n}_N]: \frac{2\pi}{L} \sum_i \bm{n}_i = \bm{P} \Big\}.
\end{equation}
For each total momentum $\bm{P}$, the associated \emph{little group} is defined as the subgroup $\littlegroup \subgroupeq O_h$ that leaves $\bm{P}$ invariant,
\begin{align}
\littlegroup \equiv \{ R \in O_h : R\bm{P}=\bm{P}\}.
\label{eq:little_group}
\end{align}
The possible little groups, up to equivalence under rotation of the total momentum, are denoted $C_{4v}$, $C_{3v}$, $C_{2v}$, $C_2^R$, $C_2^P$, and $C_1$ following the notation of Ref.~\cite{Dresselhaus:2008}, where $C_{nv}$ is the symmetry group of the $n$-gon with $2n$ elements and $C_{n}$ is the cyclic group with $n$ elements. The little groups $C_2^R$ and $C_2^P$ both have the group structure of $C_2$, but appear as little groups for inequivalent choices of total momenta. Details of these groups are presented in \Cref{app:group_conventions}; note that to match these definitions the little group may need to be rotated (by conjugation) so that the total momentum aligns with the present conventions.

By construction, the space $\orbit_{\bm{P}}$ is closed under the action of the little group $\littlegroup$. However, it can be further partitioned into minimal invariant subsets that are closed under the action of $\littlegroup$.
These momentum orbits, denoted by $\orbit_{\bm P}^{(s)}$, are constructed by computing the sets
\begin{equation}
\begin{split}
  \orbit_{\bm P}^{(s)} &\equiv \{ R \cdot [\wvnum_1,\ldots,\wvnum_N]: R \in \littlegroup\} \subseteq \orbit_{\bm P},
  \end{split} \label{eq:momentum_orbit}
\end{equation}
where $R \cdot [\wvnum_1,\ldots,\wvnum_N] = [R \wvnum_1,\ldots,R \wvnum_N]$. 
Here, $s$ is an abstract label distinguishing inequivalent subsets; the label $s$ can also be defined concretely with one representative or \emph{fiducial} list of wavevectors per orbit. 
Two examples of $K_{\bm{P}}^{(s)}$ are given by the orbits of the wavevectors $[\bm{n}_1,\bm{n}_2]=[(0,0,1),(0,0,-1)]$ and $[\bm{n}_1,\bm{n}_2]=[(0,2,1),(0,-2,-1)]$ with $G_{\bm{P}} = O_h$ as these cases both have vanishing total momentum.

It is useful to provide a conventional way of ordering the elements of a given momentum orbit.
First, select a fiducial arrangement of momenta $[\wvnum_1,\ldots,\wvnum_N]$.
With $\bm{P}$ being the total momentum defined by the fiducial momenta, let $[R_1, ..., R_n]$, with $n=|\littlegroup|$, denote the fixed ordering of the elements of little group $\littlegroup$ given in \Cref{app:group_conventions}.\footnote{Through the present work, $|X|$ indicates the number of elements in the set $X$.}
The orbit $\orbit_{\bm{P}}^{(s)}$ then inherits its ordering from the little group via \cref{eq:momentum_orbit}, with the group elements acting in order on the fiducial arrangement of momenta.
The $m$th element of $\orbit_{\bm{P}}^{(s)}$ in this conventional ordering will be denoted by $[\wvnum_1, \dots, \wvnum_N]^{(s)}_m$ so that the set can be equivalently written as
\begin{equation}
  \orbit_{\bm{P}}^{(s)} = \left\{[\wvnum_1,\ldots,\wvnum_N]^{(s)}_{m} : m \in \{1, 2, \ldots,  |\orbit_{\bm{P}}^{(s)}|\} \right\}.
\end{equation}

The partitioning into momentum orbits gives rise to a corresponding decomposition of the space of interpolating operators.
For convenience, the plane-wave vector in $\mathcal{V}$ associated with the $m$th list of wavevectors within the orbit labelled by $s$ is denoted by
\begin{equation}
    \ket{s,m} \equiv \ket{ [\wvnum_1,\ldots,\wvnum_N]^{(s)}_{m}}.
\end{equation}
It bears emphasis that each momentum orbit $\orbit_{\bm{P}}^{(s)}$ is defined in terms of a particular total momentum $\bm{P}$ with little group $\littlegroup$; this dependence is left implicit in the notation $\ket{s, m}$.
Each orbit defines an invariant subspace
\begin{equation}
\begin{gathered}
\mathcal{V}_{\bm{P}}^{(s)} \equiv \vecspan\left\{ \ket{s,m} : m \in \{ 1, 2, \dots, |\orbit_{\bm{P}}^{(s)}| \} \right\} \; \subset \; \mathcal{V}
\end{gathered}
\end{equation}
satisfying 
\begin{align}
\mathcal{D}(R) \mathcal{V}_{\bm{P}}^{(s)} = \mathcal{V}_{\bm{P}}^{(s)}, \text{ for } R \in \littlegroup. \label{eq:orbit_invariance_property}
\end{align}
These subspaces are linearly independent and jointly compose the full space,
\begin{align}
\mathcal{V} = \bigoplus_{\bm{P},s} \mathcal{V}_{\bm{P}}^{(s)},
\end{align}
where the sum ranges over all possible total momenta $\bm{P}$ and (for a given $\bm{P}$) over all distinct orbits labelled by $s$.

The vectors in $\mathcal{V}_{\bm{P}}^{(s)}$ inherit their transformation properties from the plane waves, \begin{equation}
\mathcal{D}(R) \ket{s,m} = \ket{R \cdot [ \wvnum_1, \dots,  \wvnum_N]_{m}^{(s)}} = \ket{s,m'},
\end{equation}
which extends by linearity to the entire space.
Each element $R \in \littlegroup$ permutes the basis vectors of $\mathcal{V}_{\bm{P}}^{(s)}$ due to the invariance property \cref{eq:orbit_invariance_property} and therefore acts as a linear operator $\mathcal{V}_{\bm{P}}^{(s)} \rightarrow \mathcal{V}_{\bm{P}}^{(s)}$ with matrix elements
\begin{equation}
  D^{(s)}_{m'm}(R) \equiv \amp{s,m'}{\mathcal{D}(R)}{s,m} \; \in \; \{0, 1\}. \label{eq:Dbraket}
\end{equation}
These matrices are precisely the representation matrices $D^{(s)}$ appearing in \cref{eq:block_diag}.
The superscript on $D^{(s)}$ emphasizes the block-diagonal nature of this representation within the larger space $\mathcal{V}$, since distinct momentum orbits do not mix under the cubic group.
Each such space therefore forms a (generally reducible) representation which will be denoted by $\Gamma^{(s)}$.

\subsection{Irreducible representations} \label{subsec:irrep-bases}
The irreps $\Gamma$ of $O_h$ and the irreps of its subgroups have been previously catalogued in many places. 
Following Ref.~\cite{Dresselhaus:2008}, such representations are completely specified by basis vectors $\ket{\basis^{(\Gamma)}_\mu}$, where $\mu \in \{1, \dots, |\Gamma|\}$ labels the rows of the irrep $\Gamma$.
The basis vectors can be written in an abstract coordinate space as basis
functions $\braket{\bm{r}}{\basis^{(\Gamma)} _\mu} = \basis^{(\Gamma)}_\mu(\bm{r})$ on the unit sphere, in terms of the normalized three-vector coordinate $\bm{r} \equiv (x,y,z)^T \in \mathbb{R}^3$, with $|\bm{r}| = 1$.
The basis functions used in this work, for the irreps of $O_h$ and its subgroups, are specified in \cref{table:basis_functions}.
Additional details related to the choice of basis functions are described in \cref{app:group_conventions}.

For the present work, the primary utility of the basis functions is in defining the irrep matrices $D^{(\Gamma)}(R)$ appearing in the block diagonalization of \cref{eq:block_diag}.
For instance, given the basis functions, the matrix representation associated with $R \in \littlegroup$ in irrep $\Gamma$ can be computed as
\begin{equation}
  \begin{split}
D^{(\Gamma)}_{\mu'\mu}(R) 
    &= {\frac{1}{N^{(\Gamma)}}} \bra{\basis^{(\Gamma)}_{\mu'}}\mathcal{D}(R)\ket{\basis^{(\Gamma)}_\mu}\\
    &= {\frac{1}{N^{(\Gamma)}}} \int d\Omega\, \braket{\basis^{(\Gamma)}_{\mu'}}{\bm{r}}\braket{\bm{r}}{\mathcal{D}(R)|\basis^{(\Gamma)}_{\mu}} \\
    &= \frac{1}{N^{(\Gamma)}}\int d\Omega \; \basis^{(\Gamma)*}_{\mu'}(\bm{r}) \basis^{(\Gamma)}_{\mu}(R^{-1} \bm{r}),
    \label{eq:irrep-matrix-elements}
  \end{split}
\end{equation}
where the integration is over the solid-angle measure on the unit sphere, i.e., $d\Omega \equiv d\bm{r}\, \delta(|\bm{r}|-1)$, and the normalization constant is
\begin{equation}
   N^{(\Gamma)} = \int d\Omega \; \basis^{(\Gamma)*}_{\mu}(\bm{r}) \basis^{(\Gamma)}_{\mu}(\bm{r}),
\end{equation}
with no summation over $\mu$ (by construction, all rows of a given irrep are identically normalized so $N^{(\Gamma)}$ does not carry a row index).
The action of a little group transformation therefore corresponds to right multiplication of basis vectors,
\begin{equation} \label{eq:irrep-basis-states}
\mathcal{D}(R) \ket{\basis^{(\Gamma)}_\mu} = \sum_{\mu'}  \ket{\basis^{(\Gamma)}_{\mu'}} D^{(\Gamma)}_{\mu'\mu}(R).
\end{equation}

In explicit calculations, it is convenient to recast the integral in Eq.~\eqref{eq:irrep-matrix-elements} algebraically using polarization tensors as discussed in \cref{app:polarization}.

\bgroup
\def\arraystretch{1.2}
\begin{table*}[t]
    \centering
    \caption{Basis functions $\braket{\bm{r}}{\basis^{(\Gamma)} _\mu} = \basis^{(\Gamma)}_\mu(\bm{r})$ used in this work for the irreps of the cubic group $O_h$ and its subgroups.
    As indicated, basis functions for the irreps of the subgroups $C_{4v}$, $C_{3v}$, $C_{2v}$, $C_2^R$, $C_2^P$, and $C_1$ can be written in terms of the basis functions of irreps of $O_h$. 
    The final column indicates the $z$ component of the $SO(3)$ angular momentum operator modulo $4$, denoted by $\ell_z$.
    \label{table:basis_functions}
    }
    \begin{tabular}{ccccc}
\hline\hline
Group  & Irrep    & $\mu$ & Basis function $B_\mu^{(\Gamma)}({\bf r})$ & Notes\\
\hline
$O_h$   & $A_1^+$ & 1     & $\frac{1}{\sqrt{3}}(x^2 + y^2 + z^2)$ & $\ell_z = 0$ \\
$O_h$   & $A_2^+$ & 1     & $\frac{1}{\sqrt{6}}\left[ x^4(y^2 - z^2)+ y^4(z^2-x^2) + z^4(x^2-y^2)\right]$ & $\ell_z = 2$ \\
$O_h$   & $E^+$   & 1     & $\frac{1}{\sqrt{6}}\left( 2z^2 - x^2 - y^2 \right)$ & $\ell_z = 0$ \\
$O_h$   & $E^+$   & 2     & $\frac{1}{\sqrt{2}}(x^2 - y^2)$ & $\ell_z = 2$ \\
$O_h$   & $T_1^+$ & 1     &$\frac{1}{\sqrt{2}}xy(x^2-y^2)$ & $\ell_z = 0$ \\
$O_h$   & $T_1^+$ & 2     & $\frac{1}{2}\left[- yz(y^2-z^2) + i zx(z^2-x^2)\right]$ & $\ell_z = 1$ \\
$O_h$   & $T_1^+$ & 3     & $\frac{1}{2}\left[ yz(y^2-z^2) + i zx(z^2-x^2) \right]$ & $\ell_z = 3$ \\
$O_h$   & $T_2^+$ & 1     & $\frac{1}{\sqrt{2}}(-zx + i yz)$ & $\ell_z = 1$ \\
$O_h$   & $T_2^+$ & 2     & $-ixy$ & $\ell_z = 2$ \\
$O_h$   & $T_2^+$ & 3     & $\frac{1}{\sqrt{2}}(zx + i yz)$ & $\ell_z = 3$ \\
$O_h$   & $A_1^-$ & 1     & $\frac{1}{\sqrt{6}} xyz\left[ x^4(y^2 - z^2)+ y^4(z^2-x^2) + z^4(x^2-y^2) \right]$ & $\ell_z = 0$ \\
$O_h$   & $A_2^-$ & 1     & $xyz$ & $\ell_z = 2$ \\
$O_h$   & $E^-$   & 1     & $\frac{1}{\sqrt{2}}xyz(x^2 - y^2)$ & $\ell_z = 0$ \\
$O_h$   & $E^-$   & 2     & $-\frac{1}{\sqrt{6}}xyz \left[ 2z^2 - x^2 - y^2\right]$ & $\ell_z = 2$ \\
$O_h$   & $T_1^-$ & 1     & $z$ & $\ell_z = 0$ \\
$O_h$   & $T_1^-$ & 2     & $\frac{1}{\sqrt{2}}(-x + i y)$ & $\ell_z = 1$ \\
$O_h$   & $T_1^-$ & 3     & $\frac{1}{\sqrt{2}}(x + i y)$ & $\ell_z = 3$ \\
$O_h$   & $T_2^-$ & 1     & $\frac{1}{2}\left[-y(z^2 - x^2) + i x (y^2 - z^2)\right]$ & $\ell_z = 1$ \\
$O_h$   & $T_2^-$ & 2     & $-\frac{i}{ \sqrt{2} } z (x^2 - y^2)$ & $\ell_z = 2$ \\
$O_h$   & $T_2^-$ & 3     & $\frac{1}{2}\left[y (z^2 - x^2) + i x (y^2 - z^2)\right]$ & $\ell_z = 3$ \\
\hline
$C_{4v}$& $A_1$   & 1   & $\basis^{(O_h,A_1^+)}_1(\bm{r})$ & \\
$C_{4v}$& $A_2$   & 1   & $\basis^{(O_h,A_1^-)}_1(\bm{r})$ & \\
$C_{4v}$& $B_1$   & 1   & $\basis^{(O_h,E^+)}_2(\bm{r})$ & \\
$C_{4v}$& $B_2$   & 1   & $\basis^{(O_h,T_2^+)}_2(\bm{r})$ & \\
$C_{4v}$& $E$     & 1   & $-xz = \frac{1}{\sqrt{2}}\left( \basis^{(O_h,T_2^+)}_1(\bm{r}) - \basis^{(O_h,T_2^+)}_3(\bm{r}) \right)$ & \\
$C_{4v}$& $E$     & 2   & $-yz = \frac{i}{\sqrt{2}}\left( \basis^{(O_h,T_2^+)}_1(\bm{r}) + \basis^{(O_h,T_2^+)}_3(\bm{r}) \right)$ & \\
\hline
$C_{3v}$& $A_1$ & 1 & $\basis^{(O_h,A_1^+)}_1(\bm{r})$ &\\
$C_{3v}$& $A_2$ & 1 & $\basis^{(O_h,A_2^+)}_1(\bm{r})$ &\\
$C_{3v}$& $E$   & 1 & $\basis^{(O_h,E^+)}_1(\bm{r})$ &\\
$C_{3v}$& $E$   & 2 & $\basis^{(O_h,E^+)}_2(\bm{r})$ &\\
\hline
$C_{2v}$& $A_1$ & 1 & $\basis^{(O_h,A_1^+)}_1(\bm{r})$ & \\
$C_{2v}$& $A_2$ & 1 & $\basis^{(O_h,A_1^-)}_1(\bm{r})$ & \\
$C_{2v}$& $B_1$ & 1 & $\basis^{(O_h,A_2^-)}_1(\bm{r})$ & \\
$C_{2v}$& $B_2$ & 1 & $\basis^{(O_h,A_2^+)}_1(\bm{r})$ & \\
\hline
$C_2^R$ & $A$ & 1 & $\basis^{(O_h,A_1^+)}_1(\bm{r})$ & \\
$C_2^R$ & $B$ & 1 & $\basis^{(O_h,A_1^-)}_1(\bm{r})$ & \\
\hline
$C_2^P$ & $A$ & 1 & $\basis^{(O_h,A_1^+)}_1(\bm{r})$ & \\
$C_2^P$ & $B$ & 1 & $\basis^{(O_h,A_2^+)}_1(\bm{r})$ & \\
\hline
$C_1$   & $A$ & 1 & $\basis^{(O_h,A_1^+)}_1(\bm{r})$ & \\
\hline\hline
\end{tabular}

\end{table*}
\egroup

\subsection{Block diagonalization \label{subsec:cob-construction}}

Elements of the momentum orbit $\orbit_{\bm{P}}^{(s)}$ transform in the reducible representation $\Gamma^{(s)}$, which can be decomposed into a direct sum of irreps
\begin{align}\label{eq:kappa_degeneracy}
\Gamma^{(s)} = \underbrace{\Gamma_1 \oplus \dots \oplus \Gamma_1}_{k_1 \, \text{copies}} \oplus \underbrace{\Gamma_2 \oplus \dots \oplus \Gamma_2}_{k_2 \, \text{copies}} \oplus \cdots .
\end{align}
Above, $\Gamma_1, \Gamma_2, \dots$ label the distinct irreps of the group $\littlegroup$, with each $\Gamma_i$ appearing with multiplicity $k_i \geq 0$ in the decomposition of the momentum-orbit representation.
This decomposition into irreps induces an associated decomposition of the vector space {$\mathcal{V}_{\bm{P}}^{(s)}$ associated with $\orbit_{\bm{P}}^{(s)}$ into subspaces that each transform according to a given irrep:
\begin{align} 
\mathcal{V}_{\bm{P}}^{(s)} =
    \mathcal{V}^{(\Gamma_1, s)} \oplus
    \mathcal{V}^{(\Gamma_2, s)} \oplus
    \dotsb, \label{eq:hilbert-space-ds}
\end{align}
where dependence on $\bm{P}$ is left implicit on the right-hand side.
Due to the potential for degenerate copies of irreps,\footnote{Here and below, `degenerate' is used to describe situations where multiple copies of an irrep appear in a given momentum-orbit representation. 
In Hilbert space, the physical states associated with these operators are not necessarily degenerate in the sense of having the same energy.} each summand $\mathcal{V}^{(\Gamma_i, s)}$ has dimension $k_i \cdot |\Gamma_i|$.

To separate the $k_i$ degenerate copies of each irrep $\Gamma_i$, it is helpful first to decompose each space $\mathcal{V}^{(\Gamma_i, s)}$ into sectors associated with the individual rows $\mu$ of the irrep $\Gamma_i$ as $\mathcal{V}^{(\Gamma_i,s)} = \bigoplus_\mu \mathcal{V}^{(\Gamma_i,s)}_\mu$.
Any set of $k_i$ linearly independent vectors in $\mathcal{V}^{(\Gamma_i,s)}_\mu$, which can be labeled as
\begin{equation}
    \left\{ \ket{s, \Gamma_i, \kappa, \mu} : \kappa \in \{1, \dots, k_i\}  \right\}, \label{eq:irrep_vectors}
\end{equation}
provide a basis for this space,
\begin{equation}
\begin{aligned}
\mathcal{V}^{(\Gamma_i,s)}_\mu
= \vecspan\left\{ \ket{s, \Gamma_i, \kappa, \mu} : \kappa \in \{1, \dots, k_i\} \right\}.
\end{aligned}
\end{equation}
If a consistent set of $k_i$ basis vectors is chosen for all $\mu$, the transformation properties of the basis vectors follow from Eq.~\eqref{eq:irrep-basis-states},
\begin{equation}
    \mathcal{D}(R) \ket{s, \Gamma_i, \kappa, \mu} = \sum_{\mu'} \ket{s, \Gamma_i, \kappa, \mu'}  D^{(\Gamma_i)}_{\mu' \mu}(R),
    \label{eq:shell_transform}
\end{equation}
where the irrep matrices $D^{(\Gamma_i)}_{\mu' \mu}(R)$ for $R\in \littlegroup$ are defined as described in Sec.~\ref{subsec:irrep-bases}.
The explicit construction of such a basis is detailed below.
Note that the labeling of degenerate copies of irreps by $\kappa$ is not unambiguously specified. A particular choice will be made in the following construction.

The unitary matrix $U^{(s)}$ that transforms from the reducible momentum-orbit basis 
$\{ \ket{s,m} \}$ to the irrep basis $\{ \ket{s,\Gamma_i,\kappa,\mu} \}$ is defined by
its elements 
\begin{equation}
\begin{aligned}
    U^{(\Gamma_i, \kappa, s)}_{m \mu } \equiv \braket{s,m}{s, \Gamma_i, \kappa, \mu},
    \label{eq:change_of_basis_states}
\end{aligned}
\end{equation}
where $\kappa \in \{ 1, \dots, k_i \}$, $\mu \in \{ 1, \dots, |\Gamma_i| \}$, and the total momentum $\bm{P}$ is left implicit.
In particular, these matrices enact the block diagonalization anticipated by \cref{eq:block_diag},
\begin{align}
    D^{(\Gamma_i)}_{\mu \mu'}(R) 
    &= \sum_{m, m'} [U^{(\Gamma_i,\kappa,s)}_{m \mu }]^* D_{mm'}^{(s)}(R) U^{(\Gamma_i,\kappa,s)}_{m' \mu' }\\
    &= \left([U^{(s)}]^\dagger D^{(s)}(R) U^{(s)} \right)_{\mu\mu'}.
\end{align}
Although the labels $\Gamma_i$ and $\kappa$ appear as superscripts labeling the choice of irrep, $U^{(s)}$ is indeed a unitary $|\orbit_{\bm{P}}^{(s)}| \times |\orbit_{\bm{P}}^{(s)}|$ matrix if $m$ is taken as a row index and the indices $\Gamma_i$, $\kappa$, and $\mu$ are enumerated jointly as a column index.

The change-of-basis matrix $U^{(s)}$ for each momentum orbit can be constructed explicitly using Schur's lemma, which says\footnote{Written in this form, Schur's lemma is sometimes referred to as the Wonderful Orthogonality Theorem~\cite{Dresselhaus:2008}.}
\begin{align} \label{eq:schur-lemma}
\frac{|\Gamma|}{|\littlegroup|} \sum_{R \in \littlegroup}
D^{(\Gamma)}_{\mu\nu}(R)^*
D^{(\Gamma')}_{\mu'\nu'}(R) 
= 
\delta_{\Gamma\Gamma'}
\delta_{\mu\mu'}
\delta_{\nu\nu'}
\end{align}
for arbitrary irreps $\Gamma$ and $\Gamma^\prime$.
For each irrep $\Gamma_i$ of the little group $\littlegroup$, it is convenient to define a projection operator 
\begin{align}
\Pi^{(\Gamma_i,s)}_\mu &: \mathcal{V}^{(s)}_{\bm{P}} \to \mathcal{V}_\mu^{(\Gamma_i, s)},\\
\Pi^{(\Gamma_i,s)}_{\mu} &\equiv \frac{|\Gamma_i|}{|\littlegroup|} \sum_{R\in \littlegroup} D^{(\Gamma_i)}_{\mu\mu}(R)^* \ \mathcal{D}(R),
\end{align}
with no sum on $\mu$ implied and with matrix elements 
\begin{align}
\left(\Pi^{(\Gamma_i,s)}_{\mu}\right)_{m'm} 
&= \bra{s,m'}\Pi^{(\Gamma_i,s)}_{\mu} \ket{s,m}\\
&= \frac{|\Gamma_i|}{|\littlegroup|} \sum_{R\in \littlegroup} D^{(\Gamma_i)}_{\mu\mu}(R)^* D^{(s)}_{m'm}(R).
\label{eq:schur}
\end{align}
This projector onto a given row $\mu$ of the irrep $\Gamma_i$ is closely related to the desired change-of-basis matrices:
\begin{align}
    \left(\Pi^{(\Gamma_i,s)}_{\mu}\right)_{ m' m} &= 
    \sum_{\kappa=1}^{k_i} \braket{s, m'}{s, \Gamma_i, \kappa, \mu} \braket{s, \Gamma_i, \kappa, \mu}{s, m}  \\
    =&
\sum_{\kappa=1}^{k_i} U^{(\Gamma_i,\kappa,s)}_{m' \mu} \left[ U^{(\Gamma_i,\kappa,s)}_{m \mu} \right]^*.
\label{eq:schur-f-projection}
\end{align}
The first equality follows from Schur's lemma after insertions of the identity, $\id=\sum_{\Gamma,\kappa, \mu} \ket{s,\Gamma,\kappa,\mu}\bra{s,\Gamma,\kappa,\mu}$.
By unitarity of the change-of-basis matrices, the projection operators $\Pi_\mu^{(\Gamma_i,s)}$ are idempotent.

The change-of-basis matrix elements $U^{(\Gamma_i, \kappa, s)}_{m \mu}$ can be extracted by suitable orthogonalization of the rows of $\big(\Pi^{(\Gamma_i,s)}_{\mu}\big)_{m'm}$.
The choice of an orthogonalization scheme, e.g., a Gram-Schmidt procedure, fixes the otherwise ambiguous $\kappa$ labeling of degenerate copies of each irrep.

Degenerate copies must be orthogonalized consistently for all rows $\mu$ within each irrep in order to achieve the simple transformation rule in Eq.~\eqref{eq:shell_transform}.\footnote{For example, applying the Gram--Schmidt procedure to each row of $\big(\Pi^{(\Gamma_i, s)}_{\mu}\big)_{m' m}$ independently would lead to a more cumbersome transformation rule in which states $\ket{s, \Gamma, \kappa, \mu}$ are mapped by the action of $R \in \littlegroup$ to linear combinations of states $\ket{s, \Gamma, \kappa', \mu'}$ including those with $\kappa ' \neq \kappa$.
The coefficients of these linear combinations would need to be computed separately for each momentum orbit and would generically depend upon $s$, $\kappa$, and $\kappa'$ as well as $\Gamma$, $\mu$, and $\mu'$. }
A natural prescription is first to orthonormalize the $\mu = 1$ rows within each irrep and then to use \emph{transition operators} to move between the remaining rows.
Schur's lemma applied to the off-diagonal elements of the rotational irrep matrices furnishes the transition operators,
\begin{align}
    T_{\mu\nu}^{(\Gamma_i,s)} &\equiv \frac{|\Gamma_i|}{|\littlegroup|}\sum_{R\in \littlegroup} D_{\mu\nu}^{(\Gamma_i)}(R)^* \mathcal{D}(R),\\
    \left( T_{\mu\nu}^{(\Gamma_i,s)} \right)_{m'm}
    &\equiv \frac{|\Gamma_i|}{|\littlegroup|}\sum_{R\in \littlegroup} D_{\mu\nu}^{(\Gamma_i)}(R)^* D^{(s)}_{m'm}(R), \label{eq:transition_defined}\\
    &= \sum_{\kappa=1}^{k_i} U^{(\Gamma_i, \kappa, s)}_{m' \mu} \left[ U^{(\Gamma_i, \kappa, s)}_{m\nu}\right]^*. \label{eq:transition}
\end{align}
By definition, $T_{\mu\mu}^{(\Gamma_i,s)} \equiv \Pi_\mu^{(\Gamma_i,s)}$.
The transition operators are Hermitian, $[T_{\mu\nu}^{(\Gamma_i,s)}]^\dagger = T_{\nu\mu}^{(\Gamma_i,s)}$, and satisfy 
\begin{align}
\sum_{m'} \left(T^{(\Gamma_i,s)}_{\mu\nu}\right)_{mm'}
\left(T^{(\Gamma_i,s)}_{\rho\sigma}\right)_{m'n} = 
\delta_{\nu\rho} \left(T_{\mu \sigma}^{(\Gamma_i,s)}\right)_{mn},
\label{eq:cubic_transition_fusion}
\end{align}
which follows from unitarity of the change-of-basis matrices.
Likewise, it follows from Eq.~\eqref{eq:transition} and unitarity of the change-of-basis matrices that the transition operators relate the different rows $\mu$ and $\nu$,
\begin{align}
U_{m\mu }^{(\Gamma_i,\kappa, s)}
= \sum_{n}\left(T^{(\Gamma_i),s}_{\mu\nu}\right)_{mn} U_{n\nu}^{(\Gamma_i,\kappa, s)} \label{eq:transition_applied},
\end{align}
for fixed $\mu,\nu$, which allows all elements of the change-of-basis matrices to be determined once those with $\mu=1$ are known.

\subsection{Stabilizer subgroups\label{ssec:stabilizer_classification}}

The block-diagonalization matrices for many different momentum orbits can be demonstrated to be identical using general arguments from group theory.
The construction and ordering of a momentum orbit $\orbit_{\bm{P}}^{(s)}$ are always performed in terms of a fiducial list of plane waves, with the associated set of wavevectors denoted by $\fiducialmom \equiv [\wvnum_1, \dots, \wvnum_N]$. 
The stabilizer subgroup of the little group $\littlegroup$ can then be defined as the subgroup of rotations $\stabilizer \subgroupeq \littlegroup$ that leave $\fiducialmom$ invariant.
For any particular $\fiducialmom$, the stabilizer is easily identified by acting with all elements of the (finite) little group $\littlegroup$.
By construction, each stabilizer subgroup is one of the finite groups listed after Eq.~\eqref{eq:little_group}.

The Orbit-Stabilizer theorem then implies that elements of the momentum orbit are in one-to-one-correspondence with the left cosets of $\stabilizer$~\cite{Artin:2011},
\begin{align} \label{eq:stabilizer-coset-form}
    \begin{split}
    \orbit_{\bm{P}}^{(s)} &= \{ R \cdot \fiducialmom : R \in \littlegroup \} \\ &\longleftrightarrow  
    \left\{ R_j \stabilizer : j \in \{1,2,\dots |\orbit_{\bm{P}}^{(s)}|\}\right\},
    \end{split}
\end{align}
because the rotations in each coset map the fiducial arrangement $\fiducialmom$ to a single, unique element in $\orbit_{\bm{P}}^{(s)}$.
In \cref{eq:stabilizer-coset-form}, the $j$th coset is defined as $R_j \stabilizer \equiv \{ R_j h : h \in \stabilizer \}$ where $R_j \in \littlegroup$ is a representative element of the coset.
Acting from the left with a group element $R'$ permutes the cosets on the right of Eq.~\eqref{eq:stabilizer-coset-form} in the same way as the states on the left of Eq.~\eqref{eq:stabilizer-coset-form}.
Once $\stabilizer$ is determined from $\bm{i}$, the full structure of the momentum orbit is encoded in the right-hand side and the details of the fiducial state $\fiducialmom$ no longer matter.
This means that if two distinct momentum orbits, labelled by $s$ and $s'$, share the same little group $\littlegroup$ and stabilizer group $\stabilizerS{s} = \stabilizerS{s'}$, they transform identically under $\littlegroup$, i.e.,
\begin{equation}
    D^{(s)}_{m' m}(R) = D^{(s')}_{m' m}(R)
\end{equation}
for all $m,m' \in \{1,\ldots, |\orbit_{\bm{P}}^{(s)}|\}$ and each element $R \in \littlegroup$. As such, the block diagonalization to irreps of the cubic group can also be achieved with the same set of matrices.

Conjugating $\stabilizer$ by some element $R \in \littlegroup$, $\stabilizer \rightarrow R^{-1} \stabilizer R$, is equivalent to rotating the basis states to pick a different fiducial state in the same orbit, which amounts to reordering the basis states.
This means that irrep decompositions need only be performed for one subgroup within each conjugacy class of subgroups of $\littlegroup$.
The block-diagonalization matrices for other subgroups in the same class are related by reordering the columns.

When all operators are distinguishable, the structure of the stabilizer group $\stabilizer$ is also severely restricted.
The stabilizer group $\stabilizer$ associated with products of $N$ distinguishable operators must be the intersection of a set of individual little groups $\Littlegroup{1}, \dots \Littlegroup{N} \subset O_h$, where each $\Littlegroup{j}$ is the little group of the wavevector of the $j$th operator.
These intersections can be shown to  give other---identical or smaller---finite groups in every case. Cataloguing all two-operator cases, for which all of the finite groups listed after Eq.~\eqref{eq:little_group}  appear as stabilizer groups, thus already determines the wavefunctions for all possible distinguishable-particle operators.
For cases with $N > 2$ operators, identifying the stabilizer group $\stabilizer$ allows one to select the appropriate block-diagonalization matrices already constructed from specific examples in the two-operator case.
The fact that two-operator cases already give rise to all possible change-of-basis matrices is specific to the case of distinguishable spin-zero operators and does not hold in the cases of non-zero spin; however, analogous stabilizer group considerations still restrict the distinct change-of-basis matrices to a finite number once the spin and permutation properties of all operators are specified, as described in \cref{sec:spin,sec:identical} below.

\subsection{
Complete classification for \texorpdfstring{$N$}{N} distinguishable spin-zero particles}
\label{subsec:examples} 

As discussed in \cref{ssec:stabilizer_classification}, for distinguishable spin-zero particles, solving the block-diagonalization problem for two operators in fact solves the generic $N$-operator problem.
Solution of the two-operator problem is divided into two steps:
\begin{enumerate}
    \item Enumerate the little groups and stabilizer groups associated with all possible two-body momenta.
    \item Compute the block-diagonalization matrices $U^{(s)}$ in each case (cf. \cref{eq:block_diag}).
\end{enumerate}
The remainder of this section classifies possible two-body momentum configurations and their associated little groups and stabilizers.
The results of this classification are summarized in \cref{tab:pip_Kip}.
The block-diagonalization matrices $U^{(s)}$ can be computed using the method described in \cref{subsec:cob-construction}.
Subsets of these results have previously been presented in Refs.~\cite{Luu:2011ep,Amarasinghe:2021lqa}.

\begin{table*}
\caption{
    The complete solution to the $N$-body block diagonalization problem for distinguishable spin-zero operators.
    The solution follows from classifying the possible combinations of little groups $\littlegroup$ and stabilizers $\stabilizer$ arising in the two-body case. Example states are denoted by $\ket{\pi(\bm{n}_1), K(\bm{n}_2)}$, as $\pi K$ operator construction provides a simple example of two distinguishable, spin-zero operators.
    \label{tab:pip_Kip}
    }
\begin{ruledtabular}
    \begin{tabular}{>{\centering\arraybackslash}ccccc@{\hskip 0.1in}}
$\littlegroup$ & $\stabilizer$ & Example state & Orbit dim & Irrep decomposition\\ \hline
$O_{h}$ & $O_{h}$ & $\Ket{\pi(0,0,0),\ K(0,0,0)}$ & 1 & $A_1^+$ \\ 
$O_{h}$ & $C_{4v}$ & $\Ket{\pi(0,0,1),\ K(0,0,-1)}$ & 6 & $A_1^+ \oplus E^+ \oplus T_1^-$ \\ 
$O_{h}$ & $C_{2v}$ & $\Ket{\pi(0,1,1),\ K(0,-1,-1)}$ & 12 & $A_1^+ \oplus E^+ \oplus T_2^+ \oplus T_1^- \oplus T_2^-$ \\ 
$O_{h}$ & $C_2^R$ & $\Ket{\pi(2,1,0),\ K(-2,-1,0)}$ & 24 & $A_1^+ \oplus A_2^+ \oplus 2 E^+ \oplus T_1^+ \oplus T_2^+ \oplus 2 T_1^- \oplus 2 T_2^-$ \\ 
$O_{h}$ & $C_{3v}$ & $\Ket{\pi(1,1,1),\ K(-1,-1,-1)}$ & 8 & $A_1^+ \oplus T_2^+ \oplus A_2^- \oplus T_1^-$ \\ 
$O_{h}$ & $C_2^P$ & $\Ket{\pi(1,1,2),\ K(-1,-1,-2)}$ & 24 & $A_1^+ \oplus E^+ \oplus T_1^+ \oplus 2 T_2^+ \oplus A_2^- \oplus E^- \oplus 2 T_1^- \oplus T_2^-$ \\ 
$O_{h}$ & $C_{1}$ & $\Ket{\pi(3,2,1),\ K(-3,-2,-1)}$ & 48 & $A_1^+ \oplus A_2^+ \oplus 2 E^+ \oplus 3 T_1^+ \oplus 3 T_2^+ \oplus A_1^- \oplus A_2^- \oplus 2 E^- \oplus 3 T_1^- \oplus 3 T_2^-$ \\ 
$C_{4v}$ & $C_{4v}$ & $\Ket{\pi(0,0,1),\ K(0,0,0)}$ & 1 & $A_1$ \\ 
$C_{4v}$ & $C_2^R$ & $\Ket{\pi(1,0,1),\ K(-1,0,0)}$ & 4 & $A_1 \oplus B_1 \oplus E$ \\ 
$C_{4v}$ & $C_2^P$ & $\Ket{\pi(1,1,1),\ K(-1,-1,0)}$ & 4 & $A_1 \oplus B_2 \oplus E$ \\ 
$C_{4v}$ & $C_{1}$ & $\Ket{\pi(2,1,1),\ K(-2,-1,0)}$ & 8 & $A_1 \oplus A_2 \oplus B_1 \oplus B_2 \oplus 2 E$ \\ 
$C_{3v}$ & $C_{3v}$ & $\Ket{\pi(1,1,1),\ K(0,0,0)}$ & 1 & $A_1$ \\ 
$C_{3v}$ & $C_2^P$ & $\Ket{\pi(1,1,0),\ K(0,0,1)}$ & 3 & $A_1 \oplus E$ \\ 
$C_{3v}$ & $C_{1}$ & $\Ket{\pi(1,0,-1),\ K(0,1,2)}$ & 6 & $A_1 \oplus A_2 \oplus 2 E$ \\ 
$C_{2v}$ & $C_{2v}$ & $\Ket{\pi(0,1,1),\ K(0,0,0)}$ & 1 & $A_1$ \\ 
$C_{2v}$ & $C_2^R$ & $\Ket{\pi(0,0,1),\ K(0,1,0)}$ & 2 & $A_1 \oplus B_2$ \\ 
$C_{2v}$ & $C_2^P$ & $\Ket{\pi(2,1,1),\ K(-2,0,0)}$ & 2 & $A_1 \oplus B_1$ \\ 
$C_{2v}$ & $C_{1}$ & $\Ket{\pi(-2,0,1),\ K(2,1,0)}$ & 4 & $A_1 \oplus A_2 \oplus B_1 \oplus B_2$ \\ 
$C_2^R$ & $C_2^R$ & $\Ket{\pi(1,2,0),\ K(0,0,0)}$ & 1 & $A$ \\ 
$C_2^R$ & $C_{1}$ & $\Ket{\pi(1,0,1),\ K(0,2,-1)}$ & 2 & $A \oplus B$ \\ 
$C_2^P$ & $C_2^P$ & $\Ket{\pi(1,2,2),\ K(0,0,0)}$ & 1 & $A$ \\ 
$C_2^P$ & $C_{1}$ & $\Ket{\pi(1,0,0),\ K(0,1,2)}$ & 2 & $A \oplus B$ \\ 
$C_{1}$ & $C_{1}$ & $\Ket{\pi(1,2,3),\ K(0,0,0)}$ & 1 & $A_1$ \\ 
    \end{tabular}
\end{ruledtabular}

\end{table*}

\subsubsection{Rest-frame systems}

In the rest frame, all seven conjugacy classes of little groups can act as stabilizer subgroups $\stabilizer$ within the total symmetry group $O_h$.
The stabilizer subgroups and representative choices of rest-frame momenta $[\bm{n}]$ or $[\bm{n}_1, \bm{n}_2]$ corresponding to $N\in \{1,2\}$ distinguishable particles stabilized by the group are given by:
\begin{itemize}
    \item $\stabilizer = O_h$: $[\bm{n}] = [(0,0,0)]$
    \item $\stabilizer = C_{4v}$: $[\bm{n}_1, \bm{n}_2] = [(0,0,n), (0,0,-n)]$
    \item $\stabilizer = C_{3v}$: $[\bm{n}_1, \bm{n}_2] = [(n,n,n), (-n,-n,-n)]$
    \item $\stabilizer = C_{2v}$: $[\bm{n}_1, \bm{n}_2] = [(0,n,n), (0,-n,-n)]$
    \item $\stabilizer = C_2^R$: $[\bm{n}_1, \bm{n}_2] = [(0,m,n), (0,-m,-n)]$
    \item $\stabilizer = C_2^P$: $[\bm{n}_1, \bm{n}_2] = [(m,n,n), (-m,-n,-n)]$
    \item $\stabilizer = C_1$: $[\bm{n}_1, \bm{n}_2] = [(m,l,n), (-m,-l,-n)]$.
\end{itemize}
Examples of explicit block-diagonalization matrices for rest-frame systems are tabulated in \cref{app:cob}, following the conventions described in \cref{app:group_conventions}.

\subsubsection{Boosted systems}

Boosted systems can be analyzed similarly to rest-frame systems.
First, one identifies for each little group $\littlegroup$ a representative total momentum $\bm{P}$.
Second, one identifies all (conjugacy classes of) subgroups $\stabilizer$ of the little group $\littlegroup$ which are compatible with stabilizing a set of plane waves.
Representative examples for all valid choices of $\littlegroup$ and $\stabilizer$ can be chosen as follows:
\begin{enumerate}
\item $\littlegroup = C_{4v}$: $\bm{P} = \frac{2\pi}{L}(0,0,n)$ 
\begin{itemize}
    \item $\stabilizer = C_{4v}$: $[\bm{n}] = [(0,0,n)]$
    \item $\stabilizer = C_2^R$: $[\bm{n}_1, \bm{n}_2] = [(0,m,n), (0,-m,0)]$ \\
    Note that $C_2^R$ appears in non-canonical form here as the set of reflections of the $x$ axis. 
    \item $\stabilizer = C_2^P$:\\ $[\bm{n}_1, \bm{n}_2] = [(m,m,n), (-m,-m,0)]$
    \item $\stabilizer = C_1$: $[\bm{n}_1, \bm{n}_2] = [(m,l,n), (-m,-l,0)]$,
\end{itemize}
\item $\littlegroup=C_{3v}$: $\bm{P} = \frac{2\pi}{L}(n,n,n)$
\begin{itemize}
    \item $\stabilizer = C_{3v}$: $[\bm{n}] = [(n,n,n)]$
    \item $\stabilizer = C_{2}^P$: $[\bm{n}_1, \bm{n}_2] = [(n,n,0), (0,0,n)]$
    \item $\stabilizer = C_1$: \\$[\bm{n}_1, \bm{n}_2] = [(n+m,n,0), (-m,0,n)]$,
\end{itemize}
\item $\littlegroup=C_{2v}$: $\bm{P} = \frac{2\pi}{L}(0,n,n)$
\begin{itemize}
    \item $\stabilizer = C_{2v}$: $[\bm{n}] = [(0,n,n)]$
    \item $\stabilizer = C_{2}^R$: $[\bm{n}_1, \bm{n}_2] = [(0,0,n), (0,n,0)]$
    \item $\stabilizer = C_{2}^P$: $[\bm{n}_1, \bm{n}_2] = [(-m,n,n), (m,0,0)]$
    \item $\stabilizer = C_1$: $[\bm{n}_1, \bm{n}_2] = [(-m,0,n), (m,n,0)]$,
\end{itemize}
\item $\littlegroup=C_2^R$: $\bm{P} = \frac{2\pi}{L}(n,m,0)$
\begin{itemize}
    \item $\stabilizer = C_2^R$: $[\bm{n}_1] = [(n,m,0)]$
    \item $\stabilizer = C_1$: $[\bm{n}_1, \bm{n}_2] = [(n,0,l), (0,m,-l)]$,
\end{itemize}
\item $\littlegroup=C_2^P$: $\bm{P} = \frac{2\pi}{L}(n,n,m)$
\begin{itemize}
    \item $\stabilizer = C_2^P$: $[\bm{n}_1] = [(n,n,m)]$
    \item $\stabilizer = C_1$: $[\bm{n}_1, \bm{n}_2] = [(n,0,0), (0,n,m)]$,
\end{itemize}
\item $\littlegroup=C_1$: $\bm{P} = \frac{2\pi}{L}(n,m,p)$
\begin{itemize}
    \item $\stabilizer = C_1$: $[(n,m,p)]$.
\end{itemize}
\end{enumerate}

\section{Operators with spin \label{sec:spin}}

This section extends the discussion of the previous section to operators with nonzero spin.
Since operators with half-integer spin transform in representations of $SU(2)$ (the double cover of $SO(3)$) instead of $SO(3)$ in an unbounded, continuous three-dimensional space, nonzero spin introduces the complication that the relevant symmetry group in a cubic lattice is $O_h^D$ (the double cover of $O_h$), rather than $O_h$. Consequently, the procedures discussed above must be generalized.
The approach taken here is to decompose operators into irreps of the cubic group (and its double cover) using the projection method applied to the full momentum-spin space.\footnote{
An alternative approach for incorporating spin, not pursued in this work, would treat the group representation as a tensor product of the representation under transformation of the spatial coordinates $\bm{x}_i$ and the internal spin representations.
The spatial representation associated with $N$ coordinates can be decomposed as above.
The problem then reduces to decomposing tensor products of the rotational-symmetry irreps with the spin representations of each operator. 
Computation of the Clebsch--Gordan coefficients required for this strategy is straightforward~\cite{Rykhlinskaya}, and this approach has been used in practice for constructing two-nucleon operators~\cite{Amarasinghe:2021lqa}.
However, a drawback is that the number of tensor products grows rapidly with the number of operators included.
The number of terms in, and complexity of, the resulting block-diagonalization matrices also grow rapidly with the number of operators.}

Extended operators involving fields evaluated at multiple lattice sites can also have non-trivial ``internal'' cubic transformation properties in addition to the transformation of the coordinate $\bm{x}_i$ of each operator.
The same formalism presented in this section for particles with spin can be applied to the case of extended operators with such properties by replacing spinor representation matrices with the appropriate representation matrices for such extended operators. 

The remainder of this section is organized as follows.
\Cref{ssec:spinor_irreps} discusses the irreps of the double-cover group $O_h^D$.
\Cref{ssec:spinor_operators} discusses the transformation properties of typical spinor operators.
\Cref{ssec:spinor_orbits} describes the construction of representation matrices associated with momentum-spin orbits.
\Cref{ssec:spin_examples} presents several examples of irrep decompositions for distinguishable operators with nonzero spin.

\subsection{Double-cover irreps and basis vectors}\label{ssec:spinor_irreps}

For bosonic irreps, the irreps of $O_h$ immediately furnish irreps of $O_h^D$ (see \Cref{app:group_conventions} for a concrete specification of $O_h^D$).
The key observation is that $2\pi$ rotations act trivially on states in bosonic irreps.
Therefore, the $O_h^D$-irrep matrices for group elements differing by rotations of $2\pi$ can be identified with the relevant $O_h$-irrep matrix.

For fermionic irreps, note first that the Dirac spinor representation used to define the group $O^D_h$ is a reducible representation that can be decomposed into a two-dimensional positive-parity irrep $G_1^{+}$ and a two-dimensional negative-parity irrep $G_1^{-}$ of $O^D_h$.
By convention, the Dirac spinor representation is defined in the parity eigenbasis, known as the Dirac-Pauli basis, with explicit basis states given by
\begin{equation}
\begin{split}
    \ket{1/2,+1/2,+} = \begin{pmatrix} 1 & 0 & 0 & 0 \end{pmatrix}^T,\\     
    \ket{1/2,-1/2,+} = \begin{pmatrix} 0 & 1 & 0 & 0 \end{pmatrix}^T,\\
    \ket{1/2,-1/2,-} = \begin{pmatrix} 0 & 0 & 1 & 0 \end{pmatrix}^T,\\ \ket{1/2,+1/2,-} = \begin{pmatrix} 0 & 0 & 0 & 1 \end{pmatrix}^T. \label{eq:spin_vec}
  \end{split}
\end{equation}
where the basis states are labeled as $\ket{J, J_z, \pm}$ in terms of their eigenvalues of total spin $J$, the spin $z$-projection $J_z$, and parity.
These basic irreps can be used to construct the full set of fermionic irreps.
Concrete basis vectors for the irreps of $O_h^D$ and relevant subgroups are given in terms of these spin--$1/2$ and higher-spin basis states in \cref{table:basis_vectors_double}. 
Higher-spin vectors appearing in \cref{table:basis_vectors_double} are constructed in the usual way, e.g., $\Ket{\frac{3}{2},\frac{3}{2},\pm}$ follows from the tensor product of three spin-$\frac{1}{2}$ vectors.
Additional details related to the choice of basis vectors are given in \cref{app:group_conventions}.

\bgroup
\def\arraystretch{1.2}
\begin{table*}[t]
    \centering
    \caption{Basis vectors used in this work for the fermionic irreps  of the double-cover group $O_h^D$ and its subgroups.
    As discussed in the main text, bosonic irreps of $O_h^D$ follow immediately from those of $O_h$ given in \cref{table:basis_functions}.
    The basis vectors are identical to those used in Ref.~\cite{Basak:2005ir}, and they lead to identical representation matrices to those in Ref.~\cite{Morningstar:2013bda} for all fermionic irreps.
    \label{table:basis_vectors_double}
    }
    \begin{tabular}{cccc}
\hline\hline
Group   & Irrep     & $\mu$ & Basis vector \\
\hline
$O_h^D$     & $G_1^\pm$ & 1     & $\ket{1/2,+1/2,\pm}$ \\
$O_h^D$     & $G_1^\pm$ & 2     & $\ket{1/2,-1/2,\pm}$ \\
$O_h^D$     & $H^\pm$   & 1     & $\ket{3/2,+3/2,\pm}$ \\
$O_h^D$     & $H^\pm$   & 2     & $\ket{3/2,+1/2,\pm}$ \\
$O_h^D$     & $H^\pm$   & 3     & $\ket{3/2,-1/2,\pm}$ \\
$O_h^D$     & $H^\pm$   & 4     & $\ket{3/2,-3/2,\pm}$ \\
$O_h^D$     & $G_2^\pm$ & 1     & $\sqrt{1/6}\ket{5/2,-5/2,\pm} - \sqrt{5/6}\ket{5/2,+3/2,\pm}$\\
$O_h^D$     & $G_2^\pm$ & 2     & $\sqrt{1/6}\ket{5/2,+5/2,\pm} - \sqrt{5/6}\ket{5/2,-3/2,\pm}$\\
\hline
$\Dic_{4}$  & $G_1$     & 1     & $\ket{1/2,+1/2,+}$ \\
$\Dic_{4}$  & $G_1$     & 2     & $\ket{1/2,-1/2,+}$ \\
$\Dic_{4}$  & $G_2$     & 1     & $\sqrt{1/6}\ket{5/2,-5/2,+} - \sqrt{5/6}\ket{5/2,+3/2,+}$\\
$\Dic_{4}$  & $G_2$     & 2     & $\sqrt{1/6}\ket{5/2,+5/2,+} - \sqrt{5/6}\ket{5/2,-3/2,+}$\\
\hline
$\Dic_{3}$  & $G$       & 1     & $\ket{1/2,+1/2,+}$ \\
$\Dic_{3}$  & $G$       & 2     & $\ket{1/2,-1/2,+}$ \\
$\Dic_{3}$  & $F_1$     & 1     & \cref{eq:Dic3_F1}\\ 
$\Dic_{3}$  & $F_2$     & 2     & \cref{eq:Dic3_F2} \\ 
\hline
$\Dic_{2}$  & $G$       & 1     & $\ket{1/2,+1/2,+}$ \\
$\Dic_{2}$  & $G$       & 2     & $\ket{1/2,-1/2,+}$ \\
\hline
$C_4^R$     & $F_1$     & 1     & $\ket{1/2,+1/2,+}$ \\
$C_4^R$     & $F_2$     & 1     & $\ket{1/2,-1/2,+}$ \\
\hline
$C_4^P$     & $F_1$     & 1     & $\sqrt{1/2}\ket{1/2,1/2,+} + (1-i)/2\ket{1/2,-1/2,+}$\\
$C_4^P$     & $F_2$     & 1     & $\sqrt{1/2}\ket{1/2,1/2,+} - (1-i)/2\ket{1/2,-1/2,+}$\\
\hline
$C_1^D$     & $F$       & 1     & $\ket{1/2,1/2,+}$\\
\hline\hline
\end{tabular}

\end{table*}
\egroup

\subsection{Operator representations\label{ssec:spinor_operators}}
Having defined the double-cover group structure and irreps for $O^D_h$ and its little groups, it is useful to record the transformation properties of typical spinor interpolating operators.

Creation and annihilation operators for spin-1/2 particles, denoted  $\overline{\psi}(\bm{x})$ and $\psi(\bm{x})$ respectively, transform as the spinor representation and its conjugate: 
\begin{equation}
\begin{aligned}
  \hat{U}(R^D) \, \overline{\psi}(\bm{x})_{\alpha} \, \hat{U}^\dag(R^D) &= \sum_{\beta} \overline{\psi}_{\beta}(R \bm{x}) S_{\beta\alpha}(R^D), \label{eq:qbar} \\
    \hat{U}(R^D) \,  \psi(\bm{x})_{\alpha} \, \hat{U}^\dag(R^D) &= \sum_{\beta} \psi_{\beta}(R \bm{x}) S_{\beta\alpha}(R^D)^*.
  \end{aligned}
\end{equation}
Here $\hat{U}(R^D)$ indicates the quantum operator that implements the $O^D_h$ transformation $R^D$ and 
$S(R^D)$ is the spinor representation matrix associated with the group element $R^D$,
\begin{align}
S(R^D) \equiv R^D \in O_h^D, \label{eq:spinor_rep_matrices} 
\end{align}
since this is the defining representation.

Since products of $\overline{\psi}_{\alpha}(\bm{x})$ operators act on the vacuum to create single- and multi-particle states, the same transformation rule for $\overline{\psi}_{\alpha}(\bm{x})$ operators is chosen as for irrep basis vectors in Eq.~\eqref{eq:irrep-basis-states}.
Spinor operators with larger spin (e.g., spin $3/2$) can be constructed using tensor products of spin-$1/2$ operators.

With these conventions, one readily confirms the invariance of the usual kinetic terms of the Hamiltonian for free relativistic spin-$1/2$ fermions,
 $   H_{\psi} = \int d^3 \bm{x} \sum_{j} \overline{\psi}(\bm{x}) \gamma^j \partial_j \psi(\bm{x})$.
This invariance also applies on a discrete spatial lattice under the subgroup $O_h$ of all rotations, meaning these conventions are compatible with familiar Hamiltonians (or actions) appearing in practical calculations in lattice gauge theory.

These transformation properties lead to a natural generalization of the space of operators introduced in \cref{sec:spinzero}.
Operators with spin can be specified in various ways; the generalized algorithm defined in the following is insensitive to this choice. 
Particularly for products of local operators, one useful representation arises from constructing operators with definite total spin $J$ under $SU(2)$.
In this case, besides wavenumbers $\bm{n}_i$ specifying the momenta, the extended space is defined by the total spin $J$, spin component $J_z$, and intrinsic parity $\pm$.
Given the operator transformation rules above, these states transform as
\begin{equation}
\begin{split}
  \mathcal{D}(R) \Ket{\bm{n},J,J_z,\pm} = \sum_{J_z' = -J}^{J} \Ket{R\bm{n},J,J_z',\pm} D^{[J]}_{J_z',J_z}(R^D) , \label{eq:spintransform}
\end{split}
\end{equation}
where $D^{[J]}(R^D)$ is the appropriate representation matrix for the spin-$J$ operator.
In general, one could choose operators in any reducible or irreducible representation of the little group instead of those defined by continuum spin $J$.
Though most of the $O_h^D$ irreps coincide with the $\ket{J, J_z, \pm}$ basis states, for large-$J$ states and extended operators, the $O_h^D$ representation is typically reducible.

\subsection{Momentum-spin orbits and their representations \label{ssec:spinor_orbits}}

For operators with spin, the notion of a momentum orbit from \cref{ssec:momentum_orbit} is extended to a combined momentum-spin orbit.
Because the spatial coordinates and spin degrees of freedom transform in distinct spaces, the combined orbit belongs to their tensor product.
The spin portion of the orbit can itself be understood as the tensor product of the individual spin-$J_1$ through spin-$J_N$ basis states.
In other words, the extended \emph{momentum-spin orbit} can be written as
\begin{align}
    \tilde{\orbit}_{\bm{P}}^{(s)} 
    &\equiv \left\{ R^D \cdot [\wvnum_1,\alpha_1 \ldots,\wvnum_N, \alpha_N] : R^D \in \littlegroup^D\} \right\}\\
    &=\left\{[\wvnum_1,\alpha_1 \ldots,\wvnum_N, \alpha_N]^{(s)}_{m} :
    m \in \{1, 2, \dots ,  |\tilde\orbit_{\bm{P}}^{(s)}|\} \right\} , \nonumber
\end{align}
where the $\alpha_i$ contain the spin and parity quantum numbers.
As above, elements within this orbit are indexed by the integer label $m$.
The value for $\alpha_i$ of the $m$th element in the orbit will be denoted $\alpha_i^{[m]}$.
The dimension of the momentum-spin orbit, $|\tilde{\orbit}_{\bm{p}}^{(s)}|$, is given by the product of the dimension of the momentum orbit with the dimensions of the individual spin representations $|J_i|=2J_i+1$, i.e., 
\begin{align}
|\tilde{\orbit}_{\bm{p}}^{(s)}| 
= |\orbit_{\bm{p}}^{(s)}| \times
\prod_{i=1}^N |J_i|.
\end{align}

Once the momentum-spin orbit has been constructed, the representation matrices $\tilde{D}^{(s)}_{m'm}(R^D)$ follow using the analog of Eq.~\eqref{eq:Dbraket} as matrix elements, labelled by $m'$ and $m$, between states of the momentum-spin operators.
For the case of $N$ spin-1/2 operators, these matrix elements are explicitly given by 
\begin{widetext}
\begin{equation}
  \begin{split}
    \tilde{D}^{(s)}_{m'm}(R^D) &\equiv \amp{ [\bm{n}_1,\alpha_1,\ldots, \bm{n}_N,\alpha_N]_{m'}}{\mathcal{D}(R^D)}{[\bm{n}_1,\alpha_1,\ldots, \bm{n}_N,\alpha_N]_m}.  \\
    &= \amp{ [ \bm{n}_1, \dots, \bm{n}_N ]_{m'}} {\mathcal{D}(R)}{[ \bm{n}_1, \dots, \bm{n}_N ]_{m}} \amp{ \alpha_1^{[m']}}{\mathcal{D}(R^D)}{\alpha_1^{[m]}} \dots \amp{ \alpha_N^{[m']}}{\mathcal{D}(R^D)}{\alpha_N^{[m]}} \\
    &= D^{(s)}_{m'm}(R) S_{\alpha_1^{[m']}\alpha_1^{[m]}}(R^D) \dotsb S_{\alpha_N^{[m']}\alpha_N^{[m]}}(R^D),
    \label{eq:DbraketSpin}
  \end{split}
\end{equation}
where $S(R^D)$ denotes the spinor representation matrices defined in \cref{eq:spinor_rep_matrices} and $R$ denotes the restriction of $R^D$ to $O_h$.
Expressions involving a mixture of spin-$0$ and spin-$1/2$ operators are obtained simply by removing spin labels $\alpha_i$ from spin-0 states and removing the corresponding transformation factors of $S(R^D)$. 

For operators (e.g., spin-$J$ operators or operators describing spatially extended objects) in a generic representation $\tilde{\Gamma}$ of the little group, the associated representation matrices are denoted $D^{[\tilde{\Gamma}]}_{\alpha \beta}(R^D)$, 
with $\alpha,\beta \in \{1, 2, \dots |\tilde{\Gamma}|\}$.
Given these representations for the individual operator transformations, products of $N$ such plane-wave operators transform with
\begin{equation}
  \begin{split}
    \tilde{D}^{(s)}_{m'm}(R^D) 
    &=  D^{(s)}_{m'm}(R) 
        D^{[\tilde{\Gamma}_1]}_{\alpha_1^{[m']}\alpha_1^{[m]}}(R^D) 
        \dotsb 
        D^{[\tilde{\Gamma}_N]}_{\alpha_N^{[m']}\alpha_N^{[m]}}(R^D).
    \label{eq:DbraketGeneric}
  \end{split}
\end{equation}
\end{widetext}

Once the representation matrices  $\tilde{D}^{(s)}_{m'm}(R^D)$ are determined, precisely the same steps described in Sec.~\ref{sec:spinzero} are used to determine block-diagonalization matrices of \cref{eq:block_diag}: project against the irrep matrices $D^{(\Gamma)}_{\mu'\mu}(R^D)$ using Schur's lemma, orthogonalize degenerate irreps, and fill out the remaining rows using the transition operators $T_{\mu\nu}^{(\Gamma,s)}$.

\subsection{Examples \label{ssec:spin_examples}}

Unlike the spin-zero case,
the irrep decompositions for products of $N$ plane-wave operators with non-zero spin depend on the spins of the operators and cannot be catalogued completely in terms of two-body results.
For a given set of spins, however, the irrep decompositions that can arise are severely restricted and can be fully classified for each of the little groups $\doublelittlegroup$ and stabilizer groups $\doublestabilizer$, where $\doublelittlegroup \subgroupeq O_h^D$ and $\doublestabilizer \subgroupeq \doublelittlegroup$ indicate the little group and stabilizer group constructed from the momenta as before, now within the double cover group $O_h^D$.
The possible $\doublelittlegroup$ and $\doublestabilizer$ can then be enumerated, and their irrep decompositions and block-diagonalization matrices can be tabulated analogously to the case of $N$ spin-zero plane-wave operators.
This section collects several explicit examples of phenomenologically relevant systems described by products of operators with non-zero spin and presents the irrep decompositions that arise during their block diagonalization.
Explicit change-of-basis matrices are constructed in the accompanying code package \cite{code}.

\subsubsection{\texorpdfstring{
$n p$
}{
n p
}}

\begin{table*}
\caption{
    Combinations of irreps arising in decompositions of $np$ operator orbits. Details are as in Table~\ref{tab:pip_Kip}. 
    \label{tab:n_p}
    }
    \begin{ruledtabular}
    \begin{tabular}{>{\centering\arraybackslash}ccccc@{\hskip 0.1in}}
$\littlegroup$ & $\stabilizer$ & Example state & Orbit dim & Irrep decomposition\\ \hline
$O_{h}$ & $O_{h}$ & $\Ket{n(0,0,0),\ p(0,0,0)}$ & 4 & $A_1^+ \oplus T_1^+$ \\ 
$O_{h}$ & $C_{4v}$ & $\Ket{n(0,0,1),\ p(0,0,-1)}$ & 24 & $A_1^+ \oplus E^+ \oplus 2 T_1^+ \oplus T_2^+ \oplus A_1^- \oplus E^- \oplus 2 T_1^- \oplus T_2^-$ \\ 
$O_{h}$ & $C_{2v}$ & $\Ket{n(0,1,1),\ p(0,-1,-1)}$ & 48 & $A_1^+ \oplus A_2^+ \oplus 2 E^+ \oplus 3 T_1^+ \oplus 3 T_2^+ \oplus A_1^- \oplus A_2^- \oplus 2 E^- \oplus 3 T_1^- \oplus 3 T_2^-$ \\ 
$O_{h}$ & $C_{3v}$ & $\Ket{n(1,1,1),\ p(-1,-1,-1)}$ & 32 & $A_1^+ \oplus A_2^+ \oplus E^+ \oplus 2 T_1^+ \oplus 2 T_2^+ \oplus A_1^- \oplus A_2^- \oplus E^- \oplus 2 T_1^- \oplus 2 T_2^-$ \\ 
$O_{h}$ & $C_2^R$ & $\Ket{n(2,1,0),\ p(-2,-1,0)}$ & 96 & $2 A_1^+ \oplus 2 A_2^+ \oplus 4 E^+ \oplus 6 T_1^+ \oplus 6 T_2^+ \oplus 2 A_1^- \oplus 2 A_2^- \oplus 4 E^- \oplus 6 T_1^- \oplus 6 T_2^-$ \\ 
$O_{h}$ & $C_2^P$ & $\Ket{n(2,1,1),\ p(-2,-1,-1)}$ & 96 & $2 A_1^+ \oplus 2 A_2^+ \oplus 4 E^+ \oplus 6 T_1^+ \oplus 6 T_2^+ \oplus 2 A_1^- \oplus 2 A_2^- \oplus 4 E^- \oplus 6 T_1^- \oplus 6 T_2^-$ \\ 
$O_{h}$ & $C_{1}$ & $\Ket{n(3,2,1),\ p(-3,-2,-1)}$ & 192 & $4 A_1^+ \oplus 4 A_2^+ \oplus 8 E^+ \oplus 12 T_1^+ \oplus 12 T_2^+ \oplus 4 A_1^- \oplus 4 A_2^- \oplus 8 E^- \oplus 12 T_1^- \oplus 12 T_2^-$ \\ 
$C_{4v}$ & $C_{4v}$ & $\Ket{n(0,0,1),\ p(0,0,0)}$ & 4 & $A_1 \oplus A_2 \oplus E$ \\ 
$C_{4v}$ & $C_2^R$ & $\Ket{n(1,0,1),\ p(-1,0,0)}$ & 16 & $2 A_1 \oplus 2 A_2 \oplus 2 B_1 \oplus 2 B_2 \oplus 4 E$ \\ 
$C_{4v}$ & $C_2^P$ & $\Ket{n(1,1,1),\ p(-1,-1,0)}$ & 16 & $2 A_1 \oplus 2 A_2 \oplus 2 B_1 \oplus 2 B_2 \oplus 4 E$ \\ 
$C_{4v}$ & $C_{1}$ & $\Ket{n(2,1,1),\ p(-2,-1,0)}$ & 32 & $4 A_1 \oplus 4 A_2 \oplus 4 B_1 \oplus 4 B_2 \oplus 8 E$ \\ 
$C_{3v}$ & $C_{3v}$ & $\Ket{n(1,1,1),\ p(0,0,0)}$ & 4 & $A_1 \oplus A_2 \oplus E$ \\ 
$C_{3v}$ & $C_2^P$ & $\Ket{n(1,1,0),\ p(0,0,1)}$ & 12 & $2 A_1 \oplus 2 A_2 \oplus 4 E$ \\ 
$C_{3v}$ & $C_{1}$ & $\Ket{n(1,0,-1),\ p(0,1,2)}$ & 24 & $4 A_1 \oplus 4 A_2 \oplus 8 E$ \\ 
$C_{2v}$ & $C_{2v}$ & $\Ket{n(0,1,1),\ p(0,0,0)}$ & 4 & $A_1 \oplus A_2 \oplus B_1 \oplus B_2$ \\ 
$C_{2v}$ & $C_2^R$ & $\Ket{n(0,0,1),\ p(0,1,0)}$ & 8 & $2 A_1 \oplus 2 A_2 \oplus 2 B_1 \oplus 2 B_2$ \\ 
$C_{2v}$ & $C_2^P$ & $\Ket{n(2,1,1),\ p(-2,0,0)}$ & 8 & $2 A_1 \oplus 2 A_2 \oplus 2 B_1 \oplus 2 B_2$ \\ 
$C_{2v}$ & $C_{1}$ & $\Ket{n(-2,0,1),\ p(2,1,0)}$ & 16 & $4 A_1 \oplus 4 A_2 \oplus 4 B_1 \oplus 4 B_2$ \\ 
$C_2^R$ & $C_2^R$ & $\Ket{n(1,2,0),\ p(0,0,0)}$ & 4 & $2 A \oplus 2 B$ \\ 
$C_2^R$ & $C_{1}$ & $\Ket{n(1,0,1),\ p(0,2,-1)}$ & 8 & $4 A \oplus 4 B$ \\ 
$C_{1}$ & $C_{1}$ & $\Ket{n(1,2,3),\ p(0,0,0)}$ & 4 & $4 A$ \\ 
    \end{tabular}
\end{ruledtabular}

\end{table*}

The $np$ system provides an example of a system with the quantum numbers of two distinguishable spin-$\frac{1}{2}$ particles.
The cubic irreps of the $n$ and $p$ operators both correspond to $G_1^+$.
Consideration of isospin will be deferred until discussion of internal symmetry groups in \cref{sec:identical}.

\Cref{tab:n_p} classifies the different orbit patterns in terms of little groups and stabilizer groups and shows the associated irrep decompositions of $np$, extending the results presented in Ref.~\cite{Amarasinghe:2021lqa}.
Several general features arising in decompositions for operators with non-zero spin are illustrated by this example. In particular, the orbit dimensions are four times larger than in the spin-zero case because the two-nucleon spin space $(\frac{1}{2}\otimes \frac{1}{2}= 0 \oplus 1)$ has dimension four.
Different patterns of irreps arise for $np$ than for the case of distinguishable spin-zero particles.

\subsubsection{\texorpdfstring{
$p\pi^+$
}{
p pi
}}

\begin{table*}
    \caption{Combinations of irreps arising in decompositions of $p\pi^+$ operator orbits. Details are as in Table~\ref{tab:pip_Kip}. 
    \label{tab:N_pi}
    }
    \begin{ruledtabular}
    \begin{tabular}{>{\centering\arraybackslash}ccccc@{\hskip 0.1in}}
$\littlegroup$ & $\stabilizer$ & Example state & Orbit dim & Irrep decomposition\\ \hline
$O_{h}$ & $O_{h}$ & $\Ket{p(0,0,0),\ \pi^+(0,0,0)}$ & 2 & $G_1^-$ \\ 
$O_{h}$ & $C_{4v}$ & $\Ket{p(0,0,1),\ \pi^+(0,0,-1)}$ & 12 & $G_1^+ \oplus H^+ \oplus G_1^- \oplus H^-$ \\ 
$O_{h}$ & $C_{2v}$ & $\Ket{p(0,1,1),\ \pi^+(0,-1,-1)}$ & 24 & $G_1^+ \oplus G_2^+ \oplus 2 H^+ \oplus G_1^- \oplus G_2^- \oplus 2 H^-$ \\ 
$O_{h}$ & $C_{3v}$ & $\Ket{p(1,1,1),\ \pi^+(-1,-1,-1)}$ & 16 & $G_1^+ \oplus G_2^+ \oplus H^+ \oplus G_1^- \oplus G_2^- \oplus H^-$ \\ 
$O_{h}$ & $C_2^R$ & $\Ket{p(2,1,0),\ \pi^+(-2,-1,0)}$ & 48 & $2 G_1^+ \oplus 2 G_2^+ \oplus 4 H^+ \oplus 2 G_1^- \oplus 2 G_2^- \oplus 4 H^-$ \\ 
$O_{h}$ & $C_2^P$ & $\Ket{p(2,1,1),\ \pi^+(-2,-1,-1)}$ & 48 & $2 G_1^+ \oplus 2 G_2^+ \oplus 4 H^+ \oplus 2 G_1^- \oplus 2 G_2^- \oplus 4 H^-$ \\ 
$O_{h}$ & $C_{1}$ & $\Ket{p(3,2,1),\ \pi^+(-3,-2,-1)}$ & 96 & $4 G_1^+ \oplus 4 G_2^+ \oplus 8 H^+ \oplus 4 G_1^- \oplus 4 G_2^- \oplus 8 H^-$ \\ 
$C_{4v}$ & $C_{4v}$ & $\Ket{p(0,0,1),\ \pi^+(0,0,0)}$ & 2 & $G_1$ \\ 
$C_{4v}$ & $C_2^R$ & $\Ket{p(1,0,1),\ \pi^+(-1,0,0)}$ & 8 & $2 G_1 \oplus 2 G_2$ \\ 
$C_{4v}$ & $C_2^P$ & $\Ket{p(1,1,1),\ \pi^+(-1,-1,0)}$ & 8 & $2 G_1 \oplus 2 G_2$ \\ 
$C_{4v}$ & $C_{1}$ & $\Ket{p(2,1,1),\ \pi^+(-2,-1,0)}$ & 16 & $4 G_1 \oplus 4 G_2$ \\ 
$C_{3v}$ & $C_{3v}$ & $\Ket{p(1,1,1),\ \pi^+(0,0,0)}$ & 2 & $G$ \\ 
$C_{3v}$ & $C_2^P$ & $\Ket{p(1,1,0),\ \pi^+(0,0,1)}$ & 6 & $F_1 \oplus F_2 \oplus 2 G$ \\ 
$C_{3v}$ & $C_{1}$ & $\Ket{p(1,0,-1),\ \pi^+(0,1,2)}$ & 12 & $2 F_1 \oplus 2 F_2 \oplus 4 G$ \\ 
$C_{2v}$ & $C_{2v}$ & $\Ket{p(0,1,1),\ \pi^+(0,0,0)}$ & 2 & $G$ \\ 
$C_{2v}$ & $C_2^R$ & $\Ket{p(0,0,1),\ \pi^+(0,1,0)}$ & 4 & $2 G$ \\ 
$C_{2v}$ & $C_2^P$ & $\Ket{p(2,1,1),\ \pi^+(-2,0,0)}$ & 4 & $2 G$ \\ 
$C_{2v}$ & $C_{1}$ & $\Ket{p(-2,0,1),\ \pi^+(2,1,0)}$ & 8 & $4 G$ \\ 
$C_2^R$ & $C_2^R$ & $\Ket{p(1,2,0),\ \pi^+(0,0,0)}$ & 2 & $F_1 \oplus F_2$ \\ 
$C_2^R$ & $C_{1}$ & $\Ket{p(1,0,1),\ \pi^+(0,2,-1)}$ & 4 & $2 F_1 \oplus 2 F_2$ \\ 
$C_{1}$ & $C_{1}$ & $\Ket{p(1,2,3),\ \pi^+(0,0,0)}$ & 2 & $2 F$ \\ 
    \end{tabular}
\end{ruledtabular}

\end{table*}

The $p\pi^+$ system is an example of a fermionic system with distinguishable operators transforming in different irreps.
The operator spins in this example correspond to the $G_1^+$ irrep for the proton and the $A_1^-$ irrep for the pion.

\Cref{tab:N_pi} shows the distinct irrep decompositions of $p\pi^+$ orbits, which are again classified by the corresponding little groups and stabilizer groups.
This extends the results presented in Ref.~\cite{Prelovsek:2016iyo}.
The orbit dimensions are twice as large as in the spin-zero particle case because of the nucleon spin and can be decomposed into direct sums of fermionic irreps.

\subsubsection{\texorpdfstring{
$p \pi^+ \pi^0$
}{
p pi+ pi0
}}

\begin{table*}
\caption{Combinations of irreps arising in decompositions of $p\pi^+\pi^0$ operator orbits. Details are as in Table~\ref{tab:pip_Kip}. 
    \label{tab:p_pip_pi0}
    }
\begin{ruledtabular}
    \begin{tabular}{>{\centering\arraybackslash}ccccc@{\hskip 0.1in}}
      $\littlegroup$ & $\stabilizer$ & Example state & Orbit dim & Irrep decomposition \\\hline
      $O_h$     &   $O_h$   & $\Ket{p(0,0,0),\ \pi^+(0,0,0),\ \pi^0(0,0,0)}$ & $2$  & $G_1^+$ \\
      $O_h$     &   $C_{4v}$   & $\Ket{p(0,0,1),\ \pi^+(0,0,-1),\ \pi^0(0,0,0)}$ & $12$  & $G_1^+ \oplus H^+ \oplus G_1^- \oplus H^- $ \\
      $O_h$     &   $C_{2v}$   & $\Ket{p(0,1,1),\ \pi^+(0,-1,-1),\ \pi^0(0,0,0)}$ & $24$  & $G_1^+ \oplus G_2^+ \oplus 2 H^+ \oplus G_1^- \oplus G_2^- \oplus 2H^- $  \\
      $O_h$     &   $C_{3v}$   & $\Ket{p(1,1,1),\ \pi^+(-1,-1,-1),\ \pi^0(0,0,0)}$ & $16$  & $G_1^+ \oplus G_2^+ \oplus H^+ \oplus G_1^- \oplus G_2^- \oplus H^- $ \\
      $O_h$     &   $C_2^R$   & $\Ket{p(2,1,0),\ \pi^+(-2,-1,0),\ \pi^0(0,0,0)}$ & $48$  & $2 G_1^+ \oplus 2 G_2^+ \oplus 4 H^+ \oplus 2 G_1^- \oplus 2 G_2^- \oplus 4H^- $  \\
      $O_h$     &   $C_2^P$   & $\Ket{p(2,1,1),\ \pi^+(-2,-1,-1),\ \pi^0(0,0,0)}$ & $48$  & $2 G_1^+ \oplus 2 G_2^+ \oplus 4 H^+ \oplus 2 G_1^- \oplus 2 G_2^- \oplus 4H^- $  \\
      $O_h$     &   $C_1$     & $\Ket{p(3,2,1),\ \pi^+(-3,-2,-1),\ \pi^0(0,0,0)}$ & $96$  & $4 G_1^+ \oplus 4 G_2^+ \oplus 8 H^+ \oplus 4 G_1^- \oplus 4 G_2^- \oplus 8H^- $  \\
      $C_{4v}$  &   $C_{4v}$   & $\Ket{p(0,0,1),\ \pi^+(0,0,0),\ \pi^0(0,0,0)}$ & $2$  & $G_1$ \\
      $C_{4v}$  &   $C_2^R$     & $\Ket{p(1,0,1),\ \pi^+(-1,0,0),\ \pi^0(0,0,0)}$ & $8$  & $2G_1 \oplus 2G_2$ \\
      $C_{4v}$  &   $C_2^P$     & $\Ket{p(1,1,1),\ \pi^+(-1,-1,0),\ \pi^0(0,0,0)}$ & $8$  & $2G_1 \oplus 2G_2$ \\
      $C_{4v}$  &   $C_1$     & $\Ket{p(2,1,1),\ \pi^+(-2,-1,0),\ \pi^0(0,0,0)}$ & $16$  & $4G_1 \oplus 4G_2$ \\
      $C_{3v}$  &   $C_{3v}$   & $\Ket{p(1,1,1),\ \pi^+(0,0,0),\ \pi^0(0,0,0)}$ & $2$  & $G$ \\
      $C_{3v}$  &   $C_2^P$     & $\Ket{p(1,1,0),\ \pi^+(0,0,1),\ \pi^0(0,0,0)}$ & $6$  & $F_1 \oplus F_2 \oplus 2G$ \\
      $C_{3v}$  &   $C_1$     & $\Ket{p(1,0,-1),\ \pi^+(0,1,2),\ \pi^0(0,0,0)}$ & $12$  & $2F_1 \oplus 2F_2 \oplus 4G$ \\
      $C_{2v}$  &   $C_{2v}$   & $\Ket{p(0,1,1),\ \pi^+(0,0,0),\ \pi^0(0,0,0)}$ & $2$  & $G$ \\
      $C_{2v}$  &   $C_2^R$     & $\Ket{p(0,0,1),\ \pi^+(0,1,0),\ \pi^0(0,0,0)}$ & $4$  & $2G$ \\
      $C_{2v}$  &   $C_2^P$     & $\Ket{p(2,1,1),\ \pi^+(-2,0,0),\ \pi^0(0,0,0)}$ & $4$  & $2G$ \\
      $C_{2v}$  &   $C_1$     & $\Ket{p(-2,0,1),\ \pi^+(2,1,0),\ \pi^0(0,0,0)}$ & $8$  & $4G$ \\
      $C_{2}$   &   $C_2$     & $\Ket{p(1,2,0),\ \pi^+(0,0,0),\ \pi^0(0,0,0)}$ & $2$  & $F_1 \oplus F_2$ \\
      $C_{2}$   &   $C_1$     & $\Ket{p(1,0,1),\ \pi^+(0,2,-1),\ \pi^0(0,0,0)}$ & $4$  & $2F_1 \oplus 2F_2$ \\
      $C_{1}$   &   $C_1$     & $\Ket{p(1,2,3),\ \pi^+(0,0,0),\ \pi^0(0,0,0)}$ & $2$  & $2F$ \\
    \end{tabular}
\end{ruledtabular}
\end{table*}

The $p \pi^+ \pi^0$ system provides an example with more than two distinguishable operators and nonzero spin.
States with definite isospin ($I\in\{\tfrac{1}{2}, \tfrac{3}{2}\}$) involve linear combinations with $n\pi^+\pi^+$ operators and can be treated using the methods of \cref{sec:identical} below.

\Cref{tab:p_pip_pi0} shows the irrep decomposition of $p\pi^+\pi^0$ orbits classified by the corresponding little groups and stabilizer groups.
The patterns of little groups and stabilizer groups occurring are identical to those in Table~\ref{tab:N_pi}.
The only difference in the irreps appearing in the $p\pi^+$ and $p\pi^+\pi^0$ decompositions is for the case of all operators at rest, which for $p\pi^+$ corresponds to $G_1^-$ but for $p\pi^+\pi^0$ corresponds to $G_1^+$.

\section{Internal symmetries and identical particles} \label{sec:identical}

In addition to the rotational transformation properties discussed so far, physical states also carry quantum numbers such as charge and flavor.
Moreover, for states including identical particles, exchanging such particles leaves the state unchanged up to a possible sign.
More precisely, as already exploited in Ref.~\cite{Hansen:2020zhy}, little-group irreps must be paired appropriately with irreps under other quantum numbers such that their combined transformations under particle exchanges are symmetric (antisymmetric) with respect to all possible exchanges of identical bosons (fermions), and are otherwise unconstrained.
By identifying the definite operator-exchange properties of cubic-group irreps, the framework described above can thus be readily extended to be compatible with other quantum numbers.

\subsection{Labeling by exchange-group irreps}

Since the same rotations are applied to all operators in an $N$-operator basis, rotation and permutation operations commute, and Schur--Weyl duality guarantees that the rotational group irreps can be simultaneously labeled by specific irreps of the symmetric group $S_N$~\cite{Fulton:2004uyc}. This naturally divides any space that is closed under rotations and permutations into \emph{blocks} described by the pair $(\Gamma, \lambda)$ of a rotational-group irrep $\Gamma$ and an $S_N$ irrep $\lambda$.
In the following, Young diagrams will be used to identify particular choices of $\lambda$ \cite{Georgi:2000vve}. 
Note that multiple blocks may have the same $\Gamma$ or the same $\lambda$, i.e., there is no one-to-one correspondence between the rotational irreps and permutation irreps.

In some cases, it is not necessary to consider definite representations under the full space of permutations.
For example, operators may be distinguishable by having different total isospin or other flavor quantum number.
To make this identification concrete and automatic, it is assumed that the operators $\mathcal{O}_1$, \dots, $\mathcal{O}_N$ can be respectively associated with \emph{internal labels} $\internallabel_1$, \dots, $\internallabel_N$.
In this case, the \emph{exchange group} is taken to be the subgroup of permutations
\begin{equation}
    S \equiv S_{N_1} \times S_{N_2} \times \cdots \subgroupeq S_N
\end{equation}
corresponding to exchanges among subsets of identically labeled particles of size $N_1$, $N_2$, $\cdots$ with $N_1 + N_2 + \dots = N$.
In this case, the categorization of rotational irreps can be given by the rotational representation and the individual $S_{N_1}, S_{N_2}, \dots$ irreps as $(\Gamma, \lambda_1, \lambda_2, \dots)$.

\subsection{Extended orbits}

Orbits constructed as in \Cref{ssec:momentum_orbit,ssec:spinor_orbits} may not necessarily be closed under the exchange group $S$.
To ensure states can always be constructed with definite permutation properties, the orbit must be extended to include all states generated by applying permutations in $S$.
The labeling $s$ of orbits and the indices $m$ of basis states will continue to be used for these \emph{extended orbits}, which will be denoted as $\hat{\orbit}_{\bm{P}}^{(s)}$:
\begin{align}
    \hat{\orbit}_{\bm{P}}^{(s)} &\equiv 
    \left\{ g \cdot [\bm{n}_1,\alpha_1, \internallabel_1, \dots, \bm{n}_N,\alpha_N,\internallabel_N]: g\in\littlegroup^D \times S
    \right\} \nonumber \\
    \begin{split}
    &= 
    \left\{ [\bm{n}_1,\alpha_1, \internallabel_1, \dots, \bm{n}_N,\alpha_N,\internallabel_N]_m^{(s)}: \right.\\
    & \quad\quad\quad\quad m \in \{1,2,\dots, \left. |\hat{\orbit}_{\bm{P}}^{(s)}| \} \right\},
    \end{split} \label{eq:extended_orbit}
\end{align}
where $\alpha_i$ label the spin and parity of the $i$th particle and where $\internallabel_i$ labels its internal quantum numbers.
Orbits then furnish representations of exchange-group elements $\permutation \in S$ in the usual way
\begin{equation}
  D_{m'm}^{(s)}(\permutation) = \amp{ s,m' }{\mathcal{D}(\mathcal{\permutation})}{ s,m }.
\end{equation}
Since $\permutation$ corresponds to identical-operator exchange, $D_{m'm}^{(s)}(\permutation) \in \{0, 1\}$.

\subsection{Projection to exchange-group irreps}
\label{sec:exchangegroupirreps}

Just as Schur's lemma was applied in Eq.~\eqref{eq:schur-f-projection} to project into a basis with specific rotational-group irrep $\Gamma$, it can be applied to project simultaneously into a basis with a specific permutation-group irrep.
The representation theory of the permutation group has been well studied
(see Ref.~\cite{Keppeler:2013yla} for guidance on standard textbooks).
Each irrep of $S_n$ is labeled by a Young diagram consisting of $n$ boxes in left-justified rows with row lengths in non-increasing order.
Closely related is the notion of a Young tableau, in which the $n$ boxes in a given Young diagram are filled with the numbers $\{1,\dots,n\}$ distributed such that each row and each column is strictly ascending.
Let $\mathcal{Y}_n$ denote the set of Young tableaux with $n$ boxes.
For example, 
$\mathcal{Y}_3 = \{ 
\scalebox{.7}{\begin{ytableau}1 & 2 & 3 \end{ytableau}},
\scalebox{.7}{\begin{ytableau}1 & 2 \\ 3 \end{ytableau}},
\scalebox{.7}{\begin{ytableau}1 & 3 \\ 2 \end{ytableau}},
\scalebox{.7}{\begin{ytableau}1 \\ 2 \\ 2 \end{ytableau}}
\}$. 

The connection to the internal symmetry group follows from the fact that products of operators can be taken to transform as (a row of) an irrep of the exchange group, labelled by a Young tableau $\Theta \in \mathcal{Y}_n$.
For each Young tableau $\Theta$, there exists a projection operator in the group algebra $\Pi_\Theta$ which projects onto the relevant row~\cite{Alcock-Zeilinger:2016cva,Alcock-Zeilinger:2016sxc,Keppeler:2013yla}.
In fact, mirroring the construction in \cref{subsec:cob-construction}, the group algebra can be decomposed completely into a basis of idempotents $\Pi_\Theta$ and Hermitian ($T_{\Theta\Phi}^\dagger = T_{\Phi\Theta}$)
transition operators~\cite{Alcock-Zeilinger:2016cva} $T_{\Theta\Phi}$ such that, for fixed $\Theta, \Theta', \Phi, \Phi'$,
\begin{align}
\Pi_\Theta &\equiv T_{\Theta \Theta},\\
T_{\Theta\Phi} T_{\Phi'\Theta'} &= \delta_{\Phi\Phi'}T_{\Theta\Theta'}, \label{eq:sn_transition_fusion}\\
\sum_{\Theta\in\mathcal{Y}_n} \Pi_\Theta &= \id.
\end{align}
\Cref{eq:sn_transition_fusion} is analogous to the product rule for the transition operators of the cubic group in \cref{eq:cubic_transition_fusion}. 

When multiple instances of the same irrep row $\Theta$ appear in the decomposition of a space acted on by} the exchange group, they can often be distinguished by letting $T_{\Theta\Phi}$ act on a fiducial vector for different $\Phi$.
All such products are left invariant by $\Pi_\Theta$.
In this way, the transition operators acting on fiducial vectors give a way to construct relevant multi-particle operators. 
Concrete examples are discussed below.

For arbitrary $S_n$, recursive formulas for Hermitian projection are given in Refs.~\cite{Alcock-Zeilinger:2016cva,Alcock-Zeilinger:2016sxc,Keppeler:2013yla} and for Hermitian transition operators in Ref.~\cite{Alcock-Zeilinger:2016cva}.
Generic projection and transition operators are elements of the group algebra,
\begin{align}\label{eq:S_projector}
     \Pi_\Theta &= \sum_{\permutation \in S_n} c_\permutation^{(\Theta)} \, \permutation,\\
     T_{\Theta\Phi} &= \sum_{\permutation \in S_n} c_\permutation^{(\Theta\Phi)} \, \permutation,
\end{align}
with $c_\sigma^{(\Theta\Theta)} \equiv c_\sigma^{(\Theta)}$.
As indicated, the coefficients $c_\sigma^{(\Theta\Phi)}$ depend explicitly on $\Theta, \Phi \in \mathcal{Y}_n$.
A concrete example illustrates the important features of the general case.
Consider the permutation group $S_3$, for which Table~\ref{table:S3irreps} summarizes the irreps.
The projectors $\Pi_\Theta$, with $\Theta \in \mathcal{Y}_3$, are~\cite{Alcock-Zeilinger:2016cva}
\begin{subequations}
\begin{align}
  \begin{split}\label{eq:3particle_projectors_S}
\ytproj_{\ytsub{
    1 & 2 & 3
}} =
    & \frac{1}{6}\left[(1) + (1,2,3) + (1,3,2) \right.\\
    &\quad \left. + (1,2) + (1,3) + (2, 3) \right],
\end{split}\\
\begin{split}\label{eq:3particle_projectors_A}
\ytproj_{\ytsub{
    1 \\ 2 \\ 3
}} =
    & \frac{1}{6} \left[(1) + (1,2,3) + (1,3,2) \right.\\
    &\quad \left. - (1,2) - (1,3) - (2, 3) \right],
\end{split}\\
\begin{split}\label{eq:3particle_projectors_M1}
\ytproj_{\ytsub{
1 & 2 \\
3
}} =
    & \frac{1}{6} \left[2\, (1) - (1,2,3) - (1,3,2) \right. \\
    &\quad \left. + 2\, (1,2) - (1,3) - (2,3) \right],
\end{split}\\
\begin{split}\label{eq:3particle_projectors_M2}
\ytproj_{\ytsub{
1 & 3 \\
2
}} =
    & \frac{1}{6} \left[2\, (1) - (1,2,3) - (1,3,2) \right. \\
    &\quad \left. - 2\, (1,2) + (1,3) + (2,3) \right].
\end{split}
\end{align}
\end{subequations}
Here and below, cycle notation $(i, j, k, ..., z)$ is used to indicate permutations mapping the elements cyclically $i \to j \to k \dots \to z \to i$.
The transition operators between the two rows of the two-dimensional standard irrep are~\cite{Alcock-Zeilinger:2016cva}
\begin{subequations}
\begin{align}
    T_{\ytsub{1&2\\3}\,\ytsub{1&3\\2}} = \frac{1}{\sqrt{12}}\left[(2,3) + (1,2,3) - (1,3) - (1,3,2)\right],\\
    T_{\ytsub{1&3\\2}\,\ytsub{1&2\\3}} = \frac{1}{\sqrt{12}}\left[(2,3) - (1,2,3) - (1,3) + (1,3,2)\right],
\end{align}
\end{subequations}
where the normalization factors follow from \cref{eq:sn_transition_fusion}.

The projection matrix acting on the extended orbit then follows from linearity, with components:
\begin{equation}
  D_{m'm}^{(s)}(\Pi_\Theta) = \sum_{\permutation\in S} c_\permutation^{(\Theta)} D_{m'm}^{(s)}(\permutation)
  \label{eq:Dmm_symmetrization}.
\end{equation}
The symmetric-group projection matrices are applied to the orbit representation matrices in~\cref{eq:DbraketGeneric} to construct projected orbit representation matrices
\begin{align}
    \begin{split}
    \hat{D}^{(s,\Theta)}(R^D) & \equiv
    D^{(s)}(\Pi_\Theta) \cdot
    \tilde{D}^{(s)}(R^D) \cdot
    D^{(s)}(\Pi_\Theta)\\
    &= \tilde{D}^{(s)}(R^D) \cdot
    D^{(s)}(\Pi_\Theta).
    \end{split}
    \label{eq:apply_Dmm_symmetrization}
\end{align}
The equality in the second line follows from the fact that the permutations commute with rotations and that $D^{(s)}(\Pi_\Theta)$ is idempotent.
Subsequent application of block diagonalization using Schur's lemma, orthogonalization, and rotational transition operators---needed to construct the block-diagonalization matrices in the analogue of \cref{eq:block_diag}---remains unchanged. 

\begin{table}[t!]
    \centering
    \caption{Irreps of the symmetric group $S_3$.
    \label{table:S3irreps}
    }
    \begin{ruledtabular}
    \begin{tabular}{l c c c}
        Name    & Dimension & Young diagram & Young tableaux\\
        \hline
        \vphantom{${\tiny \ydiagram{1,1,1}}$}Trivial & 1 & ${\tiny \ydiagram{3}}$ & ${\tiny \begin{ytableau} 1 & 2 & 3 \end{ytableau}}$\\
        Sign    & 1 & ${\tiny \ydiagram{1,1,1}}$& 
        ${\tiny \begin{ytableau} 1 \\ 2 \\ 3 \end{ytableau}}$
        \\[1em]
        \vphantom{${\tiny \ydiagram{1,1,1}}$}Standard& 2 & ${\tiny \ydiagram{2,1}}$ & ${\tiny \begin{ytableau} 1 & 2 \\ 3 \end{ytableau}}$, 
        ${\tiny \begin{ytableau} 1 & 3 \\ 2 \end{ytableau}}$ 
    \end{tabular}
    \end{ruledtabular}
\end{table}

\subsection{Examples}

\subsubsection{Identical fermions: \texorpdfstring{$nn$}{nn} and \texorpdfstring{$nnn$}{nnn}}

\begin{table*}[t]
  \caption{
    Combinations of irreps arising in decompositions of $nn$ operator orbits. Details are as in Table~\ref{tab:pip_Kip}. }
    \begin{ruledtabular}
    \begin{tabular}{>{\centering\arraybackslash}cccccc@{\hskip 0.1in}}
$\littlegroup$ & $\stabilizer$ & $S_2$ irrep & Example state & Orbit dim & Irrep decomposition\\ \hline
$O_{h}$ & $O_{h}$ & $\scalebox{.5}{\ydiagram{1,1}}$ & $\Ket{n(0,0,0),\ n(0,0,0)}$ & 1 & $A_1^+$ \\ 
$O_{h}$ & $C_{4v}$ & $\scalebox{.5}{\ydiagram{1,1}}$ & $\Ket{n(0,0,1),\ n(0,0,-1)}$ & 12 & $A_1^+ \oplus E^+ \oplus A_1^- \oplus E^- \oplus T_1^- \oplus T_2^-$ \\ 
$O_{h}$ & $C_{2v}$ & $\scalebox{.5}{\ydiagram{1,1}}$ & $\Ket{n(0,1,1),\ n(0,-1,-1)}$ & 24 & $A_1^+ \oplus E^+ \oplus T_2^+ \oplus A_1^- \oplus A_2^- \oplus 2 E^- \oplus 2 T_1^- \oplus 2 T_2^-$ \\ 
$O_{h}$ & $C_{3v}$ & $\scalebox{.5}{\ydiagram{1,1}}$ & $\Ket{n(1,1,1),\ n(-1,-1,-1)}$ & 16 & $A_1^+ \oplus T_2^+ \oplus A_1^- \oplus E^- \oplus T_1^- \oplus 2 T_2^-$ \\ 
$O_{h}$ & $C_2^R$ & $\scalebox{.5}{\ydiagram{1,1}}$ & $\Ket{n(2,1,0),\ n(-2,-1,0)}$ & 48 & $A_1^+ \oplus A_2^+ \oplus 2 E^+ \oplus T_1^+ \oplus T_2^+ \oplus 2 A_1^- \oplus 2 A_2^- \oplus 4 E^- \oplus 4 T_1^- \oplus 4 T_2^-$ \\ 
$O_{h}$ & $C_2^P$ & $\scalebox{.5}{\ydiagram{1,1}}$ & $\Ket{n(2,1,1),\ n(-2,-1,-1)}$ & 48 & $A_1^+ \oplus E^+ \oplus T_1^+ \oplus 2 T_2^+ \oplus 2 A_1^- \oplus A_2^- \oplus 3 E^- \oplus 4 T_1^- \oplus 5 T_2^-$ \\ 
$O_{h}$ & $C_{1}$ & $\scalebox{.5}{\ydiagram{1,1}}$ & $\Ket{n(3,2,1),\ n(-3,-2,-1)}$ & 96 & $A_1^+ \oplus A_2^+ \oplus 2 E^+ \oplus 3 T_1^+ \oplus 3 T_2^+ \oplus 3 A_1^- \oplus 3 A_2^- \oplus 6 E^- \oplus 9 T_1^- \oplus 9 T_2^-$ \\ 
$C_{4v}$ & $C_{4v}$ & $\scalebox{.5}{\ydiagram{1,1}}$ & $\Ket{n(0,0,1),\ n(0,0,1)}$ & 1 & $A_1$ \\ 
$C_{4v}$ & $C_{4v}$ & $\scalebox{.5}{\ydiagram{1,1}}$ & $\Ket{n(0,0,1),\ n(0,0,0)}$ & 4 & $A_1 \oplus A_2 \oplus E$ \\ 
$C_{4v}$ & $C_2^R$ & $\scalebox{.5}{\ydiagram{1,1}}$ & $\Ket{n(1,0,1),\ n(-1,0,1)}$ & 8 & $2 A_1 \oplus A_2 \oplus 2 B_1 \oplus B_2 \oplus E$ \\ 
$C_{4v}$ & $C_2^R$ & $\scalebox{.5}{\ydiagram{1,1}}$ & $\Ket{n(1,0,1),\ n(-1,0,0)}$ & 16 & $2 A_1 \oplus 2 A_2 \oplus 2 B_1 \oplus 2 B_2 \oplus 4 E$ \\ 
$C_{4v}$ & $C_2^P$ & $\scalebox{.5}{\ydiagram{1,1}}$ & $\Ket{n(1,1,1),\ n(-1,-1,1)}$ & 8 & $2 A_1 \oplus A_2 \oplus B_1 \oplus 2 B_2 \oplus E$ \\ 
$C_{4v}$ & $C_2^P$ & $\scalebox{.5}{\ydiagram{1,1}}$ & $\Ket{n(1,1,1),\ n(-1,-1,0)}$ & 16 & $2 A_1 \oplus 2 A_2 \oplus 2 B_1 \oplus 2 B_2 \oplus 4 E$ \\ 
$C_{4v}$ & $C_{1}$ & $\scalebox{.5}{\ydiagram{1,1}}$ & $\Ket{n(2,1,1),\ n(-2,-1,1)}$ & 16 & $3 A_1 \oplus 3 A_2 \oplus 3 B_1 \oplus 3 B_2 \oplus 2 E$ \\ 
$C_{4v}$ & $C_{1}$ & $\scalebox{.5}{\ydiagram{1,1}}$ & $\Ket{n(2,1,1),\ n(-2,-1,0)}$ & 32 & $4 A_1 \oplus 4 A_2 \oplus 4 B_1 \oplus 4 B_2 \oplus 8 E$ \\ 
$C_{3v}$ & $C_{3v}$ & $\scalebox{.5}{\ydiagram{1,1}}$ & $\Ket{n(1,1,1),\ n(1,1,1)}$ & 1 & $A_1$ \\ 
$C_{3v}$ & $C_{3v}$ & $\scalebox{.5}{\ydiagram{1,1}}$ & $\Ket{n(1,1,1),\ n(0,0,0)}$ & 4 & $A_1 \oplus A_2 \oplus E$ \\ 
$C_{3v}$ & $C_2^P$ & $\scalebox{.5}{\ydiagram{1,1}}$ & $\Ket{n(1,1,0),\ n(0,0,1)}$ & 12 & $2 A_1 \oplus 2 A_2 \oplus 4 E$ \\ 
$C_{3v}$ & $C_{1}$ & $\scalebox{.5}{\ydiagram{1,1}}$ & $\Ket{n(1,0,-1),\ n(0,1,2)}$ & 24 & $4 A_1 \oplus 4 A_2 \oplus 8 E$ \\ 
$C_{2v}$ & $C_{2v}$ & $\scalebox{.5}{\ydiagram{1,1}}$ & $\Ket{n(0,1,1),\ n(0,1,1)}$ & 1 & $A_1$ \\ 
$C_{2v}$ & $C_{2v}$ & $\scalebox{.5}{\ydiagram{1,1}}$ & $\Ket{n(0,1,1),\ n(0,0,0)}$ & 4 & $A_1 \oplus A_2 \oplus B_1 \oplus B_2$ \\ 
$C_{2v}$ & $C_2^R$ & $\scalebox{.5}{\ydiagram{1,1}}$ & $\Ket{n(0,0,1),\ n(0,1,0)}$ & 4 & $2 A_1 \oplus A_2 \oplus B_1$ \\ 
$C_{2v}$ & $C_{2v}$ & $\scalebox{.5}{\ydiagram{1,1}}$ & $\Ket{n(0,2,2),\ n(0,-1,-1)}$ & 4 & $A_1 \oplus A_2 \oplus B_1 \oplus B_2$ \\ 
$C_{2v}$ & $C_2^P$ & $\scalebox{.5}{\ydiagram{1,1}}$ & $\Ket{n(2,1,1),\ n(-2,1,1)}$ & 4 & $2 A_1 \oplus A_2 \oplus B_2$ \\ 
$C_{2v}$ & $C_2^P$ & $\scalebox{.5}{\ydiagram{1,1}}$ & $\Ket{n(2,1,1),\ n(-2,0,0)}$ & 8 & $2 A_1 \oplus 2 A_2 \oplus 2 B_1 \oplus 2 B_2$ \\ 
$C_{2v}$ & $C_{1}$ & $\scalebox{.5}{\ydiagram{1,1}}$ & $\Ket{n(-2,0,1),\ n(2,1,0)}$ & 8 & $3 A_1 \oplus 3 A_2 \oplus B_1 \oplus B_2$ \\ 
$C_{2v}$ & $C_{1}$ & $\scalebox{.5}{\ydiagram{1,1}}$ & $\Ket{n(1,2,1),\ n(-1,-1,0)}$ & 16 & $4 A_1 \oplus 4 A_2 \oplus 4 B_1 \oplus 4 B_2$ \\ 
$C_2^R$ & $C_2^R$ & $\scalebox{.5}{\ydiagram{1,1}}$ & $\Ket{n(1,2,0),\ n(1,2,0)}$ & 1 & $A$ \\ 
$C_2^R$ & $C_2^R$ & $\scalebox{.5}{\ydiagram{1,1}}$ & $\Ket{n(1,2,0),\ n(0,0,0)}$ & 4 & $2 A \oplus 2 B$ \\ 
$C_2^R$ & $C_{1}$ & $\scalebox{.5}{\ydiagram{1,1}}$ & $\Ket{n(1,0,1),\ n(0,2,-1)}$ & 8 & $4 A \oplus 4 B$ \\ 
$C_{1}$ & $C_{1}$ & $\scalebox{.5}{\ydiagram{1,1}}$ & $\Ket{n(1,2,3),\ n(1,2,3)}$ & 1 & $A$ \\ 
$C_{1}$ & $C_{1}$ & $\scalebox{.5}{\ydiagram{1,1}}$ & $\Ket{n(1,2,3),\ n(0,0,0)}$ & 4 & $4 A$ \\ 
    \end{tabular}
\end{ruledtabular}

    \label{tab:n_n}
\end{table*}

\begin{table*}[t]
    \caption{ 
    Combinations of irreps arising in decompositions of $nnn$ operator orbits. Details are as in Table~\ref{tab:pip_Kip}. }
    \begin{ruledtabular}
    \begin{tabular}{>{\centering\arraybackslash}cccccc@{\hskip 0.1in}}
$\littlegroup$ & $\stabilizer$ & $S_3$ irrep & Example state & Orbit dim & Irrep decomposition\\ \hline
$O_{h}$ & $C_{4v}$ & $\scalebox{.4}{\ydiagram{1,1,1}}$ & $\Ket{n(0,0,1),\ n(0,0,-1),\ n(0,0,0)}$ & 24 & $G_1^+ \oplus H^+ \oplus 2 G_1^- \oplus G_2^- \oplus 3 H^-$ \\ 
$O_{h}$ & $C_{4v}$ & $\scalebox{.4}{\ydiagram{1,1,1}}$ & $\Ket{n(0,0,1),\ n(0,0,2),\ n(0,0,-3)}$ & 48 & $3 G_1^+ \oplus G_2^+ \oplus 4 H^+ \oplus 3 G_1^- \oplus G_2^- \oplus 4 H^-$ \\ 
$O_{h}$ & $C_{2v}$ & $\scalebox{.4}{\ydiagram{1,1,1}}$ & $\Ket{n(0,1,1),\ n(0,-1,-1),\ n(0,0,0)}$ & 48 & $G_1^+ \oplus G_2^+ \oplus 2 H^+ \oplus 3 G_1^- \oplus 3 G_2^- \oplus 6 H^-$ \\ 
$O_{h}$ & $C_{2v}$ & $\scalebox{.4}{\ydiagram{1,1,1}}$ & $\Ket{n(0,-1,-1),\ n(0,-2,-2),\ n(0,3,3)}$ & 96 & $4 G_1^+ \oplus 4 G_2^+ \oplus 8 H^+ \oplus 4 G_1^- \oplus 4 G_2^- \oplus 8 H^-$ \\ 
$O_{h}$ & $C_{3v}$ & $\scalebox{.4}{\ydiagram{1,1,1}}$ & $\Ket{n(1,1,1),\ n(-1,-1,-1),\ n(0,0,0)}$ & 32 & $G_1^+ \oplus G_2^+ \oplus H^+ \oplus 2 G_1^- \oplus 2 G_2^- \oplus 4 H^-$ \\ 
$O_{h}$ & $C_{3v}$ & $\scalebox{.4}{\ydiagram{1,1,1}}$ & $\Ket{n(1,1,1),\ n(2,2,2),\ n(-3,-3,-3)}$ & 64 & $3 G_1^+ \oplus 3 G_2^+ \oplus 5 H^+ \oplus 3 G_1^- \oplus 3 G_2^- \oplus 5 H^-$ \\ 
$O_{h}$ & $C_2^R$ & $\scalebox{.4}{\ydiagram{1,1,1}}$ & $\Ket{n(2,1,0),\ n(-2,-1,0),\ n(0,0,0)}$ & 96 & $2 G_1^+ \oplus 2 G_2^+ \oplus 4 H^+ \oplus 6 G_1^- \oplus 6 G_2^- \oplus 12 H^-$ \\ 
$O_{h}$ & $C_2^R$ & $\scalebox{.4}{\ydiagram{1,1,1}}$ & $\Ket{n(0,1,1),\ n(0,-1,0),\ n(0,0,-1)}$ & 96 & $4 G_1^+ \oplus 4 G_2^+ \oplus 8 H^+ \oplus 4 G_1^- \oplus 4 G_2^- \oplus 8 H^-$ \\ 
$O_{h}$ & $C_2^R$ & $\scalebox{.4}{\ydiagram{1,1,1}}$ & $\Ket{n(2,1,0),\ n(-2,0,0),\ n(0,-1,0)}$ & 192 & $8 G_1^+ \oplus 8 G_2^+ \oplus 16 H^+ \oplus 8 G_1^- \oplus 8 G_2^- \oplus 16 H^-$ \\ 
$O_{h}$ & $C_2^P$ & $\scalebox{.4}{\ydiagram{1,1,1}}$ & $\Ket{n(2,1,1),\ n(-2,-1,-1),\ n(0,0,0)}$ & 96 & $2 G_1^+ \oplus 2 G_2^+ \oplus 4 H^+ \oplus 6 G_1^- \oplus 6 G_2^- \oplus 12 H^-$ \\ 
$O_{h}$ & $C_2^P$ & $\scalebox{.4}{\ydiagram{1,1,1}}$ & $\Ket{n(1,1,1),\ n(-1,-1,0),\ n(0,0,-1)}$ & 192 & $8 G_1^+ \oplus 8 G_2^+ \oplus 16 H^+ \oplus 8 G_1^- \oplus 8 G_2^- \oplus 16 H^-$ \\ 
$O_{h}$ & $C_{1}$ & $\scalebox{.4}{\ydiagram{1,1,1}}$ & $\Ket{n(0,-1,1),\ n(2,1,0),\ n(-2,0,-1)}$ & 192 & $8 G_1^+ \oplus 8 G_2^+ \oplus 16 H^+ \oplus 8 G_1^- \oplus 8 G_2^- \oplus 16 H^-$ \\ 
$O_{h}$ & $C_{1}$ & $\scalebox{.4}{\ydiagram{1,1,1}}$ & $\Ket{n(2,1,1),\ n(-2,-1,0),\ n(0,0,-1)}$ & 384 & $16 G_1^+ \oplus 16 G_2^+ \oplus 32 H^+ \oplus 16 G_1^- \oplus 16 G_2^- \oplus 32 H^-$ \\ 
    \end{tabular}
\end{ruledtabular}

    \label{tab:n_n_n}
\end{table*}

Identical fermions provide a first example with internal symmetry.
Operators constructed from products of identical fermions such as neutrons, e.g., $nn$ or $nnn$, must be totally antisymmetric under simultaneous exchange of the spins and momenta of any two particles.
For $nn$, the exchange group is $S_2$.
Fermion antisymmetry implies $nn$ operators transform in the sign irrep ($\tiny{\ydiagram{1,1}}$) of $S_2$.
Irrep decomposition proceeds as in the distinguishable $np$ case summarized Table~\ref{tab:n_p}, with the additional projection step of \cref{eq:apply_Dmm_symmetrization} using the projection operator for the sign irrep of $S_2$,
\begin{align}  \begin{split}\label{eq:2particle_projectors_A}
\Pi_{\ytsub{
1 \\ 2
}}=
    & \frac{1}{2}\left[(1) - (1,2) \right].
\end{split}
\end{align}
\Cref{tab:n_n} summarizes the irrep decompositions for $nn$ operators.

Operators constructed from products of more neutrons can be decomposed analogously.
For $nnn$, the exchange group is $S_3$.
Fermion antisymmetry requires that $nnn$ operators transform in the sign irrep of $S_3$, corresponding to the projection operator in \cref{eq:3particle_projectors_A}.
\Cref{tab:n_n_n} summarizes the irrep decompositions for $nnn$ operators.
Note that, in this case, specification of the little group and stabilizer group does \emph{not} suffice to specify the irrep decomposition uniquely---whether certain permutations correspond to little-group transformations affects the resulting irrep decomposition.
These results illustrate the general fact that the irrep decomposition of a multi-particle operator depends on the little group $\littlegroup$, the stabilizer group $\stabilizer$, and on the irrep of the exchange group $\exchangegroup$.
It is also noteworthy that in this case the application of the exchange group projects away the orbit with $\bm{n}_1 = \bm{n}_2 = \bm{n}_3 = 0$; fermion antisymmetry dictates that this orbit vanishes from the irrep decomposition.

\subsubsection{Three pions with isospin}

The three-pion system provides an example of the interplay between the cubic group and non-trivial internal symmetries.
Since each pion transforms as an isotriplet with $I=1$, the three-pion system has the isospin decomposition
\begin{align}
1\otimes 1 \otimes 1= 3\oplus 2\oplus 2\oplus 1\oplus 1\oplus 1 \oplus 0.
\end{align}
The relevant exchange group is $S_3$, for which the irreps and projectors have been summarized above.
As in \cref{sec:spinzero}, Schur's Lemma provides the means to decompose the product-state isospin representation into irreps of $S_3$.
Since $I_z$ is conserved by permutations, it suffices to work at fixed $I_z$.
Permutations of the states with a given $I_z$ furnish the reducible representation matrices $D_{m'm}^{(I_z)}(\sigma)$.
For example, the seven states with $I_z=0$ yield $7\times 7$ representation matrices $D_{m'm}^{(I_z=0)}(\sigma)$.
Irrep matrices $D^{(\lambda)}$ for $S_3$ are also readily obtained.
For the trivial irrep they are simply unity, while for the sign representation they are equal to the permutation signature.
Irrep matrices for the two-dimensional standard representation of $S_3$ are given in Ref.~\cite{Hansen:2020zhy}.

\Cref{table:pipipi_S3decomp} shows the result of applying Schur's Lemma, i.e., using \cref{eq:schur-lemma} in terms of the $D^{(\lambda)}$ and $D^{(I_z)}$ matrices,
to extract the overlap of the reducible $I_z$ representation onto the irrep $\lambda$.
As expected, $I=3$ corresponds to the trivial representation of $S_3$ for each $I_z \in \{-3, \dots, 3\}$, which is totally symmetric.
The doublets of states with $I=2$ for each $I_z$ fall in the standard representation of $S_3$.
The three copies with $I=1$ split into a doublet from the standard representation and a trivial representation for each $I_z$.
Finally, the isosinglet transforms in the sign representation of $S_3$, which is totally antisymmetric.
Explicit expressions for the associated states have been given in Ref.~\cite{Hansen:2020zhy}.

Combining these results with block diagonalization for the cubic group amounts to applying \cref{eq:Dmm_symmetrization}, where 
$\Pi_\Theta$ is selected to project onto the rows of \Cref{table:pipipi_S3decomp} with the desired isospin.
Applying these projectors in \cref{eq:apply_Dmm_symmetrization}
reproduces the cubic-group irrep decompositions given in Appendix D of Ref.~\cite{Hansen:2020zhy} for the rest frame. 

As discussed in Sec.~\ref{sec:exchangegroupirreps}, the transition operators $T_{\Theta\Phi}$ can be used to construct multi-particle operators with definite $S_n$ transformation properties.
For the case at hand, consider three-pion operators with $I_z=0$ and $I\in \{0, 1,2, 3\}$.
These operators can be built from permutations of the fiducial ordering of single-pion operators, say, $\ket{\pi^+\pi^-\pi^0}$.
Letting permutations act in the natural way, e.g., $(1,2,3)\ket{\pi^+\pi^-\pi^0} = \ket{\pi^0\pi^+\pi^-}$, gives the seven three-pion states with $I_z=0$ constructed in Ref.~\cite{Hansen:2020zhy}:
\begin{subequations}
\begin{align}
    \Pi_{\ytsub{1&2&3}}\left(6\ket{\pi^+\pi^-\pi^0} +2 \ket{\pi^0\pi^0\pi^0}\right)&\propto \ket{\chi_s}_{I=3}, \label{eq:Tsym_psi1}\\    
    \Pi_{\ytsub{1&2&3}}\left(6\ket{\pi^+\pi^-\pi^0} -3 \ket{\pi^0\pi^0\pi^0}\right)&\propto \ket{\chi_s}_{I=1}, \label{eq:Tsym_psi2}\\    
    \Pi_{\ytsub{1&2\\3}}\ket{\pi^+\pi^-\pi^0} &\propto \ket{\chi_1}_{I=1}, \label{eq:T11psi}\\
    T_{\ytsub{1&2\\3}\,\ytsub{1&3\\2}} \ket{\pi^+\pi^-\pi^0} &\propto \ket{\chi_1}_{I=2}, \label{eq:T12psi}\\
    \Pi_{\ytsub{1&3\\2}}\ket{\pi^+\pi^-\pi^0} &\propto \ket{\chi_2}_{I=2}, \label{eq:T22psi}\\
    T_{\ytsub{1&3\\2}\,\ytsub{1&2\\3}}\ket{\pi^+\pi^-\pi^0} &\propto \ket{\chi_2}_{I=1}, \label{eq:T21psi}\\
    \Pi_{\ytsub{1\\2\\3}}\ket{\pi^+\pi^-\pi^0} &\propto \ket{\chi_a}_{I=0}.
    \label{eq:Tasym_psi}
\end{align}
\end{subequations}
The totally symmetric cases (\cref{eq:Tsym_psi1,eq:Tsym_psi2}) contain a component proportional to $\ket{\pi^0\pi^0\pi^0}$, which vanishes in all other representations; the relative coefficient between the $\ket{\pi^+\pi^-\pi^0}$ and $\ket{\pi^0\pi^0\pi^0}$ terms is related to the isospin Clebsch--Gordan decomposition.
As in Ref.~\cite{Hansen:2020zhy}, the two-dimensional standard irrep is spanned by a basis denoted by $\ket{\chi_1}$ and $\ket{\chi_2}$, with a final subscript outside the ket giving the total isospin.
As expected, \cref{eq:T11psi,eq:T12psi} are invariant under \cref{eq:3particle_projectors_M1} and associated with degenerate copies of the first row $\ket{\chi_1}$ of the irrep ${\tiny \ydiagram{2,1}}$.
Similarly, \cref{eq:T21psi,eq:T22psi} are invariant under \cref{eq:3particle_projectors_M2} and associated with degenerate copies of the second row $\ket{\chi_2}$.

Three pions in boosted frames provide examples where the full $S_3$ exchange group influences the cubic-group irrep decomposition.
A minimal example is the system with momenta $[\bm{n}_1, \bm{n}_2, \bm{n}_3] = [(1,0,0), (0,1,0), (0,0,1)]$, for which the momenta can all be permuted by little group operations.
The little group of the total momentum is $C_{3v}$, while the stabilizer is $C_1$.
Without projecting under the exchange group, the resulting irreps can be seen from the relevant row of \Cref{tab:pip_Kip} to be
\begin{equation}
    A_1 \oplus A_2 \oplus 2 E.
\end{equation}
Applying the various $S_3$ projectors $\Pi_\Theta$ with $\Theta\in\mathcal{Y}_3$ restricts to the following irreps:
\begin{equation}
\begin{aligned}
{\ytsub{1 & 2 & 3}}:    && A_1 \oplus A_2 \oplus 2E \mapsto A_1, \\
{\ytsub{1 \\ 2 \\ 3}}:  && A_1 \oplus A_2 \oplus 2E \mapsto A_2, \\
{\ytsub{1 & 2 \\ 3}}:   && A_1 \oplus A_2 \oplus 2E \mapsto E, \\
{\ytsub{1 & 3 \\ 2}}:   && A_1 \oplus A_2 \oplus 2E \mapsto E.
\end{aligned}
\end{equation}
Comparison with \Cref{table:pipipi_S3decomp}  determines which combinations of cubic group and isospin irreps are compatible with bosonic statistics for this set of momenta.
For example, total $I = 0$ must be combined with the cubic group irrep $A_2$, while total $I = 1$ or $I = 3$ must be combined with the cubic group irrep $A_1$. Both $I = 1$ and $I = 2$ may be combined with the cubic group irrep $E$, but the correct degenerate copy of this irrep must be chosen.

\begin{table}
    \centering
    \caption{Decomposition of the $\pi\pi\pi$ system into irreps of $S_3$ given in \cref{table:S3irreps}.
    For each column of fixed $I_z$, the number of check marks equals the sum of the dimensions in the irrep decomposition.
    \label{table:pipipi_S3decomp}
    } 
    \begin{ruledtabular}
    \begin{tabular}{c|ccccccc|l }
    \diagbox[height=1.25\line]{{\scriptsize $I$}}{{\scriptsize $I_z$}}
        & $+3$ & $+2$ & $+1$ & 0 & $-1$ & $-2$ & $-3$ & \hspace{2pt} $S_3$ irrep \hspace{2pt} \\
    \hline
    $3$& \checkmark & \checkmark & \checkmark & \checkmark & \checkmark & \checkmark & \checkmark & $\hspace{14pt} {\tiny \ydiagram{3}}$
    \\[1.5ex]
    $2$& & \checkmark & \checkmark & \checkmark & \checkmark & \checkmark & & \multirow{2}{*}{$\Big\} \hspace{10pt} {\tiny \ydiagram{2,1}}$} \\
    $2$& & \checkmark & \checkmark & \checkmark & \checkmark & \checkmark & & \\
    $1$& & & \checkmark & \checkmark & \checkmark & & & \multirow{2}{*}{$\Big\} \hspace{10pt} {\tiny \ydiagram{2,1}}$} \\
    $1$& & & \checkmark & \checkmark & \checkmark & & & \\[1.5ex]
    $1$& & & \checkmark & \checkmark & \checkmark & & & $\hspace{14pt} {\tiny \ydiagram{3}}$ \\[1.5ex]
    $0$& & & & \checkmark & & & & $\hspace{17pt} {\tiny \ydiagram{1,1,1}}$ \\
    \end{tabular}
    \end{ruledtabular}
\end{table}

\subsubsection{\texorpdfstring{$DD\pi$}{DDpi}}

\begin{table*}[t]
\centering
\caption{
    Combinations of irreps arising in decompositions of $DD\pi$ operator orbits.
    In the example states, the label $D$ refers collectively to $D^0$ and $D^+$ and the pion is either $\pi^0$ or $\pi^+$.
    Details are as in Table~\ref{tab:pip_Kip}. 
    \label{table:ddpi_isospin}
   }
\begin{ruledtabular}
    \begin{tabular}{>{\centering\arraybackslash}cccccc@{\hskip 0.1in}}
$\littlegroup$ & $\stabilizer$ & $S_2$ Irrep & Example state & Orbit dim & Irrep decomposition \\
\hline
$O_h$ & $O_h$ & $\scalebox{0.5}{\ydiagram{2}}$ & $\left|D(0, 0, 0),D(0, 0, 0),\pi(0, 0, 0)\right\rangle$ & 1 & $A_1^-$\\
$O_h$ & $C_4^v$ & $\scalebox{0.5}{\ydiagram{2}}$ & $\left|D(0, 0, 1),D(0, 0, -1),\pi(0, 0, 0)\right\rangle$ & 3 & $A_1^- \oplus E^-$\\
$O_h$ & $C_4^v$ & $\scalebox{0.5}{\ydiagram{1,1}}$ & $\left|D(0, 0, 1),D(0, 0, -1),\pi(0, 0, 0)\right\rangle$ & 3 & $T_1^+$\\
$O_h$ & $C_2^v$ & $\scalebox{0.5}{\ydiagram{2}}$ & $\left|D(0, 1, 1),D(0, -1, -1),\pi(0, 0, 0)\right\rangle$ & 6 & $A_1^- \oplus E^- \oplus T_2^-$\\
$O_h$ & $C_2^v$ & $\scalebox{0.5}{\ydiagram{1,1}}$ & $\left|D(0, 1, 1),D(0, -1, -1),\pi(0, 0, 0)\right\rangle$ & 6 & $T_1^+ \oplus T_2^+$\\
$O_h$ & $C_3^v$ & $\scalebox{0.5}{\ydiagram{2}}$ & $\left|D(1, 1, 1),D(-1, -1, -1),\pi(0, 0, 0)\right\rangle$ & 4 & $A_1^- \oplus T_2^-$\\
$O_h$ & $C_3^v$ & $\scalebox{0.5}{\ydiagram{1,1}}$ & $\left|D(1, 1, 1),D(-1, -1, -1),\pi(0, 0, 0)\right\rangle$ & 4 & $A_2^+ \oplus T_1^+$\\
$O_h$ & $C_2^R$ & $\scalebox{0.5}{\ydiagram{2}}$ & $\left|D(2, 1, 0),D(-2, -1, 0),\pi(0, 0, 0)\right\rangle$ & 12 & $A_1^- \oplus A_2^- \oplus 2E^- \oplus T_1^- \oplus T_2^-$\\
$O_h$ & $C_2^R$ & $\scalebox{0.5}{\ydiagram{1,1}}$ & $\left|D(2, 1, 0),D(-2, -1, 0),\pi(0, 0, 0)\right\rangle$ & 12 & $2T_1^+ \oplus 2T_2^+$\\
$O_h$ & $C_2^P$ & $\scalebox{0.5}{\ydiagram{2}}$ & $\left|D(2, 1, 1),D(-2, -1, -1),\pi(0, 0, 0)\right\rangle$ & 12 & $A_1^- \oplus E^- \oplus T_1^- \oplus 2T_2^-$\\
$O_h$ & $C_2^P$ & $\scalebox{0.5}{\ydiagram{1,1}}$ & $\left|D(2, 1, 1),D(-2, -1, -1),\pi(0, 0, 0)\right\rangle$ & 12 & $A_2^+ \oplus E^+ \oplus 2T_1^+ \oplus T_2^+$\\
$O_h$ & $C_1$ & $\scalebox{0.5}{\ydiagram{2}}$ & $\left|D(3, 2, 1),D(-3, -2, -1),\pi(0, 0, 0)\right\rangle$ & 24 & $A_1^- \oplus A_2^- \oplus 2E^- \oplus 3T_1^- \oplus 3T_2^-$\\
$O_h$ & $C_1$ & $\scalebox{0.5}{\ydiagram{1,1}}$ & $\left|D(3, 2, 1),D(-3, -2, -1),\pi(0, 0, 0)\right\rangle$ & 24 & $A_1^+ \oplus A_2^+ \oplus 2E^+ \oplus 3T_1^+ \oplus 3T_2^+$
\end{tabular}
\end{ruledtabular}
\end{table*}

Decay channels with resonances are also categorized by isospin and provide examples of internal symmetry where not all particles are identical.
For instance, the doubly charmed tetraquark $T_{cc}(3875)^+$ has been observed just below threshold for $D^{\star+}D^0$ in the decay mode $D^0 D^0 \pi^+$ with charmness $C=+2$ and charge $Q=+1$, corresponding to $I_z=0$~\cite{LHCb:2021vvq}.
The isospin decomposition of the $DD\pi$ system is 
\begin{align}
\tfrac{1}{2} \otimes \tfrac{1}{2} \otimes 1 = 2 \oplus 1 \oplus 1 \oplus 0,
\end{align}
where the $D$-meson isodoublet is $(D^+, D^0)^T$.
Similar to the preceding example, for each fixed $I_z$ the direct sum is decomposed into irreps of $S_2$ (permutations of the two $D$-meson operators) using projectors
$\Pi_{\ytsub{1 \\ 2}}$ (defined in \cref{eq:2particle_projectors_A} above) and 
\begin{align}  \begin{split}\label{eq:2particle_projectors_S}
\Pi_{\ytsub{
1 & 2
}}=
    & \frac{1}{2}\left[(1) + (1,2) \right].
\end{split}
\end{align}

The four states with $I_z=0$ are constructed from linear combinations of the states $\ket{D^+ D^+ \pi^-}$, $\ket{D^+ D^0 \pi^+}$, $\ket{D^0 D^+ \pi^+}$, and $\ket{D^0 D^0 \pi^+}$.
In terms of fiducial orderings,
\begin{equation}
\begin{aligned}
\ket{\psi_{00}} &\equiv \ket{D^0 D^0 \pi^+},\\
\ket{\psi_{++}} &\equiv \ket{D^+ D^+ \pi^-},\\
\ket{\psi_{+0}} &\equiv \ket{D^+ D^0 \pi^+},
\end{aligned}
\end{equation}
states of definite isospin with $I_z=0$ are given by \begin{equation}
\begin{aligned}
    \ket{(DD)_1\pi}_2    &\propto \Pi_{\ytsub{1 & 2}} \left(\ket{\psi_{00}} + \ket{\psi_{++}} + 2\sqrt{2} \ket{\psi_{+0}}\right)\\    
    \ket{(DD)_1\pi}_{1,a} &\propto \Pi_{\ytsub{1 & 2}} \left( \ket{\psi_{++}} - \ket{\psi_{00}}\right)\\
    \ket{(DD)_1\pi}_{1,b} &\propto \Pi_{\ytsub{1\\ 2}} \ket{\psi_{+0}}\\
    \ket{(DD)_1\pi}_0    &\propto \Pi_{\ytsub{1 & 2}} \left(\sqrt{2}(\ket{\psi_{00}} + \ket{\psi_{++}}) - 2\ket{\psi_{+0}}\right),
\end{aligned}
\end{equation}
where the left-hand side uses the notation of Ref.~\cite{Hansen:2024ffk}.
In the state $\ket{(DD)_{I_{DD}}\pi}_{I}$,
$I$ is the total isospin, and $I_{DD}$ is the isospin of the $DD$ subsystem.
As expected, the states with $I=2$ and $I=0$ transform in the symmetric representation of $S_2$.
Of the two copies of $I=1$, one is symmetric, while the other is antisymmetric.
Results for irrep decompositions of $DD\pi$ operators are summarized in \cref{table:ddpi_isospin}, where the orbit dimension refers to the rank of the projected orbit-representation matrices $\hat{D}^{(s,\Theta)}$.
The trivial and sign irreps of $S_2$ correspond to odd- and even-parity irreps in the decomposition of the cubic group, respectively, such that the overall exchange of two identical $D$ mesons is symmetric.

\subsubsection{\texorpdfstring{$H$}{H}-dibaryon}

\begin{table}[t!]
    \centering
    \caption{Permutation irreps of $S_2$ or $S_3$ for  $SU(3)$-singlet operators arising from products of three meson octet $M^a$ and baryon octet $B^a$ operators.
    \label{table:adjoint_perms}
    }
    \begin{ruledtabular}
    \begin{tabular}{c c c c c}
        Flavor tensor    & $M^a M^b M^c$ & $M^a M^b B^c$ & $B^a B^b M^c$ & $B^a B^b B^c$ \\
        \hline\\ 
        \\[-2em]
        $d_{abc}$ & \begin{ytableau}1 & 2 & 3 \end{ytableau} & \begin{ytableau}1 & 2 \end{ytableau} & \begin{ytableau}1 \\ 2 \end{ytableau}  & \begin{ytableau}1 \\ 2 \\ 3 \end{ytableau} \\
        $f_{abc}$ & \begin{ytableau}1 \\ 2 \\ 3 \end{ytableau} & \begin{ytableau}1 \\ 2 \end{ytableau} & \begin{ytableau}1 & 2 \end{ytableau}  & \begin{ytableau}1 & 2 & 3 \end{ytableau} \\
    \end{tabular}
    \end{ruledtabular}
\end{table}

The $H$-dibaryon provides an example where flavor and operator-exchange symmetry come together to satisfy bosonic symmetry or fermionic antisymmetry.
The $H$-dibaryon is a hypothetical two-baryon bound state with strangeness $S=-2$ that corresponds to an $SU(3)$-flavor singlet when the up, down, and strange quark masses are equal.
Interpolating operators with these quantum numbers can be constructed from two flavor-octet baryon interpolating operators using the $SU(3)$-flavor irrep decomposition
\begin{equation}
  \mathbf{8} \otimes \mathbf{8} = \mathbf{1} \oplus \mathbf{8}_A \oplus \mathbf{8}_S \oplus \mathbf{10} \oplus \overline{\mathbf{10}} \oplus \mathbf{27},
\end{equation}
where $\mathbf{8}_A$ ($\mathbf{8}_S$) denotes an $SU(3)$ octet irrep where the two-baryon flavor state is antisymmetric (symmetric) under exchange.
The singlet irrep $\mathbf{1}$ corresponds to a symmetric flavor state associated with the operator $B^a B^b \delta_{ab}$ where $B^a$ is a baryon octet field with $SU(3)$ adjoint index $a$.
Writing explicitly the baryon spin representation indices $\alpha_1$, $\alpha_2$ and the momentum labels $\bm{n}_1,\ldots$ for each field, fermion antisymmetry implies $ B^a_{\alpha_1}(\bm{n}_1) B^b_{\alpha_2}(\bm{n}_2) = - B^b_{\alpha_2}(\bm{n}_2) B^a_{\alpha_1}(\bm{n}_1)$.
This implies that the momentum-spin states $\ket{ \bm{n}_1,\alpha_1, \bm{n}_2,\alpha_2}$ associated with $\delta_{ab} B^a_{\alpha_1}(\bm{n}_1) B^b_{\alpha_2}(\bm{n}_2)$ will be antisymmetric under the exchange of momentum-spin pairs $(\bm{n}_1,\alpha_1) \leftrightarrow ( \bm{n}_2, \alpha_2 )$.
These states therefore transform in the sign irrep of $S_2$, and the permutation projector defined in Eq.~\eqref{eq:2particle_projectors_A} should be applied.
These two-baryon operators, which are linear combinations of $\Lambda\Lambda$, $\Sigma\Sigma$, and $N \Xi$, therefore have identical irrep decompositions to the case of $nn$ operators summarized in Table~\ref{tab:n_n}.

Operators with the same quantum numbers can be constructed from products of two octet baryon operators and one pseudoscalar octet meson operator.
This corresponds to the product of $SU(3)$ irreps $\mathbf{8} \otimes \mathbf{8} \otimes \mathbf{8}$, which includes two copies of the $\mathbf{1}$ irrep relevant for the $H$-dibaryon.
Tensor operators describing these products are given by
\begin{equation}
  d_{abc} B^a B^b M^c, \hspace{10pt} f_{abc} B^a B^b M^c,
\end{equation}
where $a,b,c$ are $SU(3)$ adjoint indices, $M^a$ is a pseudoscalar octet meson operator, and $d_{abc}$ ($f_{abc}$) are totally symmetric (antisymmetric) structure constants.
Although these operators involve linear combinations of several different products of meson and baryon flavors, for example $\Lambda \Lambda \pi^0$, $\Lambda p K^-$, and $\Sigma^+ \Sigma^0 \pi^-$, their cubic irrep decompositions and block-diagonalization matrices only depend on the permutation transformation properties of the flavor tensors $d_{abc}$ and $f_{abc}$ as well as the fermionic nature of the baryon fields.

Because $B^a$ and $M^a$ fields represent distinct types of $SU(3)$-octet particles while $B^a$ and $B^b$ are identical besides their $SU(3)$ flavor indices, the relevant exchange group for this case is $S_2$.
Writing explicitly the baryon spin representation indices $\alpha_1$, $\alpha_2$ and the momentum labels $\bm{n}_1,\ldots,\bm{n}_3$ for each field, fermion antisymmetry implies $B^a_{\alpha_1}(\bm{n}_1) B^b_{\alpha_2}(\bm{n}_2) M^c(\bm{n}_3) = -B^b_{\alpha_2}(\bm{n}_2) B^a_{\alpha_1}(\bm{n}_1) M^c(\bm{n}_3)$.
When contracted with the totally symmetric tensor $d_{abc}$, momentum-spin states $\ket{ \bm{n}_1,\alpha_1, \bm{n}_2,\alpha_2, \bm{n}_3}$ associated with these operators will be antisymmetric under the exchange of momentum-spin pairs $(\bm{n}_1,\alpha_1) \leftrightarrow ( \bm{n}_2, \alpha_2 )$.
These states therefore transform in the sign irrep of $S_2$, and the permutation projector defined in Eq.~\eqref{eq:2particle_projectors_A} should be applied.
Conversely, when contracted with the totally antisymmetric tensor $f_{abc}$, momentum-spin states created by these operators will be symmetric under the exchange of momentum-spin pairs $(\bm{n}_1,\alpha_1) \leftrightarrow ( \bm{n}_2, \alpha_2 )$.
These states therefore transform in the trivial irrep of $S_2$, and the permutation projector defined in Eq.~\eqref{eq:2particle_projectors_S} should be applied.

Similar considerations apply to other $SU(3)$-singlet operators built from different combinations of three meson- and baryon-octet operators, $M^a M^b B^c$, $M^a M^b M^c$, and $B^a B^b B^c$, which provide further examples of the interplay between identical-particle labels and flavor-transformation properties.
\Cref{table:adjoint_perms} summarizes the possible combinations with irreps of the particle-exchange group $S_2$ or $S_3$.

\subsubsection{\texorpdfstring{$\pi\pi K K$}{pi pi K K}}

\begin{table*}
\caption{
    Combinations of irreps arising in the decomposition of momentum orbits with four spin-zero operators, where $\bm{n}_1 = (0,0,1)$ and $\bm{n}_2=(0,2,0)$.
    Projection from symmetrization over identical operators generically reduces the number of irreps appearing in the decomposition.
    \label{tab:four_particles}
    }
\begin{ruledtabular}
    \begin{tabular}{>{\centering\arraybackslash}cccccc@{\hskip 0.1in}}
$\littlegroup$ & $\stabilizer$ & Exchange & Example state & Orbit dim & Irrep decomposition\\ \hline
$O_h$ & $C_2^R$ & $-$ & $|\pi^+(\bm{n}_1), \pi^-(-\bm{n}_1), K^+(\bm{n}_2), K^-(-\bm{n}_2)\rangle$ & 24 & $A_1^+ \oplus A_2^+ \oplus 2E^+ \oplus T_1^+ \oplus T_2^+ \oplus 2T_1^- \oplus 2T_2^-$\\
$O_h$ & $C_2^R$ & $\scalebox{.75}{\ydiagram{2}}\times\scalebox{.75}{\ydiagram{2}}$ & $|\pi^+(\bm{n}_1), \pi^+(-\bm{n}_1), K^+(\bm{n}_2), K^+(-\bm{n}_2)\rangle$ & 6 & $A_1^+ \oplus A_2^+ \oplus 2E^+$\\
$O_h$ & $C_2^R$ & $\scalebox{.75}{\ydiagram{2}}\times\scalebox{.75}{\ydiagram{2}}$ & $|\pi^+(\bm{n}_1), \pi^+(\bm{n}_2), K^+(-\bm{n}_1), K^+(-\bm{n}_2)\rangle$ & 24 & $A_1^+ \oplus A_2^+ \oplus 2E^+ \oplus T_1^+ \oplus T_2^+ \oplus 2T_1^- \oplus 2T_2^-$\\
$O_h$ & $C_2^R$ & $\scalebox{.75}{\ydiagram{4}}$ & $|\pi^+(\bm{n}_1), \pi^+(-\bm{n}_1), \pi^+(\bm{n}_2), \pi^+(-\bm{n}_2)\rangle$ & 6 & $A_1^+ \oplus A_2^+ \oplus 2E^+$\\
    \end{tabular}
\end{ruledtabular}

\end{table*}

An illustrative example of how the exchange group and extended orbit dimension depend on the configuration of the momentum orbit is provided by the $\pi\pi KK$ and $\pi\pi\pi\pi$ systems with two pairs of particles moving back to back.
This also provides an example where the exchange group is a direct product of non-trivial subgroups.
Table~\ref{tab:four_particles} shows the irrep decomposition for four spin-zero particles moving pairwise back-to-back with momenta $\bm{n}_1 = (0,0,1)$, $\bm{n}_2 = (0,2,0)$, $-\bm{n}_1$, and $-\bm{n}_2$.
Depending on how many operators correspond to identical bosons, different permutation projectors are used and lead to different cubic-group irrep decompositions. 

The simplest case is $\pi^+ \pi^- K^+ K^-$, which has no identical particles.
The (extended) orbits for all momentum configurations have the standard irrep decomposition for distinguishable spin-zero operator products with stabilizer group $C_2^R$.

When there are two pairs of identical bosons such as $\pi^+\pi^+K^+K^+$, the exchange group is $S_2 \times S_2$. 
States are invariant under exchange of both the first two momenta and the last two momenta and therefore transform in the trivial irrep of both $S_2$ factors.
The appropriate projector is a product of the $S_2$ trivial irrep projectors given in Eq.~\eqref{eq:2particle_projectors_S},
\begin{equation}
  \begin{split}
\ytproj_{\ytsub{
1 & 2
}}
\ytproj_{\ytsub{
3 & 4
}}
    &= \left( \frac{1}{2}[(1) + (1,2)] \right) \left( \frac{1}{2}[(1) + (3,4)] \right) \\
    &= \frac{1}{4}[ (1) + (1,2) + (3,4) + (1,2)(3,4)].
    \label{eq:proj2x2}
  \end{split}
\end{equation} 
In this case, the dimensionality of the extended orbit, i.e., the rank of the projected orbit-representation matrices $\hat{D}^{(s,\Theta)}$, and the irrep decomposition depend on the momentum configuration.

With four identical bosons such as $\pi^+\pi^+\pi^+\pi^+$, states transform in the trivial representation of the exchange group $S_4$.
The appropriate permutation projector is therefore obtained from the normalized sum of all $4!=24$ elements of $S_4$, that is
\begin{align}
\begin{split}
\ytproj_{\ytsub{
1 & 2 & 3 & 4
}}
  = \tfrac{1}{4!}[(1) + (1,2) + (1,3) + (1,4) + \dotsb ].
\end{split}
\end{align}

The irrep decompositions for each of these cases are shown in Table~\ref{tab:four_particles}.
For a fixed exchange group, the size of the irrep is related to the size of the orbit. 
In the second row, identical particles are moving back-to-back, which reduces the size of the decomposition compared to the third row, where exchange group elements only affect operators with momenta that cannot be related by cubic transformations.

\section{Algorithm Summary  \label{sec:algorithm}}

To collect details spread across several sections, the steps of the full algorithm are reproduced here, including for the case of identical particles with non-zero spin.
The algorithm begins by selecting a fiducial state $\bm{i} = [\bm{n}_1, \alpha_1, \internallabel_1, \dots, \bm{n}_N, \alpha_N, \internallabel_N]$ (in terms of the momentum $\bm{n}_i$, spin and parity $\alpha_i$, and internal quantum numbers $\internallabel_i$ of the $i$th particle)
and specifying its permutation properties as corresponding to a row $\Theta$ of an exchange-group irrep.
Given the fiducial state $\bm{i}$ and desired exchange-group irrep row $\Theta$, the algorithm proceeds as follows:
\begin{enumerate}
    \item Compute the little group $\littlegroup$, defined in \cref{eq:little_group}, of the total momentum.
    \item Compute irrep matrices of the little group $D^{(\Gamma)}_{\mu'\mu}(R)$ via \cref{eq:irrep-matrix-elements-2}.
    \item Compute the extended orbit $\hat{\orbit}_{\bm{P}}^{(s)}$ of the fiducial state under the action of the little group and the exchange group via Eq.~\eqref{eq:extended_orbit}.
    \item Compute the momentum-orbit representation matrices $D^{(s)}_{m'm}(R)$, \cref{eq:Dbraket}.
    \item Compute the spin-representation matrices $D_{J_z' J_z}^{[J]}(R^D)$ associated with each interpolating operator, defined generically in \cref{eq:spintransform} or specifically for spin-half ($G_1^+$) operators in \cref{eq:spinor_rep_matrices}.    
    \item Construct the combined momentum-spin-orbit representation matrices via \cref{eq:DbraketGeneric}.
    \item Compute the exchange-group projector $\Pi_{\Theta}$, \cref{eq:S_projector}, as described in Refs.~\cite{Alcock-Zeilinger:2016cva,Alcock-Zeilinger:2016sxc,Keppeler:2013yla}.
    \item Construct projected orbit-representation matrices via \cref{eq:Dmm_symmetrization}.
    \item Apply Schur's lemma in the form of \cref{eq:schur} to the projected orbit representation matrices in order to compute the first row for each irrep in the block-diagonalization matrices. Orthogonalize any degenerate copies which appear.
    \item Construct transition operators via \cref{eq:transition_defined} and use \cref{eq:transition_applied} to fill the remaining rows of each irrep.
    The result is the complete set of block-diagonalization matrices 
    $U^{(\Gamma, \kappa, s)}_{m\mu}$.
\end{enumerate}

\section{Outlook} \label{sec:outlook}

This work presents a general algorithm with which to construct multi-particle interpolating operators for quantum field theories with cubic symmetry, including both lattice and continuum theories. 
The algorithm, together with the implementation in Ref.~\cite{code}, automates the block diagonalization to build multi-particle interpolating operators transforming under irreps of the relevant little group.
Automating this technical component allows the focus of interpolating-operator construction to shift to the design of local and extended operators to access multi-particle states of interest.
It also helps facilitate construction of large operator sets in variational calculations aiming to constrain finite-volume spectra precisely.

These or similar methods can be expected to play an increasingly important role in lattice QCD studies of multi-particle systems.
Especially for systems with multiple baryons, the field has developed rapidly over the past several years as algorithmic advances (e.g., Refs.~\cite{HadronSpectrum:2009krc,Morningstar:2011ka,Detmold:2019fbk,Li:2020hbj,10.1145/3592979.3593409,Humphrey:2022yjc,9820666})
have rendered variational studies a practical reality~\cite{Francis:2018qch,Horz:2020zvv, Green:2021qol,Amarasinghe:2021lqa}.

\acknowledgements{}

The authors gratefully acknowledge useful discussions with Fernando Romero-L{\'o}pez.

WD, WJ and PES are supported in part by the U.S. Department of Energy, Office of Science under grant Contract Number DE-SC0011090 and by the SciDAC5 award DE-SC0023116. This work is supported by the National Science Foundation under Cooperative Agreement PHY-2019786 (The NSF AI Institute for Artificial Intelligence and Fundamental Interactions, http://iaifi.org/). PES is additionally supported by Early Career Award DE-SC0021006 and by Simons Foundation grant 994314 (Simons Collaboration on Confinement and QCD Strings).
GK is supported by funding from the Swiss National Science Foundation (SNSF) through grant agreement no.\ 200020\_200424.
This manuscript has been authored by Fermi Research Alliance, LLC under Contract No. DE-AC02-07CH11359 with the U.S. Department of Energy, Office of Science, Office of High Energy Physics.

The reference implementation of this work makes use of Numpy~\cite{vanderWalt:2011bqk,Harris:2020xlr}, Scipy~\cite{Virtanen:2019joe}, and Sympy~\cite{Meurer:2017yhf}.
Preliminary work and internal verifications were performed with Wolfram Mathematica~\cite{Mathematica}.

\appendix

\section{Group Conventions \label{app:group_conventions}}

This appendix describes the conventions for the groups $O_h$, $O_h^D$, and their subgroups, giving the concrete forms used in the numerical implementation of the algorithm presented in this work, available in Ref.~\cite{code}.
Connections to conventions in the literature are also discussed.

\subsection{The cubic group \texorpdfstring{$O_h$}{Oh}}

For the cubic group $O_h$, any element $R \in O_h$ may be written as a product of a reflection $r$ and a permutation $p$~\cite{Mandula:1983ut},
\begin{equation}
  R = r p,
\end{equation}
with $r$ and $p$ given by
  \begin{equation}
    \begin{split}
      r &\in \{ e, \, r_z, \, r_y, \, r_y r_z, \, r_x, \, r_x r_z, \, r_x r_y, \, r_x r_y r_z \} \\
      p &\in \{ e, \, p_{xy}, \, p_{yz}, \, p_{xz}, \, p_{xyz}, \, p_{xzy} \},
    \end{split}\label{eq:perms_refs}
  \end{equation}
where $e$ is the identity matrix, $r_k$ acts on 3-vectors by multiplying the $k$-th component by $-1$, $p_{ij}$ acts on 3-vectors by permuting their $i$-th and $j$-th components, and the cyclic permutations are defined by $p_{xyz}~=~p_{xy} p_{yz}$ and $p_{xzy}~=~p_{yz} p_{xy}$.
For example,
\begin{equation}
  p_{xy} = \begin{pmatrix} 0&1&0 \\ 1&0&0 \\ 0&0&1 \end{pmatrix}
  \quad \text{and} \quad
  r_x = \begin{pmatrix} -1&0&0 \\ 0&1&0 \\ 0&0&1 \end{pmatrix}.
  \label{eq:Oh_3x3_rep}
\end{equation}
An ordering for the 48 elements of $O_h$ can be established by labeling permutations and reflections as $r_a$ and $p_b$ with $a\in\{1,\ldots,8\}$ and $b\in\{1,\ldots,6\}$ ordered as shown in Eq.~\eqref{eq:perms_refs} and $R_c = r_a p_b$ labeled by $c=6(a-1)+b$ with $c \in \{1,\ldots,48\}$. 
For completeness, the elements are enumerated as:
\begin{equation} \label{eq:Oh-elts}
\begin{aligned}
R_{1} &= e, \hspace{20pt} & R_{2} &=  p_{xy},\\
R_{3} &=  p_{yz}, \hspace{20pt} & R_{4} &=  p_{xz},\\
R_{5} &=  p_{xyz}, \hspace{20pt} & R_{6} &=  p_{xzy},\\
R_{7} &= r_z,  \hspace{20pt} & R_{8} &= r_z p_{xy},\\
R_{9} &= r_z p_{yz}, \hspace{20pt} & R_{10} &= r_z p_{xz},\\
R_{11} &= r_z p_{xyz}, \hspace{20pt} & R_{12} &= r_z p_{xzy},\\
R_{13} &= r_y,  \hspace{20pt} & R_{14} &= r_y p_{xy},\\
R_{15} &= r_y p_{yz}, \hspace{20pt} & R_{16} &= r_y p_{xz},\\
R_{17} &= r_y p_{xyz}, \hspace{20pt} & R_{18} &= r_y p_{xzy},\\
R_{19} &= r_y r_z,  \hspace{20pt} & R_{20} &= r_y r_z p_{xy},\\
R_{21} &= r_y r_z p_{yz}, \hspace{20pt} & R_{22} &= r_y r_z p_{xz},\\
R_{23} &= r_y r_z p_{xyz}, \hspace{20pt} & R_{24} &= r_y r_z p_{xzy},\\
R_{25} &= r_x,  \hspace{20pt} & R_{26} &= r_x p_{xy},\\
R_{27} &= r_x p_{yz}, \hspace{20pt} & R_{28} &= r_x p_{xz},\\
R_{29} &= r_x p_{xyz}, \hspace{20pt} & R_{30} &= r_x p_{xzy},\\
R_{31} &= r_x r_z,  \hspace{20pt} & R_{32} &= r_x r_z p_{xy},\\
R_{33} &= r_x r_z p_{yz}, \hspace{20pt} & R_{34} &= r_x r_z p_{xz},\\
R_{35} &= r_x r_z p_{xyz}, \hspace{20pt} & R_{36} &= r_x r_z p_{xzy},\\
R_{37} &= r_x r_y,  \hspace{20pt} & R_{38} &= r_x r_y p_{xy},\\
R_{39} &= r_x r_y p_{yz}, \hspace{20pt} & R_{40} &= r_x r_y p_{xz},\\
R_{41} &= r_x r_y p_{xyz}, \hspace{20pt} & R_{42} &= r_x r_y p_{xzy},\\
R_{43} &= r_x r_y r_z,  \hspace{20pt} & R_{44} &= r_x r_y r_z p_{xy},\\
R_{45} &= r_x r_y r_z p_{yz}, \hspace{20pt} & R_{46} &= r_x r_y r_z p_{xz},\\
R_{47} &= r_x r_y r_z p_{xyz}, \hspace{20pt} & R_{48} &= r_x r_y r_z p_{xzy}.
\end{aligned}

\end{equation}

The basis functions for the irreps of $O_h$ used in this work are specified in \cref{table:basis_functions}. Irreps are classified by their dimension and eigenvalue ($\pm 1$) under the parity operation $R_{43}: \bm{r} \mapsto -\bm{r}$.
The basis functions for the irreps $A_1^+$, $T_1^-$, $T_2^+$, $E^+$, $A_2^-$ are chosen to match those used in  Ref.~\cite{Basak:2005ir} and correspond to linear combinations of spherical harmonics with $\ell_z$ equal to 0, 1, 2, 2, and 3, respectively.
The basis vectors for the remaining irreps $A_1^-$, $T_1^+$, $T_2^-$, $E^-$, and $A_2^+$ are taken to be linear combinations of the corresponding basis vectors in Ref.~\cite{Dresselhaus:2008}.
The linear combinations are chosen so that the same Clebsch-Gordan coefficients presented in Ref.~\cite{Basak:2005ir} can be used for positive and negative parity irreps in all cases.
Note however that the rows of the $T_1^\pm$ irreps are ordered differently here than in Ref.~\cite{Basak:2005ir} and Clebsch-Gordan coefficient results must be transposed accordingly (cf. conventions in \cref{app:phases} below).
A different set of basis vectors was used for $O_h$ irreps in Ref.~\cite{Morningstar:2013bda}, and the explicit representation matrices obtained in the present work therefore differ from those in Ref.~\cite{Morningstar:2013bda} by a change of basis.

The subgroups of $O_h$ are summarized in \cref{table:group_presentations}.
The present work follows the naming scheme of Ref.~\cite{Dresselhaus:2008} which labels one-dimensional irreps as variants of $A$ or $B$ and two-dimensional irreps as variants of $E$.
It bears emphasizing that irreps of different groups may have identical names, but should be distinguished. 
The Clebsch-Gordan coefficients for little-group irreps below can be deduced from the corresponding Clebsch-Gordan results in Ref.~\cite{Basak:2005ir} for the $O_h$ basis vectors identified with the little-group basis vectors, or they can be calculated directly from the little-group irrep matrices as described for example in Ref.~\cite{Rykhlinskaya}.
As shown in \cref{table:basis_functions}, basis functions for all irreps follow from the irreps of $O_h$.

\bgroup
\def\arraystretch{1.2}
\begin{table*}
\centering
\caption{
    Explicit forms for groups appearing in this work for $O_h$, $O_h^D$, and their subgroups.
    Basis functions for irreps of $O_h$ (and its subgroups) are given in \cref{table:basis_functions}.
    Basis functions for the fermionic irreps of $O_h^D$ (and its subgroups) are given in \cref{table:basis_vectors_double}.
   Group parameterizations for the subgroups of $O_h^D$ follow from those of $O_h$ by the replacements $r_i \to r^D_i$, $p_{ij} \to p^D_{ij}$ and the inclusion of inversions. 
    \label{table:group_presentations}
}
\begin{tabular}{cccccc}
\hline\hline
Momentum & $\littlegroup$ & Order & Irreps & Group Parameterization $rp$ & Group Elements \\
\hline
$\frac{2\pi}{L}(0,0,0)$   & $O_h$     & 48    & $\{A_1^\pm, A_2^\pm , E^\pm, T_1^\pm, T_2^\pm\}$      &\cref{eq:perms_refs}    & \cref{eq:Oh-elts} \\
$\frac{2\pi}{L}(0,0,n)$   & $C_{4v}$  & 8     & $\{A_1, A_2, B_1, B_2, E\}$   & $\{e,r_x,r_y,r_x r_y\} \times \{e, p_{xy}\}$                      & $\{1,2,13,14,25,26,37,38\}$ \\
$\frac{2\pi}{L}(n,n,n)$   & $C_{3v}$  & 6     & $\{A_1, A_2, E\}$             & $\{e\} \times \{e, p_{xy}, p_{yz}, p_{zx}, p_{xyz}, p_{xzy} \}$   & $\{1,2,3,4,5,6\}$ \\
$\frac{2\pi}{L}(0,n,n)$   & $C_{2v}$  & 4     & $\{A_1, A_2, B_1, B_2\}$      & $\{e, r_x\} \times \{e, p_{yz} \}$                                & $\{1,3,25,27\}$\\
$\frac{2\pi}{L}(n,m,0)$   & $C_2^R$   & 2     & $\{A, B\}$                    & $\{e, r_z\} \times \{e\}$                                         & $\{1,7\}$\\
$\frac{2\pi}{L}(n,n,m)$   & $C_2^P$   & 2     & $\{A, B\}$                    & $\{e\} \times \{e, p_{xy}\}$                                      & $\{1,2\}$\\
$\frac{2\pi}{L}(n,m,p)$   & $C_1$     & 1     & $\{A \}$                      & $\{e\} \times \{e\}$ & $\{1\}$\\
\hline
$\frac{2\pi}{L}(0,0,0)$   & $O_h^D$   & 96    & $\{ G_1^\pm, G_2^\pm, H^\pm \}$ & $\{e^D, R^D_{2\pi}\}\times $(doubled \cref{eq:perms_refs})& $\{1,2,\dots, 96\}$\\
$\frac{2\pi}{L}(0,0,n)$   & $\Dic_4$  & 16    & $\{ G_1, G_2 \}$                & $\{e^D, R^D_{2\pi}\}\times$ (doubled $C_{4v}$)& (doubled $C_{4v}$)\\
$\frac{2\pi}{L}(n,n,n)$   & $\Dic_3$  & 12    & $\{ G, F_1, F_2 \}$             & $\{e^D, R^D_{2\pi}\}\times$ (doubled $C_{3v}$)& (doubled $C_{3v}$)\\
$\frac{2\pi}{L}(0,n,n)$   & $\Dic_2$  & 8     & $\{ G \}$                       & $\{e^D, R^D_{2\pi}\}\times$ (doubled $C_{2v}$)& (doubled $C_{2v}$)\\
$\frac{2\pi}{L}(n,m,0)$   & $C_4^R$   & 4     & $\{ F_1, F_2 \}$                & $\{e^D, R^D_{2\pi}\}\times$ (doubled $C_2^R$)& (doubled $C_2^R$)\\
$\frac{2\pi}{L}(n,n,m)$   & $C_4^P$   & 4     & $\{ F_1, F_2 \}$                & $\{e^D, R^D_{2\pi}\}\times$ (doubled $C_2^P$)& (doubled $C_2^P$)\\
$\frac{2\pi}{L}(n,m,p)$   & $C_1^D$   & 2     & $\{ F \}$                       & $\{e^D, R^D_{2\pi}\}\times$ (doubled $C_1$)& (doubled $C_1$)\\
\hline\hline
\end{tabular}
\end{table*}
\egroup
It is useful to make several notes regarding the conventions in \cref{table:basis_functions}. \begin{itemize}
    \item For $C_{4v}$, the irrep names and the coefficients appearing in the $E$ irrep definition are chosen so that identical representation matrices for little group transformations are obtained as those presented in Ref.~\cite{Morningstar:2013bda}. 
    \item For $C_{4v}$, the basis functions for irreps of $O_h$ in \cref{table:basis_functions} are eigenstates of $L_z$, which singles out the $\hat{e}_z$--axis.
    Other choices for the reference momentum, e.g., $\bm{P}' = \frac{2\pi}{L}(n,0,0)$, remain valid but less convenient, since the associated basis functions for $C_{4v}$ must then be permuted.
    \item For $C_{3v}$, the $A_1$ and $A_2$ representation matrices built from these basis vectors using Eq.~\eqref{eq:irrep-matrix-elements} match those explicitly presented in Ref.~\cite{Morningstar:2013bda}. 
    The $E$ representation matrices corresponding to this definition differ from those of Ref.~\cite{Morningstar:2013bda} by interchange of the rows/columns (equivalent to $\basis^{(C_{3v},E)}_1(\bm{r}) \leftrightarrow \basis^{(C_{3v},E)}_2(\bm{r})$) for consistency with the convention of increasing $\ell_z$ with $\mu$ applied here to irreps of $O_h$.
    The conventions adopted here permit the Clebsch-Gordan coefficients for the $A_1$, $A_2$, and $E$ irreps of Ref.~\cite{Basak:2005ir} to be applied to the corresponding irreps of $C_{3v}$.
    \item For $C_{2v}$, the irrep names are chosen so that the representation matrices match those explicitly presented in Ref.~\cite{Morningstar:2013bda}.
    Clebsch-Gordan coefficients for this and other little groups with only one-dimensional irreps are equal to the  Clebsch-Gordan coefficients in Ref.~\cite{Basak:2005ir} for the irreps corresponding to the same basis vectors.
\end{itemize}

\subsection{The double-cover group \texorpdfstring{$O_h^D$}{OhD}
\label{ssec:OhD}}

In the present work, the group $O_h^D$ is defined using the Dirac spinor representation consisting of the direct sum of a positive-parity and a negative-parity spin-$1/2$ state, which provides a faithful representation of the full group of spatial transformations. 

The group elements of $O_h$ can be mapped to (half of) the group elements of $O_h^D$ by replacing rotation operators in the defining representations of $SO(3)$ with the corresponding rotation operators in the Dirac spinor representation.
To do so, first note that the explicit matrix representation of $O_h$ in terms of permutations and reflections in \cref{eq:perms_refs,eq:Oh_3x3_rep} can be (non-uniquely) related to a matrix representation in terms of rotations and the parity operator $P=\text{diag}(-1,-1,-1)$ by
\begin{equation}
  \begin{split}
    r_k &= P \cdot R(\pi \hat{e}_k), \\
    p_{ij} &= P \cdot R(\pi \hat{e}_i)  \cdot R\left(\frac{\pi}{2} \hat{e}_i \times \hat{e}_j\right) .
    \label{eq:perms_refs_rots}
  \end{split}
\end{equation}
Above, the $\hat{e}_i$ are unit vectors in the $i$th direction and $R(\vec{\omega}) \equiv R(\sum_k \omega_k \hat{e}_k)$ describes a rotation by angle $|\vec{\omega}|$ about the $\hat{\omega}$ axis,
\begin{equation}
    R(\vec{\omega}) = \exp\left( - \sum_k \omega_k t_k \right)
\end{equation}
in terms of the $\mathfrak{so}(3)$ generators $[t_k]_{ij} = \varepsilon_{ijk}$.

The Dirac spinor representation of the corresponding element of the double cover is given by
\begin{equation}
  \begin{split}
    R^{D}(\vec{\omega}) &= \exp\left( -\frac{1}{8} \sum_{i,j,k} \omega_k\varepsilon_{ijk} [\gamma_i,\gamma_j] \right), \label{eq:rots_spinor}\\ 
  \end{split}
\end{equation}
where the $\gamma_i$ are the spatial gamma matrices satisfying
$\{\gamma_i,\gamma_j\}=2\delta_{ij}$
and $\gamma_i^\dagger = \gamma_i$ (this choice coincides with both the Euclidean and mostly-positive Minkowski gamma matrices). The superscript $D$ is used to denote double-cover group elements here and below.
The definition in \cref{eq:rots_spinor} implies the transformation property
\begin{equation} \label{eq:spinor_link}
  R^D \gamma^j (R^D)^\dagger = \sum_i \gamma^i R_{ij}.
\end{equation}
The Dirac spinor representation of the parity element $P^D$ is given by the temporal gamma matrix up to an overall phase.
The Euclidean $\gamma_4$ and Minkowski $\gamma_0$ are equivalent up to a phase choice.
The present work takes $P^D = \gamma_4$, which satisfies $\{\gamma_i, \gamma_4\} = 0$, $\gamma_4^\dag = \gamma_4$, and $\gamma_4^2 = 1$.

The double-cover permutations and reflections are then defined in the Dirac spinor representation by
\begin{equation}
  \begin{split}
    r_k^D &= P^D \cdot R^D(\pi \hat{e}_k) \\
    &= \gamma_5 \gamma_k, \\
    p_{ij}^D &= P^D \cdot R^D(\pi \hat{e}_i)  \cdot R^D\left(\frac{\pi}{2} \hat{e}_i \times \hat{e}_j\right) \\
    &= \frac{1}{\sqrt{2}} \gamma_5 (\gamma_i - \gamma_j) , \label{eq:perms_refs_rots_D}
  \end{split}
\end{equation}
where $\gamma_5 = \gamma_1 \gamma_2 \gamma_3 \gamma_4$ is the fifth Euclidean gamma matrix.
The set of products of $r_k^D$ and $p_{ij}^D$ 
analogous to Eq.~\eqref{eq:Oh-elts}
provides an explicit matrix representation of the first $48$ elements of $O_h^D$.
Note that this set is not closed under group multiplication; $O_h$ is not a subgroup of $O_h^D$.
The remaining $48$ elements can be obtained by multiplying these elements by a $2\pi$ rotation,
\begin{equation}
  R^D_{2\pi} = \text{diag}(-1,-1,-1,-1),
\end{equation}
where the form above holds for a $2\pi$ rotation about any axis.

The numerical implementation in Ref.~\cite{code} uses the Dirac-Pauli basis, in which the $\gamma$-matrices are represented in $2\times 2$ block form as
\begin{equation}
\begin{gathered}
  \gamma_k = \begin{pmatrix} 0&-i\sigma_k \\ i\sigma_k&0 \end{pmatrix},
  \quad
  \gamma_4 = \begin{pmatrix} I&0 \\ 0&-I \end{pmatrix}, \\
  \gamma_5 = \begin{pmatrix} 0 & -I \\ -I & 0 \end{pmatrix},
  \label{eq:gamma}
  \end{gathered}
\end{equation}
where the $\sigma_k$ are the usual Pauli matrices and $I$ is the $2\times 2$ identity matrix; for more details and relations to other common bases see Ref.~\cite{Basak:2005ir}.\footnote{Note that the change-of-basis matrix relating the Dirac-Pauli and DeGrand-Rossi bases denoted $U^{(\text{DR})}$ in Ref.~\cite{Basak:2005ir} should be $U^{(\text{DR})} = (-i \gamma_2 + i \gamma_1\gamma_3)/\sqrt{2}$ in terms of Dirac-Pauli matrices.}

To distinguish the notion of abstract group elements $R^D \in O^D_h$ from the specific spinor representation in this basis,  the action of the group on these spinors is denoted by $S(R^D)$, following the convention of Ref.~\cite{Morningstar:2013bda}. 
Because this is the defining representation, the matrix representations are simply given by $S(R^D) = R^D$ using the definitions in \cref{eq:perms_refs_rots_D} and the Dirac-Pauli basis above.

Basis vectors for irreps of $O_h^D$ and its subgroups are summarized in \cref{table:basis_vectors_double}.
The naming convention in the present work follows
Ref.~\cite{Morningstar:2013bda} in denoting 1, 2, and 4 dimensional fermionic irreps by $F$, $G$, and $H$, respectively.
For alternative strategies involving subduction of $O_h^D$ irreps into little-group irreps, see Refs.~\cite{Moore:2005dw,Morningstar:2013bda}.

The only non-trivial basis vectors are associated with 1-dimensional irreps $F_1$ and $F_2$ of $\Dic_3$, which serves as the little group for the momentum $\bm{P} = \frac{2\pi}{L}(n,n,n)$ for any $n\in \Z\setminus\{0\}$.
Basis vectors for $F_1$ and $F_2$ may be obtained by projecting to the linear combinations of $\{ \Ket{\frac{3}{2}, \frac{3}{2}, +}, \Ket{\frac{3}{2}, \frac{1}{2}, +}, \Ket{\frac{3}{2}, -\frac{1}{2}, +}, \Ket{\frac{3}{2}, -\frac{3}{2}, +} \}$ 
transforming in these irreps using Schur's lemma (see Eq.~\eqref{eq:schur-lemma}) and the characters of $F_1$ and $F_2$ given in Ref.~\cite{Morningstar:2013bda}.
It is also possible to identify these irreps from first principles by determining the two-dimensional orthogonal complement of the $G$ irrep contained in the $J=3/2$ representation using Gram-Schmidt orthonormalization, then solving for linear combinations that diagonalize the representation matrices for this orthogonal compliment.
The basis vectors used in this work are:
\begin{align}
\basis^{(\Dic_3,F_1)}_1
    &= \frac{1}{2} \Ket{\frac{3}{2}, \frac{3}{2}, +} - \frac{(1-i)(\sqrt{2}-2i)}{4\sqrt{3}} \Ket{\frac{3}{2}, \frac{1}{2}, +} \nonumber\\
    &\hspace{10pt}  + \frac{\sqrt{2}+i}{2\sqrt{3}} \Ket{\frac{3}{2}, -\frac{1}{2}, +} - \frac{1+i}{2\sqrt{2}} \Ket{\frac{3}{2}, -\frac{3}{2}, +},\label{eq:Dic3_F1} \\
\basis^{(\Dic_3,F_2)}_1 
    &= \frac{1}{2}\Ket{\frac{3}{2}, \frac{3}{2}, +} + \frac{(1+i)(2-i\sqrt{2})}{4\sqrt{3}} \Ket{\frac{3}{2}, \frac{1}{2}, +} \nonumber\\
    &\hspace{10pt} - \frac{\sqrt{2} - i}{2\sqrt{3}} \Ket{\frac{3}{2}, -\frac{1}{2}, +} + \frac{1+i}{2\sqrt{2}} \Ket{\frac{3}{2}, -\frac{3}{2}, +} \label{eq:Dic3_F2}.
\end{align}

\subsection{Phase conventions \label{app:phases}}

The block-diagonalization matrices in \cref{eq:block_diag} are only defined up to an overall phase.
This appendix records the phase conventions used in Ref.~\cite{code}.

For each irrep \emph{except} $T_2^{\pm}$ of the cubic group, the overall phase within each irrep is selected such that the first non-zero entry of 
$U^{(\Gamma_i,\kappa,s)}_{m,\mu=1}$
is real and positive.
For $T_2^{\pm}$ of the cubic group, the overall phase is selected such that
$U^{(\Gamma_i,\kappa,s)}_{m,\mu=2}$
is purely imaginary with negative imaginary part. 
This choice matches the basis-vector conventions of Ref.~\cite{Basak:2005ir}, where the combination of spherical harmonics  $Y_2^2 - Y_2^{-2}$ is used as the $\mu=2$ basis vector for $T_2$.
This choice ensures that the Clebsch-Gordan coefficients presented in Ref.~\cite{Basak:2005ir} are applicable to operators constructed using the methods of the present work (noting the different ordering of the rows of the $T_1^\pm$ irreps discussed above).

\section{Polarization tensors \label{app:polarization}}

This appendix recasts the computation of irrep matrices $D^{(\Gamma)}_{\mu'\mu}(R)$ (see \cref{eq:irrep-matrix-elements}) algebraically using polarization tensors.
For a given irrep $\Gamma$, the basis functions are homogeneous polynomials of fixed degree $\degree$, i.e.,
$\basis^{(\Gamma)}_\mu( \lambda \bm{r}) = \lambda^\degree \basis^{(\Gamma)}_\mu(\bm{r})$ for $\lambda \in \R$.
Homogenous polynomials $\basis^{(\Gamma)}_\mu(\bm{r})$ can be expressed in terms of so-called \emph{polarizations} $\mathcal{P} \basis^{(\Gamma)}_\mu$, symmetric rank-$d$ tensors defined via~\cite{ProcesiLieGroups:2007}
\begin{equation} \label{eq:polarization-iso}
\begin{split}
& \left[ \mathcal{P}\basis^{(\Gamma)}_\mu  \right] (\bm{r}_{(1)},\dots,\bm{r}_{(\degree)})\\
  &\equiv \left(\mathcal{P}\basis^{(\Gamma)}_\mu \right)_{a_1 a_2 \dots a_\degree} r_{(1)}^{a_1} r_{(2)}^{a_2}\dots r_{(d)}^{a_\degree}\\
&\equiv 
  \left.\frac{1}{\degree!} 
\frac{\partial}{\partial \lambda_1}\cdots \frac{\partial}{\partial \lambda_\degree}
\basis^{(\Gamma)}_\mu\left(\lambda_1 \bm{r}_{(1)} + \dots + \lambda_\degree \bm{r}_{(\degree)}\right)\right|_{\lambda_i=0},
\end{split}
\end{equation}
where $\{\bm{r}_{(i)}\in\R^3, 1\leq i \leq \degree \}$ are arbitrary auxiliary vectors.\footnote{%
The identification between basis functions and symmetric tensors amounts to a map between the polynomial ring $K(\{x,y,z\})$ and the symmetric tensor space  $S(\R^{3*})$.
Such a relationship is quite general.
In fact, for any vector space $V$ with basis $B$ and dual space $V^*$, the two spaces are canonically isomorphic:
$K(B)\simeq S(V^*)$~\cite{ProcesiLieGroups:2007}.
}
In the second line, summation is implied over the repeated indices $a_i\in\{1,2,3\}$.
The polarization is symmetric and tensorial due to the symmetry and linearity of the derivatives.
Evaluated diagonally (i.e., contracted with the same vector $\bm{r}$ along all $\degree$ indices), the polarization returns the original homogeneous polynomial:
\begin{equation} \label{eq:polarization-diag}
\begin{split}
\left[\mathcal{P}\basis^{(\Gamma)}_\mu\right] (\bm{r},\dots,\bm{r}) 
  &=  \basis^{(\Gamma)}_\mu(\bm{r}).
\end{split}
\end{equation}
This relationship can provide a useful consistency check in explicit calculations.
In this work, polarizations are normalized with respect to the tensor inner product 
\begin{align}
    \tensornorm{X}{Y} \equiv \left( X^* \right)^{a_1 a_2 \dots a_d} \left( Y \right)_{a_1 a_2 \dots a_d}.
\end{align}
Polarizations normalized with respect to this inner product are denoted by $\bar{B}^{(\Gamma)}_\mu$.

The transformation of $\bar{\basis}^{(\Gamma)}_{\mu}$ under rotations follows immediately from the definition in \cref{eq:polarization-iso},
\begin{equation}
\begin{aligned}
    &\bar{\basis}^{(\Gamma)}_\mu(R^{-1} \bm{r}, \dots, R^{-1} \bm{r}) \\
    &\hspace{20pt}=
    \left( \bar{\basis}^{(\Gamma)}_{\mu}\right)_{a_1 \dots a_\degree}
    R^{\phantom{a_1}a_1}_{b_1} \cdots R^{\phantom{a_\degree}a_\degree}_{b_\degree}
    r^{b_1} \cdots {r}^{b_\degree}\\
    &\hspace{20pt}\equiv
     \left( R \circ \bar{\basis}^{(\Gamma)}_{\mu} \right)_{b_1 \dots b_\degree}
    r^{b_1} \cdots {r}^{b_\degree},
\end{aligned}
\end{equation}
where the second line uses $R^{-1}=R^T$ and the final line defines $R \circ \bar{\basis}^{(\Gamma)}_{\mu} $.
Given this transformation property, the inner product in Eq.~\eqref{eq:irrep-matrix-elements} reduces to
\begin{align}\label{eq:irrep-matrix-elements-2}
    D^{(\Gamma)}_{\mu' \mu}(R) &=
    \tensornorm
        {\bar{\basis}^{(\Gamma)}_{\mu'}}
        {R  \circ \bar{\basis}^{(\Gamma)}_{\mu}  }.
\end{align}
The tensorial method for computing the matrix elements via Eq.~\eqref{eq:irrep-matrix-elements-2} generalizes easily to particles with spin (see \cref{sec:spin}), since spin vectors can be viewed as basis vectors for the double-cover of the relevant little group with transformation properties analogous to Eq.~\eqref{eq:irrep-basis-states}. 

\begin{table*}[p]
    \caption{
    Table of $U^{(\Gamma, \kappa, s)}_{m \mu }$
    for the orbit of $[\bm{n}_1, \bm{n}_2] = [(0,0,1),(0,0,-1)]$.
    Columns of the tables are listed in order of increasing $m \in \{1, \dots, 6\}$.
    Only the momentum of the first operator is shown in the column header. 
    \label{table:basis_100}
    }
    \begin{ruledtabular}
\begin{tabular}{>{\hspace{0.5cm}}cc>{\hspace{0.75cm}}c*{6}{c}}
$\Gamma,\ \kappa$ & $\mu$ & (0, 0, 1) & (0, 1, 0) & (1, 0, 0) & (0, 0, -1) & (0, -1, 0) & (-1, 0, 0) &\\[0.8ex]
\hline\multirow{1}{*}{$A_1^+,1$} & \rule{0pt}{2.5ex}1 & $\frac{1}{\sqrt{6}}$ & $\frac{1}{\sqrt{6}}$ & $\frac{1}{\sqrt{6}}$ & $\frac{1}{\sqrt{6}}$ & $\frac{1}{\sqrt{6}}$ & $\frac{1}{\sqrt{6}}$\\[2.0ex]
\multirow{2}{*}{$E^+,1$} & 1 & $\frac{1}{\sqrt{3}}$ & $-\frac{1}{2 \sqrt{3}}$ & $-\frac{1}{2 \sqrt{3}}$ & $\frac{1}{\sqrt{3}}$ & $-\frac{1}{2 \sqrt{3}}$ & $-\frac{1}{2 \sqrt{3}}$\\
 & 2 & $0$ & $-\frac{1}{2}$ & $\frac{1}{2}$ & $0$ & $-\frac{1}{2}$ & $\frac{1}{2}$\\[2.0ex]
\multirow{3}{*}{$T_1^-,1$} & 1 & $\frac{1}{\sqrt{2}}$ & $0$ & $0$ & $-\frac{1}{\sqrt{2}}$ & $0$ & $0$\\
 & 2 & $0$ & $\frac{i}{2}$ & $-\frac{1}{2}$ & $0$ & $-\frac{i}{2}$ & $\frac{1}{2}$\\
 & 3 & $0$ & $\frac{i}{2}$ & $\frac{1}{2}$ & $0$ & $-\frac{i}{2}$ & $-\frac{1}{2}$\\
\end{tabular}
\end{ruledtabular}

    \vspace{.5cm}

    \caption{
    Table of $U^{(\Gamma, \kappa, s)}_{m \mu}$
    for the orbit of $[\bm{n}_1, \bm{n}_2] = [(0,1,1),(0,-1,-1)]$.
    Columns of the tables are listed in order of increasing $m \in \{1, \dots, 12\}$.
    Only the momentum of the first operator is shown in the column header. 
    \label{table:basis_110}
    }
    \begin{ruledtabular}
\begin{tabular}{>{\hspace{0pt}}cc>{\hspace{0pt}}c*{12}{c}}
$\Gamma,\ \kappa$ & $\mu$ & (1, 1, 0) & (1, 0, 1) & (0, 1, 1) & (1, 0, -1) & (0, 1, -1) & (1, -1, 0) & (0, -1, 1) & (0, -1, -1) & (-1, 1, 0) & (-1, 0, 1) & (-1, 0, -1) & (-1, -1, 0) &\\[0.8ex]
\hline\multirow{1}{*}{$A_1^+,1$} & \rule{0pt}{2.5ex}1 & $\frac{1}{2 \sqrt{3}}$ & $\frac{1}{2 \sqrt{3}}$ & $\frac{1}{2 \sqrt{3}}$ & $\frac{1}{2 \sqrt{3}}$ & $\frac{1}{2 \sqrt{3}}$ & $\frac{1}{2 \sqrt{3}}$ & $\frac{1}{2 \sqrt{3}}$ & $\frac{1}{2 \sqrt{3}}$ & $\frac{1}{2 \sqrt{3}}$ & $\frac{1}{2 \sqrt{3}}$ & $\frac{1}{2 \sqrt{3}}$ & $\frac{1}{2 \sqrt{3}}$\\[2.0ex]
\multirow{2}{*}{$E^+,1$} & 1 & $\frac{1}{\sqrt{6}}$ & $-\frac{1}{2 \sqrt{6}}$ & $-\frac{1}{2 \sqrt{6}}$ & $-\frac{1}{2 \sqrt{6}}$ & $-\frac{1}{2 \sqrt{6}}$ & $\frac{1}{\sqrt{6}}$ & $-\frac{1}{2 \sqrt{6}}$ & $-\frac{1}{2 \sqrt{6}}$ & $\frac{1}{\sqrt{6}}$ & $-\frac{1}{2 \sqrt{6}}$ & $-\frac{1}{2 \sqrt{6}}$ & $\frac{1}{\sqrt{6}}$\\
 & 2 & $0$ & $-\frac{1}{2 \sqrt{2}}$ & $\frac{1}{2 \sqrt{2}}$ & $-\frac{1}{2 \sqrt{2}}$ & $\frac{1}{2 \sqrt{2}}$ & $0$ & $\frac{1}{2 \sqrt{2}}$ & $\frac{1}{2 \sqrt{2}}$ & $0$ & $-\frac{1}{2 \sqrt{2}}$ & $-\frac{1}{2 \sqrt{2}}$ & $0$\\[2.0ex]
\multirow{3}{*}{$T_2^+,1$} & 1 & $0$ & $-\frac{1}{2 \sqrt{2}}$ & $\frac{i}{2 \sqrt{2}}$ & $\frac{1}{2 \sqrt{2}}$ & $-\frac{i}{2 \sqrt{2}}$ & $0$ & $-\frac{i}{2 \sqrt{2}}$ & $\frac{i}{2 \sqrt{2}}$ & $0$ & $\frac{1}{2 \sqrt{2}}$ & $-\frac{1}{2 \sqrt{2}}$ & $0$\\
 & 2 & $-\frac{i}{2}$ & $0$ & $0$ & $0$ & $0$ & $\frac{i}{2}$ & $0$ & $0$ & $\frac{i}{2}$ & $0$ & $0$ & $-\frac{i}{2}$\\
 & 3 & $0$ & $\frac{1}{2 \sqrt{2}}$ & $\frac{i}{2 \sqrt{2}}$ & $-\frac{1}{2 \sqrt{2}}$ & $-\frac{i}{2 \sqrt{2}}$ & $0$ & $-\frac{i}{2 \sqrt{2}}$ & $\frac{i}{2 \sqrt{2}}$ & $0$ & $-\frac{1}{2 \sqrt{2}}$ & $\frac{1}{2 \sqrt{2}}$ & $0$\\[2.0ex]
\multirow{3}{*}{$T_1^-,1$} & 1 & $0$ & $\frac{1}{2 \sqrt{2}}$ & $\frac{1}{2 \sqrt{2}}$ & $-\frac{1}{2 \sqrt{2}}$ & $-\frac{1}{2 \sqrt{2}}$ & $0$ & $\frac{1}{2 \sqrt{2}}$ & $-\frac{1}{2 \sqrt{2}}$ & $0$ & $\frac{1}{2 \sqrt{2}}$ & $-\frac{1}{2 \sqrt{2}}$ & $0$\\
 & 2 & $-\frac{1}{4}+\frac{i}{4}$ & $-\frac{1}{4}$ & $\frac{i}{4}$ & $-\frac{1}{4}$ & $\frac{i}{4}$ & $-\frac{1}{4}-\frac{i}{4}$ & $-\frac{i}{4}$ & $-\frac{i}{4}$ & $\frac{1}{4}+\frac{i}{4}$ & $\frac{1}{4}$ & $\frac{1}{4}$ & $\frac{1}{4}-\frac{i}{4}$\\
 & 3 & $\frac{1}{4}+\frac{i}{4}$ & $\frac{1}{4}$ & $\frac{i}{4}$ & $\frac{1}{4}$ & $\frac{i}{4}$ & $\frac{1}{4}-\frac{i}{4}$ & $-\frac{i}{4}$ & $-\frac{i}{4}$ & $-\frac{1}{4}+\frac{i}{4}$ & $-\frac{1}{4}$ & $-\frac{1}{4}$ & $-\frac{1}{4}-\frac{i}{4}$\\[2.0ex]
\multirow{3}{*}{$T_2^-,1$} & 1 & $\frac{1}{4}+\frac{i}{4}$ & $-\frac{i}{4}$ & $-\frac{1}{4}$ & $-\frac{i}{4}$ & $-\frac{1}{4}$ & $-\frac{1}{4}+\frac{i}{4}$ & $\frac{1}{4}$ & $\frac{1}{4}$ & $\frac{1}{4}-\frac{i}{4}$ & $\frac{i}{4}$ & $\frac{i}{4}$ & $-\frac{1}{4}-\frac{i}{4}$\\
 & 2 & $0$ & $-\frac{i}{2 \sqrt{2}}$ & $\frac{i}{2 \sqrt{2}}$ & $\frac{i}{2 \sqrt{2}}$ & $-\frac{i}{2 \sqrt{2}}$ & $0$ & $\frac{i}{2 \sqrt{2}}$ & $-\frac{i}{2 \sqrt{2}}$ & $0$ & $-\frac{i}{2 \sqrt{2}}$ & $\frac{i}{2 \sqrt{2}}$ & $0$\\
 & 3 & $-\frac{1}{4}+\frac{i}{4}$ & $-\frac{i}{4}$ & $\frac{1}{4}$ & $-\frac{i}{4}$ & $\frac{1}{4}$ & $\frac{1}{4}+\frac{i}{4}$ & $-\frac{1}{4}$ & $-\frac{1}{4}$ & $-\frac{1}{4}-\frac{i}{4}$ & $\frac{i}{4}$ & $\frac{i}{4}$ & $\frac{1}{4}-\frac{i}{4}$\\
\end{tabular}
\end{ruledtabular}

    \vspace{.5cm}

    \caption{
    Table of $U^{(\Gamma, \kappa, s)}_{m \mu}$ for the orbit of $[\bm{n}_1, \bm{n}_2] = [(1,1,1),(-1,-1,-1)]$.
    Columns of the tables are listed in order of increasing $m \in \{1, \dots, 8\}$.
    Only the momentum of the first operator is shown in the column header. 
    \label{table:basis_111}
    }
    \begin{ruledtabular}
\begin{tabular}{>{\hspace{0.1cm}}cc>{\hspace{0.2cm}}c*{8}{c}}
$\Gamma,\ \kappa$ & $\mu$ & (1, 1, 1) & (1, 1, -1) & (1, -1, 1) & (1, -1, -1) & (-1, 1, 1) & (-1, 1, -1) & (-1, -1, 1) & (-1, -1, -1) &\\[0.8ex]
\hline\multirow{1}{*}{$A_1^+,1$} & \rule{0pt}{2.5ex}1 & $\frac{1}{2 \sqrt{2}}$ & $\frac{1}{2 \sqrt{2}}$ & $\frac{1}{2 \sqrt{2}}$ & $\frac{1}{2 \sqrt{2}}$ & $\frac{1}{2 \sqrt{2}}$ & $\frac{1}{2 \sqrt{2}}$ & $\frac{1}{2 \sqrt{2}}$ & $\frac{1}{2 \sqrt{2}}$\\[2.0ex]
\multirow{3}{*}{$T_2^+,1$} & 1 & $-\frac{1}{4}+\frac{i}{4}$ & $\frac{1}{4}-\frac{i}{4}$ & $-\frac{1}{4}-\frac{i}{4}$ & $\frac{1}{4}+\frac{i}{4}$ & $\frac{1}{4}+\frac{i}{4}$ & $-\frac{1}{4}-\frac{i}{4}$ & $\frac{1}{4}-\frac{i}{4}$ & $-\frac{1}{4}+\frac{i}{4}$\\
 & 2 & $-\frac{i}{2 \sqrt{2}}$ & $-\frac{i}{2 \sqrt{2}}$ & $\frac{i}{2 \sqrt{2}}$ & $\frac{i}{2 \sqrt{2}}$ & $\frac{i}{2 \sqrt{2}}$ & $\frac{i}{2 \sqrt{2}}$ & $-\frac{i}{2 \sqrt{2}}$ & $-\frac{i}{2 \sqrt{2}}$\\
 & 3 & $\frac{1}{4}+\frac{i}{4}$ & $-\frac{1}{4}-\frac{i}{4}$ & $\frac{1}{4}-\frac{i}{4}$ & $-\frac{1}{4}+\frac{i}{4}$ & $-\frac{1}{4}+\frac{i}{4}$ & $\frac{1}{4}-\frac{i}{4}$ & $-\frac{1}{4}-\frac{i}{4}$ & $\frac{1}{4}+\frac{i}{4}$\\[2.0ex]
\multirow{1}{*}{$A_2^-,1$} & 1 & $\frac{1}{2 \sqrt{2}}$ & $-\frac{1}{2 \sqrt{2}}$ & $-\frac{1}{2 \sqrt{2}}$ & $\frac{1}{2 \sqrt{2}}$ & $-\frac{1}{2 \sqrt{2}}$ & $\frac{1}{2 \sqrt{2}}$ & $\frac{1}{2 \sqrt{2}}$ & $-\frac{1}{2 \sqrt{2}}$\\[2.0ex]
\multirow{3}{*}{$T_1^-,1$} & 1 & $\frac{1}{2 \sqrt{2}}$ & $-\frac{1}{2 \sqrt{2}}$ & $\frac{1}{2 \sqrt{2}}$ & $-\frac{1}{2 \sqrt{2}}$ & $\frac{1}{2 \sqrt{2}}$ & $-\frac{1}{2 \sqrt{2}}$ & $\frac{1}{2 \sqrt{2}}$ & $-\frac{1}{2 \sqrt{2}}$\\
 & 2 & $-\frac{1}{4}+\frac{i}{4}$ & $-\frac{1}{4}+\frac{i}{4}$ & $-\frac{1}{4}-\frac{i}{4}$ & $-\frac{1}{4}-\frac{i}{4}$ & $\frac{1}{4}+\frac{i}{4}$ & $\frac{1}{4}+\frac{i}{4}$ & $\frac{1}{4}-\frac{i}{4}$ & $\frac{1}{4}-\frac{i}{4}$\\
 & 3 & $\frac{1}{4}+\frac{i}{4}$ & $\frac{1}{4}+\frac{i}{4}$ & $\frac{1}{4}-\frac{i}{4}$ & $\frac{1}{4}-\frac{i}{4}$ & $-\frac{1}{4}+\frac{i}{4}$ & $-\frac{1}{4}+\frac{i}{4}$ & $-\frac{1}{4}-\frac{i}{4}$ & $-\frac{1}{4}-\frac{i}{4}$\\
\end{tabular}
\end{ruledtabular}

\end{table*}

An explicit example of polarization tensors in this context follows.
\Cref{table:basis_functions} gives the basis function for $\ell_z = 1$ row of the irrep $T_2^+$ of $O_h$: $\basis^{(T_2^+)}_{1}(\bm{r}) = (-zx + iyz)/\sqrt{2}$.
This function is a homogeneous degree-2 polynomial, so the polarization will be a rank-2 tensor.
Using the coordinates $\bm{r}_{(i)}=(x_{(i)}, y_{(i)}, z_{(i)})$ for auxiliary vectors, the polarization is
\begin{align}
    &\left[\mathcal{P} \bar{\basis}^{(T_2^+)}_{1}\right](\bm{r}_{(1)}, \bm{r}_{(2)}) = \\
    &\hspace{20pt} \frac{1}{2\sqrt{2}} \Big[
    -z_{(1)} x_{(2)} - x_{(1)} z_{(2)}
    + i y_{(1)} z_{(2)} + i z_{(1)} y_{(2)} \Big]. \nonumber
\end{align}
The nonzero components of the normalized basis tensor are
\begin{equation}
\begin{aligned}
  &\left(\bar{\basis}^{(T_2^+)}_{1}\right)_{13} = \left(\bar{\basis}^{(T_2^+)}_{1}\right)_{31} = -\frac{1}{2}, \\ &\left(\bar{\basis}^{(T_2^+)}_{1}\right)_{23} = \left(\bar{\basis}^{(T_2^+)}_{1}\right)_{32} = \frac{i}{2}.
\end{aligned}
\end{equation}

As an example of the algebraic setup, consider the transformation about the $y$-axis sending $x\to z$ and $z \to -x$:
\begin{align}
R = \begin{pmatrix}
    0 & 0 & 1 \\
    0 & 1 & 0 \\
    -1 & 0 & 0
\end{pmatrix}.
\end{align}
The matrix element of this rotation between $\bar{B}_1^{(T_2^+)}$ and itself is given by
\begin{align}
    D^{(T_2^+)}_{11}(R) 
    & = \tensornorm
        {\bar{\basis}^{(T_2^+)}_{1}}
        {R  \circ \bar{\basis}^{(T_2^+)}_{1} }\\
    & = 
    \tr \left[
    \begin{psmallmatrix}
    0 & 0 & -1/2\\
    0 & 0 & -i/2\\
    -1/2 & -i/2 & 0\\
    \end{psmallmatrix}
    \begin{psmallmatrix}
    0 & i/2 & 1/2\\
    i/2 & 0 & 0\\
    1/2 & 0 & 0
    \end{psmallmatrix}
    \right]\\
    &= -1/2.
\end{align}
The equivalent calculation in the integral setup is
\begin{align}
    &D^{(T_2^+)}_{11}(R) =
    \frac{8}{\pi^2} \int d\Omega \,
    \left( \tfrac{-z x + i y z}{\sqrt{2}}\right)^*
    \left( \tfrac{x z + i y x}{\sqrt{2}}\right) \\
    &= \frac{-4}{\pi^2} \int d\Omega \,
    e^{i \varphi } \sin^2\theta \cos\theta  \cos\varphi (\cos\theta +i\sin\theta \sin\varphi) \nonumber \\
    &= -1/2.
\end{align}
Although the integral in the second line is elementary, evaluating many such integrals is cumbersome when compared to tensor algebra.\\
\\ 

\section{Explicit block-diagonalization matrices\label{app:cob}}

This appendix presents examples of block-diagonalization matrices in the rest frame in a few illustrative cases.
The results in \cref{table:basis_100,table:basis_110,table:basis_111} employ the phase conventions in \cref{app:phases}.

\bibliography{interpolators.bib}

\end{document}